**Chemotaxing *E. coli* do not count single molecules**

Henry H. Mattingly†,*,1, Keita Kamino†,2,3,4, Jude Ong‡,5, Rafaela Kottou‡,5, Thierry Emonet*,5,6,7, Benjamin B. Machta*,6,7

[1] Center for Computational Biology, Flatiron Institute

[2] Institute of Molecular Biology, [3]Institute of Physics, Academia Sinica

[4]Institute of Physics, National Yang Ming Chiao Tung University

[5] Molecular, Cellular, and Developmental Biology, [6] Physics, and [7] QBio Institute, Yale University

† These authors contributed equally.

‡ These authors contributed equally.

* Correspondence to: Benjamin.machta@yale.edu, Thierry.emonet@yale.edu, Hmattingly@flatironinstitute.org.

**Abstract**

Organisms use specialized sensors to measure their environments, but the fundamental principles that determine their accuracy remain largely unknown. In *Escherichia coli* chemotaxis, we previously found that gradient-climbing speed is bounded by the amount of information that cells acquire from their environment, and that *E. coli* operate near this bound. However, it remains unclear what prevents them from acquiring more information. Past work argued that *E. coli*'s chemosensing is limited by the physics of molecules stochastically arriving at cells' receptors, without direct evidence. Here, we show instead that *E. coli* are far from this physical limit. To show this, we develop a theoretical approach that uses information rates to quantify how accurately behaviorally-relevant signals can be estimated from available observations: molecule arrivals for the physical limit; chemotaxis signaling activity for *E. coli*. Measuring these information rates in single-cell experiments across multiple background concentrations, we find that *E. coli* encode two orders of magnitude less information than the physical limit. Thus, *E. coli* chemosensing is limited by internal noise in signal processing rather than the physics of molecule diffusion, motivating investigation of what specific physical and biological constraints shaped the evolution of this prototypical sensory system.

**Introduction**

Evolution selects function, and therefore living systems are shaped by complex fitness objectives and constraints. This has motivated the use of normative theories, subject only to constraints of physics, to derive fundamental limits on function and to rationalize the design of biological systems (1–17). This approach has been especially successful in the context of information processing, a hallmark of living systems where theories of optimal estimation can be brought to bear (18,19). However, biology needs to implement information processing and other functions using non-ideal components, in the confines of a body, and with limited resources, which introduce additional system-specific constraints (20–27). Determining what bounds or constraints meaningfully limit information processing in a particular biological system would shed light on the forces that have shaped its evolution, and inform our understanding of biological information processing more broadly.

*Escherichia coli* chemotaxis is an ideal system for studying the limits on biological information processing (28–30). *E. coli* climb chemical gradients by alternating between straight-swimming runs and randomly-reorienting tumbles (31). As they swim, they measure the time-dependent concentration of attractant along their trajectory, $c(t)$, using transmembrane receptors, encode these measurements into the activity of intracellular, receptor-associated CheA kinase activity, $a(t)$, and act on these measurements to decide when to tumble (Fig. 1). Importantly, chemotaxis provides a fitness advantage, even above undirected motility, in structured chemical environments (32).

*E. coli* must acquire information about their chemical environment in order to climb gradients. Recently, we asked how fast an ideal bacterium can climb a gradient with the information it gets, and how *E. coli* compare to this theoretical performance bound (17). We found that although typical *E. coli* cells get very little information about chemical signals—about 0.01 bits/s in a centimeter-long gradient—they climb gradients at speeds near the theoretical maximum with the information they get. Thus, information is functionally important for chemotaxis.

This raises the question: why don't *E. coli* get more information, and thus climb gradients faster? One possibility is that they are limited by fundamental physics. The first physically-measurable quantity is the rate of ligand molecule arrivals at the cell's receptors by diffusion, $r(t)$ (Fig. 1). In a classic paper (4), Berg and Purcell demonstrated that the stochasticity of this arrival rate limits the accuracy of any estimate of chemical concentration, $c(t)$, inspiring an entire field of biophysics (20,21,33–47). They and others further argued that bacteria approach this physical limit, a widely-held understanding in the field. However, no direct comparison between bacterial chemosensing and physical limits has been made because it has remained unclear how to quantify a real cell's uncertainty about external signals. This leaves open the alternative possibility that *E. coli*'s sensory information might be limited by system-specific, internal constraints.

Directly answering whether physical limits or internal constraints prevent chemotaxing *E. coli* from acquiring more information faces several general challenges. First, not all environmental signals are useful for function. For chemotaxis in shallow gradients, we recently showed that the time derivative of (log) concentration, $s(t) = \frac{d}{dt}\log(c)$—rather than concentration, $c(t)$, itself—is the "behaviorally-relevant" signal (17). Second, cells do not need to represent their estimates of relevant signals in a straightforward way. In chemotaxis, CheA kinase activity depends on external signals, but $a(t)$ is not necessarily the cell's estimate of $s(t)$, and the variation in $a(t)$ is not the cell's uncertainty about $s(t)$. Signals are instead encoded in the dynamics of the cell's intermediate variables and decoded by downstream processing. Third, probing cells' encodings of time-varying signals requires dynamic experimental measurements of both the environment and responses in single cells, which was recently made possible by single-cell FRET (17,48–52).

Here, we address these challenges and determine whether *E. coli* chemosensing approaches the physical limits. To frame this question in an experimentally-testable way, we ask how accurately the signal $s(t)$ can be inferred from molecule arrivals, which sets the physical limit, compared to how accurately $s(t)$ can be inferred from the dynamics of kinase activity, the quantity accessible to the cell. Sensing accuracy in each case takes the form of an information rate (Fig. 1). Then, we quantify these information rates using single-cell FRET measurements in multiple background concentrations. We find, surprisingly, that a typical *E. coli* cell gets orders of magnitude less information than the physical limit—estimates of signal made from kinase activity are far less accurate than those made from molecule arrival rate. This is because *E. coli*'s signal transduction noise far exceeds molecule arrival noise, and we conclude that information processing during *E. coli* chemotaxis is *internally-limited*. We predict that the functional consequence is that *E. coli* climb gradients much slower than the physical limits on chemosensing allow, and support this with simulations. These results raise questions about what specific constraints limit *E. coli*'s chemosensing, and more broadly motivate consideration of the physical and biological constraints on information processing.

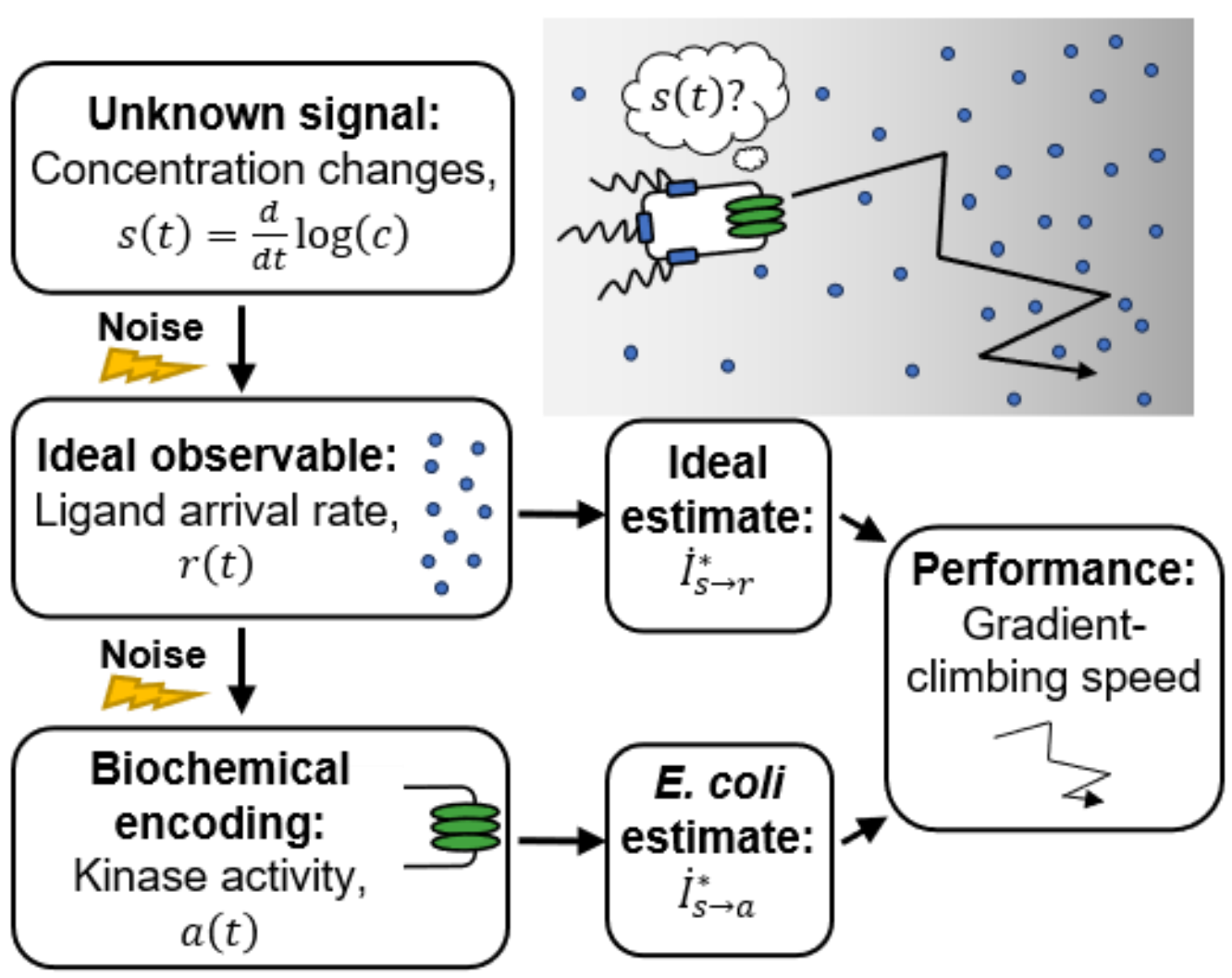

**Figure 1: Is *E. coli*'s sensing accuracy set by physical limits or internal constraints?** To climb chemical gradients, *E. coli* need to accurately estimate an unknown signal: the rate of change of attractant ligand concentration, $s(t) = \frac{d}{dt}\log\big(c(t)\big)$ (17). The first physically-observable quantity is the stochastic rate at which ligand molecules arrive at the cell's receptors, $r(t)$ (4). Thus, the physical limit on chemosensing, and in turn gradient-climbing speed, is set by how accurately $s(t)$ can be estimated from the time series of past $r(t)$, quantified by an information rate, $\dot{I}^*_{s\to r}$. *E. coli* respond to ligand arrivals with changes in the activity of intracellular CheA kinases, $a(t)$. The accuracy with which the signal can be estimated from kinase activity is quantified by another information rate, $\dot{I}^*_{s\to a}$. Since kinase activity is stochastic, *E. coli*'s sensing accuracy and gradient climbing speed must be below the physical limit, but how much less?

**Physical limit on behaviorally-relevant information due to stochastic molecule arrivals**

To climb chemical gradients, *E. coli* must encode information about the time derivative of concentration, $s(t)$, to be read out by the motors (17) (SI section "Drift speed and information rate"). The first quantity that is observable to the cell and informative of $s(t)$ is the stochastic arrival rate of ligand molecules at the cell's receptors, $r(t)$ (Fig. 1). An ideal agent would estimate $s(t)$ and make navigation decisions based on perfect observations of past particle arrivals $\{r\}$. The behaviorally-relevant information about signal, $s(t)$, thus acquired from past particle arrivals, $\{r\}$, is quantified by the following transfer entropy rate (53):

$$\dot{I}^*_{s\to r} \equiv \lim_{dt\to 0} \frac{1}{dt} I(r(t+dt); s(t)|\{r\}) \,, \tag{1}$$

where $I(X;Y|Z)$ is the mutual information between $X$ and $Y$, conditioned on $Z$ (54,55). This quantity obeys a data processing inequality (55) in the context of bacterial chemotaxis in shallow gradients, where feedback from behavior onto signals is negligible (56,57) (SI section "Data processing inequality"). Therefore this quantity sets the physical limit on information available in any downstream encoding of the signal, including *E. coli*'s kinase activity.

The form of the physical limit in Eqn. 1 is unknown. To derive it, we first need a dynamical model for the signal and the particle arrival rate. In static gradients, the signals a cell experiences are determined by their own run-and-tumble motion in the gradient. Accordingly, in a gradient of steepness $g = d\log(c)/dx$, the signal is $s(t) = g\, v_x(t)$, where $v_x$ is the cell's up-gradient velocity. In shallow gradients, where weak signals have small effects on the cell's run-tumble statistics, we can rigorously approximate $s(t)$ as Gaussian with correlation function $\langle s(t)\, s(t')\rangle = g^2\, V(t-t') \approx g^2\, \sigma_v^2 \exp\left(-\frac{|t-t'|}{\tau_v}\right)$, to leading order in $g$ (17,22). Here, $V(t)$ is the correlation function of $v_x$ in the absence of a gradient; $\sigma_v^2$ is the variance of $v_x$, which depends on the cell's swimming speed; and $\tau_v$ is the signal correlation time, which depends on the cell's mean run duration, the persistence of tumbles, and rotational diffusion (17,58).

Molecule arrival events follow a Poisson process with time-varying rate $\langle r(t)\rangle = k_D\, c(t) = 4\, D\, l\, c(t)$, where $D \approx 800\ \mu\mathrm{m}^2$/s (59,60) is the diffusivity of the ligand and $l$ is the radius of a circular sensor on the cell's surface (4,42). We choose $l \approx 60\ \mathrm{nm}$ (61) to match the size of the receptor array in *E. coli*'s cell membrane. These give $k_D \approx 1.2 \times 10^5\ \mathrm{s}^{-1}\ \mu\mathrm{M}^{-1}$, which is comparable to previous estimates (4,62). If many molecules arrive per run, $r_0\, \tau_v \gg 1$, we can approximate the Poisson process for arrival events with a Gaussian process for the number of molecule arrivals per unit time, $r(t) = k_D\, c(t) + \sqrt{r_0}\ \xi(t)$. Here, $r_0 = k_D\, c_0$ is the background molecule arrival rate, $c_0$ is the background concentration, and the noise is $\langle \xi(t)\, \xi(t')\rangle = \delta(t-t')$. We assume the sensor absorbs every molecule it senses (4), but if it cannot distinguish between new ligand arrivals and rebinding events, the limit is lower by an $O(1)$ prefactor (42,43).

We next focused on calculating the behaviorally-relevant information quantity in Eqn. 1. Towards this, we discovered that the transfer entropy rate in Eqn. 1 is equivalent to a predictive information rate (22,23,63–66) (SI section "Equivalence of transfer entropy and predictive information rates"):

$$\dot{I}^*_{s\to r} = -[\partial_\tau I(s(t+\tau); \{r\})]_{\tau=0}. \tag{2}$$

This quantifies how the ability to predict future signals $s(t+\tau)$ from past particle arrivals degrades as the forecasting interval $\tau$ increases, and is evaluated at the current moment, $\tau = 0$. Importantly, this quantity only quantifies the information that is relevant for climbing the gradient. Therefore it is different from the total information encoded by *E. coli*'s signaling pathway about *all* past signals, $\{s\}$, both relevant (current signal) and irrelevant (signal experienced in the past), that we and others studied previously (17,57,67,68).

Since $s(t)$ and $\{r\}$ are approximately Gaussian, the physical limit in Eqn. 2 only depends on the posterior variance, $\sigma^2_{s|r}$, of $s(t)$ given past particle arrivals $\{r\}$ (SI Eqn. 23), which can be derived using causal Wiener filtering theory (22,64,69–74) (SI section "Derivation of the physical limit on behaviorally-relevant information for chemotaxis"). We find that the physical limit on behaviorally-relevant information for chemotaxis in shallow gradients is:

$$\dot{I}^*_{s\to r} \approx \frac{1}{\tau_v}\rho^2_{rs} \approx \frac{1}{\tau_v}\frac{1}{4}\gamma_r, \tag{3}$$

where $\rho_{rs}$ is the Peason correlation coefficient between the true signal $s(t)$ and the optimal estimator of $s(t)$ constructed from past molecule arrivals, $\hat{s}_r(t)$. Here, we defined the dimensionless signal-to-noise ratio of molecule arrivals, $\gamma_r = 2\, r_0\, g^2\, \sigma_v^2\, \tau_v^3$. Eqn. 3 is valid when $\gamma_r \ll 1$, which sets the small-signal regime for $\dot{I}^*_{s\to r}$. We also provide a full expression for $\dot{I}^*_{s\to r}$ in the SI (SI Eqn. 46), and we validate our expression for $\rho^2_{rs}$ using simulations (SI Fig. S5). Increasing the background $r_0$, the gradient steepness $g$, or the swimming speed $\sigma_v$ increases the signal-to-noise ratio of molecule arrivals. Longer runs, $\tau_v$, also increases $\dot{I}^*_{s\to r}$ by allowing more time to average out noise. We expect spatial sensing across the cell body to be negligible compared to temporal sensing, as argued by Berg and Purcell (SI section "Comparing temporal and spatial sensing"; see also (75)). The derivation of $\dot{I}^*_{s\to r}$ also provides the optimal kernel for constructing $\hat{s}_r(t)$, which we discuss in the SI (section "Optimal kernel for estimating signal from particle arrivals").

**Relevant information encoded in *E. coli*'s CheA kinase activity**

In *E. coli*, ligand binding to receptors modulates the activity of the CheA kinases in the receptor-kinase complex. Thus, kinase acivity $a(t)$ depends on past signals $s(t)$, but is not necessarily the cell's representation of them. To compare *E. coli* to the theoretical limit, we next derive $\dot{I}^*_{s\to a}$, which quantifies how well $s(t)$ can be estimated from the dynamics of kinase activity. For this, we need models of kinase responses to ligand molecule arrivals and noise in kinase activity. In shallow gradients, our approach is to use linear, Gaussian theory, which has been validated experimentally (17,48,49) and computationally (68). For a cell with steady-state kinase activity $a_0$ in background $r_0$, kinase responses are described by linear response theory (17,76,77) as follows:

$$a(t) = a_0 - \int_{-\infty}^{t} K_r(t-t')\,(r(t') - r_0)\,dt' + \eta_n(t). \tag{4}$$

*E. coli* respond to a step increase in attractant concentration with a fast drop in kinase activity, followed by slow adaptation back to the pre-stimulus level (78). We model this phenomenologically with response function $K_r(t) = G_r\left(\frac{1}{\tau_1}\exp\left(-\frac{t}{\tau_1}\right) - \frac{1}{\tau_2}\exp\left(-\frac{t}{\tau_2}\right)\right)\Theta(t)$, where $G_r$ is the gain of the response to molecule arrival rate $r$, $\tau_1$ is the fast response time, $\tau_2$ is the slow adaptation time, and $\Theta(t)$ is the Heaviside step function. Kinase responses can equivalently be expressed in terms of past signals $s$, with a related kernel $K(t)$ that we used previously (17) ($K_r(t) = \frac{1}{r_0}\frac{d}{dt}K(t)$; SI Eqn. 78).

Noise in kinase activity is driven by a combination of stochastic molecule arrivals and internally-driven fluctuations. Previous single-cell FRET experiments have observed large, slow fluctuations in kinase activity, $\eta_n(t)$, on a time scale of 10 s (17,48,49,79). These are well-described as Gaussian, with correlation function $\langle \eta_n(t)\, \eta_n(t') \rangle = D_n\, \tau_n\, \exp\left(-\frac{|t-t'|}{\tau_n}\right)$. Here, $D_n$ is the diffusivity of internal noise in kinase activity, and $\tau_n$ is its correlation time. In addition, Eqn. 4 has additive noise arising from responses to molecule arrival noise. To date, it has not been possible to measure kinase fluctuations on time scales shorter than the CheY-CheZ relaxation time ($\tau_1$), but it cannot go below the level set by responses to molecule arrival noise. Thus, the phenomenological model above agrees with experiments at low frequencies while obeying known physics at high frequencies.

With the relation between transfer entropy and predictive information in Eqn. 2, evaluating $\dot{I}^*_{s\to a}$ again reduces to deriving the posterior variance, $\sigma^2_{s|a}$, of the signal $s(t)$ given past kinase activity $\{a\}$ (SI section "Derivation of the behaviorally-relevant information in kinase activity"). Furthermore, previous measurements (and measurements below) show that $\tau_1 \ll \tau_v$ (17,80,81) and $\tau_2 \approx \tau_n \gg \tau_1$ (17). Thus, in shallow gradients, we find that the information rate encoded in kinase activity is:

$$\dot{I}^*_{s\to a} \approx \frac{1}{\tau_v}\rho^2_{as} \approx \frac{1}{\tau_v}\frac{1}{4}\,\gamma_a \frac{\gamma_r/\gamma_a}{\left(1+\sqrt{\gamma_r/\gamma_a}\right)^2}. \quad (5)$$

where $\rho_{as}$ is the Peason correlation coefficient between the true signal $s(t)$ and the optimal estimator of $s(t)$ constructed from past kinase activity, $\hat{s}_a(t)$. Here, we define the dimensionless kinase activity signal-to-noise ratio $\gamma_a = \frac{G_r^2}{D_n}\, r_0^2\, g^2\, \sigma_v^2\, \tau_v$. Eqn. 5 is valid when $\gamma_a \ll 1$, which sets the small-signal regime for $\dot{I}^*_{s\to a}$. We also provide a full expression for $\dot{I}^*_{s\to a}$ in the SI (SI Eqn. 108), and we validate our expression for $\rho^2_{as}$ using simulations (SI Fig. S5). An ideal sensor with no internal noise corresponds to $\gamma_a \to \infty$. Taking this limit in Eqn. 5 results in the expression for $\dot{I}^*_{s\to r}$ in Eqn. 3. Conversely, internal noise degrades information about the signal, and the information rate becomes $\dot{I}^*_{s\to a} \approx \frac{1}{\tau_v}\frac{1}{4}\,\gamma_a$ as $\gamma_a/\gamma_r \to 0$. The derivation of $\dot{I}^*_{s\to a}$ also provides the optimal kernel for constructing $\hat{s}_a(t)$, which we discuss in the SI (section "Optimal kernel for estimating signal from kinase activity").

## Single-cell measurements constrain signal and kinase properties

To quantify the information rates above, we then performed single-cell tracking and FRET experiments to measure the parameters characterizing the signal statistics, kinase response function, and kinase noise statistics. As the attractant, we used aspartate (Asp), to which the *E. coli* chemotaxis signaling pathway responds with the highest sensitivity among known attractants (82).

To quantify the signal statistics, we recorded trajectories of cells swimming in multiple background concentrations of Asp: $c_0 =$ 0.1, 1, and 10 μM (Fig. 2A). Single cells in the clonal population exhibited a range of phenotypes (79,83–91). Therefore, as before (17), we focused on a typical cell by estimating the median single-cell parameter values in the population. In particular, we binned cells by the fraction of time spent running, $P_{run}$, and computed $V(t)$ among cells with the median $P_{run}$. The parameters $\sigma_v^2$ and $\tau_v$ in each background $c_0$ were then estimated by fitting $V(t)$ with a decaying exponential. These parameters depended weakly on $c_0$, and their values in $c_0 = 1$ μM were $\sigma_v^2 = 146 \pm 5$ $(\mu m/s)^2$ and $\tau_v = 1.19 \pm 0.01$ $s$ (see SI Fig. S1AB for all values).

We measured kinase response functions as before (17), using a microfluidic device in which we can deliver controlled chemical stimuli with high time resolution (~100 ms) (50). Cells immobilized in the device were delivered ten small positive and negative step changes of Asp concentration around multiple backgrounds $c_0$ (Fig. 2B). Kinase responses were measured in single cells through FRET (48–50,52,92–94) between CheZ-mYFP and CheY-mRFP1. Then we fit each cell's average response to $K_r(t)$ above, and computed the population-median parameter values. Since $\tau_1$ estimated this way includes the relatively slow dynamics of CheY-CheZ interactions, we used $\tau_1 = 0$ for calculations below, which only slightly overestimates $\dot{I}^*_{s\to a}$. The adaptation time $\tau_2$ depended weakly on $c_0$ (in $c_0 = 1$ μM, $\tau_2 = 7.4 \pm 0.3$ $s$) (Fig. S1D), but $G_r$ varied significantly: for $c_0 = \{0.1, 1, 10\}$ μM we measured $G_r = \frac{1}{k_D}\{3.2 \pm 0.1, 2.28 \pm 0.05, 0.251 \pm 0.009\}$ (Fig. S1EF).

The dependence of $G_r$ on $c_0$ was consistent with the Monod-Wyman-Changeux (MWC) model for kinase activity (29,95–97), which captures numerous experimental measurements (50,52,93,94,98). In particular, $G_r = \frac{1}{r_0} G(c_0)$, where $G(c_0) \approx G_\infty \frac{c_0}{c_0 + K_i}$ is the MWC gain, $K_i$ is the dissociation constant of two-state receptors for Asp when in their inactive state, and $G_\infty$ is a constant (SI section "Modeling kinase activity"). Thus, in the "linear-sensing" regime ($c_0 \ll K_i$), the gain is constant, $G_r = G_\infty \frac{1}{k_D K_i}$, and in the "log-sensing" regime ($c_0 \gg K_i$) (99–101), the gain decreases with background, $G_r \approx G_\infty / r_0$. Fitting the measured $G_r$ to the MWC model gave $G_\infty = 3.5 \pm 0.1$ and $K_i = 0.81 \pm 0.04$ μM.

Finally, we estimated the parameters of slow kinase fluctuations by measuring kinase activity in single cells experiencing constant Asp concentrations $c_0$ (Fig. 2C). The diffusivity $D_n$ and time scale $\tau_n$ of these fluctuations were extracted from each time series using Bayesian filtering (17,102). We then computed the population-median parameter values. Both of these parameters depended weakly on $c_0$, and their values in $c_0 = 1$ μM were $D_n = 8.1 \pm 0.9 \times 10^{-4}$ $s^{-1}$ and $\tau_n = 8.7 \pm 0.9$ $s$ (see Fig. S1CD for all values).

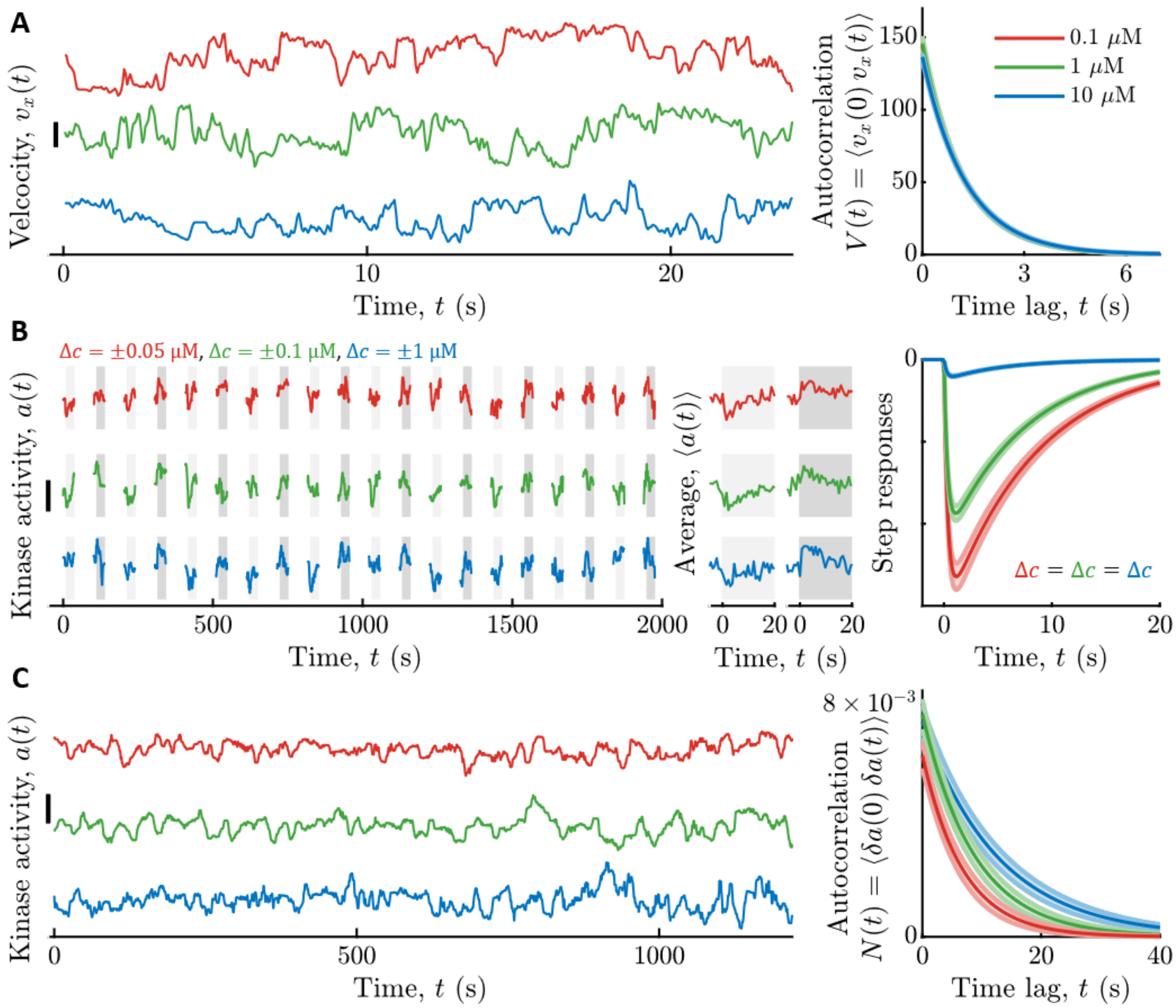

**Figure 2: Measured signal statistics and kinase responses and fluctuations in different background ligand concentrations**. **A)** Signal statistics. Left: Representative time series of up-gradient velocity $v_x$ from three individual cells are shown, one in each aspartate (Asp) concentration $c_0$. Scale bar is 20 μm/s. Cells were binned by the fraction of time spent running, $P_{run}$, and the velocity autocorrelation function $V(t)$ was computed by averaging over cells with the median $P_{run}$. The parameters of $V(t)$ were extracted by fitting a decaying exponential to the data. Right: $V(t)$ model fits for each $c_0$. The curves are on top of each other. Vertical axis units are $(\mu m/s)^2$. Throughout, shading is standard error of the mean (SEM), and line colors indicate $c_0$: Red: 0.1 μM; Green: 1 μM; Blue: 10 μM. **B)** Linear responses. Left: Kinase activity was measured by FRET in blocks of 25 seconds, separated by 65 seconds without illumination. In each block, after 5 s, concentration was stepped up (light gray shading) or down (dark gray shading) around $c_0$, then maintained for 20 s, then returned to $c_0$. Concentration step sizes $\Delta c$ were different for each $c_0$ (shown above the panel). Shown are three representative cells, one from each $c_0$. Scale bar is 0.3. Middle: Average responses of the cells in the left panel to steps up (light gray) and steps down (dark gray). Single-cell responses were fit to extract parameters of the response function $K_r(t)$. Right: Model fits for kinase responses to a steps size $\Delta c$, using population-median parameters. The gain $G_r$ decreases with $c_0$. **C)** Noise statistics. Left: Fluctuations in kinase activity were measured in constant background concentrations.

Representative time series from three cells are shown, one from each $c_0$. Scale bar height is 0.3. Parameters of the slow noise autocorrelation function were fit to single-cell traces using Bayesian filtering (17,102). Right: Estimated noise autocorrelation functions with population median parameters. Vertical axis units are kinase activity squared.

**Comparing *E. coli* to the physical limit**

Both *E. coli*'s information rate, $\dot{I}^*_{s\to a}$, and the physical limit, $\dot{I}^*_{s\to r}$, are proportional to $g^2$ in shallow gradients. Therefore, using the measured parameters, we plotted the information rates per $g^2$ as functions of $c_0$ (Fig. 3A), for values of $g$ in which we previously measured *E. coli*'s gradient-climbing speeds (17). Doing so reveals that *E. coli* are surprisingly far from the physical limit: in shallow gradients, $\dot{I}^*_{s\to a}$ is at least two orders of magnitude below $\dot{I}^*_{s\to r}$ across all background concentrations.

To quantify this comparison, we computed the ratio of *E. coli*'s information rate and the physical limit, $\eta \equiv \dot{I}^*_{s\to a}/\dot{I}^*_{s\to r}$ (Fig. 3B, small error bars). In vanishingly small gradients (black curve), $\eta$ is independent of $g$. In this regime, $\dot{I}^*_{s\to r} \propto c_0$ in all background concentrations, and the shape of $\eta$ is determined by the gain of kinase response, $G_r$. When $c_0 \ll K_i$, the gain is constant, and $\eta$ increases with background, $\eta \propto c_0$. When $c_0 \gg K_i$, $G_r$ decreases and cancels out increasing $c_0$, so $\eta \propto 1/c_0$. These two regimes are separated by a peak at $c_0 = K_i$, where $\eta \approx 0.014 \pm 0.002$ at our closest measurement. As the gradient gets steeper, $\eta$ increases, up to $\eta \approx 0.1$ when $g = 0.4 \text{ mm}^{-1}$. This larger value of $\eta$ does not mean that *E. coli* count nearly every molecule in steeper gradients. Instead, the physical limit saturates (solid lines decreasing with $g$ in Fig. 3A). Thus, in a steep gradient, even a poor sensor can infer the signal with decent accuracy.

Although typical cells in a population are far from the sensing limit, individual cells exhibit non-genetic diversity in sensing and swimming phenotypes (49,50,52,83,90,98), which could cause a significant fraction of the population to approach the limit. Our experimental setup did not allow us to measure all parameters in the same single cells, limiting our ability to answer this question. However, we do have single-cell parameters from different cells. Assuming that swimming, kinase responses, and kinase noise parameters are uncorrelated across cells, we use a maximum-likelihood approach to estimate the variability of $\eta$ in the population (SI section "Estimating population variability in $\eta$"). This analysis indicates that although the 95th percentile of the population can be ~5 times closer to the physical limit, they are still far from it (Fig. 3B , large error bars).

In Fig. 3C, we show the power spectral density (PSD) of slow noise in kinase activity (green line) compared to the PSD of filtered molecule arrival noise (blue line) in $c_0 = 1$ μM. If *E. coli* were close to the physical limit, nearly all noise in kinase activity would come from filtered molecule arrivals. Instead, slow kinase fluctuations are much larger over the range of frequencies observable in the experiment (Fig. 3C, outside the pink region). Thus, *E. coli*'s chemosensing is limited by constraints on its internal signal processing, rather than the external physics of ligand diffusion.

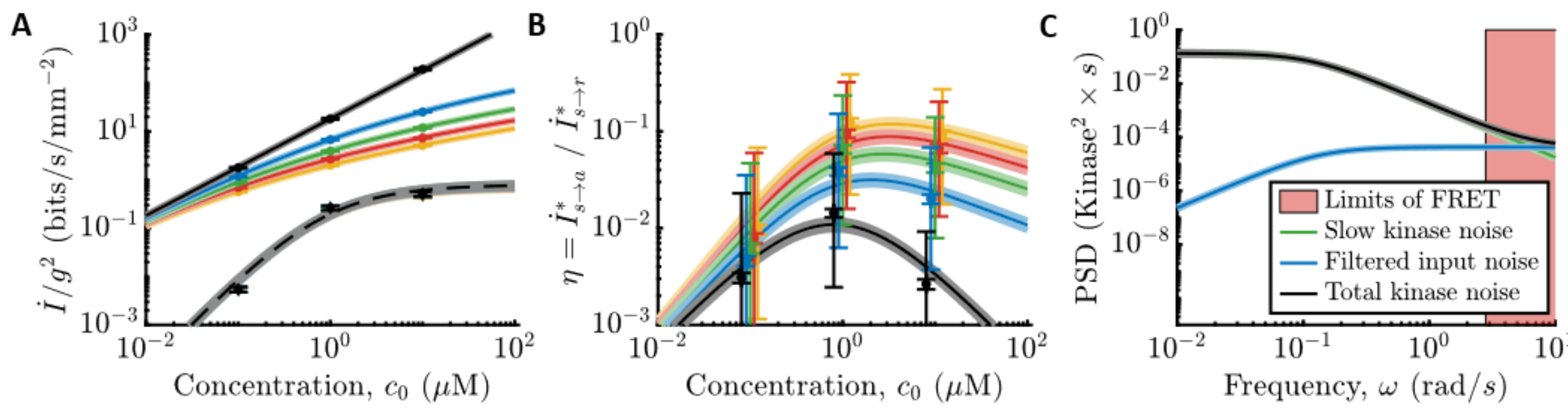

**Figure 3: Comparing *E. coli*'s sensing accuracy to the physical limit. A)** Information rates per gradient steepness squared, $g^2$, in molecule arrival rate, $\dot{I}^*_{s\to r}$ (SI Eqn. 46; solid lines), and in kinase activity, $\dot{I}^*_{s\to a}$ (SI Eqn. 108; dashed lines use the MWC model gain $G(c_0)$ and remaining parameters measured in $c_0 = 1\ \mu\text{M}$) for gradients of varying steepness, $g \in \{0^+, 0.1, 0.2, 0.3, 0.4\}\ \text{mm}^{-1}$ in black, blue, green, red, yellow, where $0^+$ is the limit of an infinitely shallow gradient. Dots are experimental measurements. Error bars and shading are the SEM. *E. coli* are far from the physical limit when signals are weak and sensor quality matters. **B)** $\eta = \dot{I}^*_{s\to a}/\dot{I}^*_{s\to r}$ versus $c_0$. Colors and markers are the same as in (A). Shading and small error bars on the dots are the SEM. Large error bars on the dots are estimates of 95% confidence intervals of population variation in $\eta$, assuming that swimming, kinase response, and kinase noise parameters are uncorrelated. Dots are shifted slightly for visual clarity. **C)** Fit models for the PSD's of noise sources in $c_0 = 1\ \mu\text{M}$. Green: Slow noise in kinase activity. Blue: Molecule arrival noise filtered through the kinase response function. Black: Sum of green and blue. Red shading: Experimentally-inaccessible time scales using CheY-CheZ FRET. See also SI Fig. S3 and the SI section "Modeling kinase activity."

In Fig. 4, we demonstrate what this means for *E. coli* by simulating run-tumble motion in a gradient and constructing the optimal signal estimates (see SI section "Simulation details"). The top panels of Fig. 4A show the observed quantities: molecule arrival rate for an ideal cell, and kinase activity for *E. coli*. The bottom panels show the optimal estimates of the signal in each case, $\hat{s}_r(t)$ and $\hat{s}_a(t)$, overlaid on the true signal. The estimate from kinase activity, $\hat{s}_a(t)$, is visibly lower-quality than $\hat{s}_r(t)$. Quantitatively, $\hat{s}_a(t)$ is less correlated with the true signal by nearly a factor of 10, and likewise kinase activity encodes about 10 times less information about signals than the physical limit. This figure shows the best-case scenario among those in Fig. 3AB; in shallower gradients or other background concentrations, this discrepancy increases to 100-fold or more.

Since information is needed for chemotaxis, this result implies that an ideal cell with the same swimming speed and run duration as *E. coli* (e.g. same $\gamma_r$) could climb gradients much faster than *E. coli*. To support this, we simulated chemotaxis of ideal cells and *E. coli*-like cells in gradients of varying steepness. Fig. 4B indeed shows that ideal cells (blue), which directly observe particle arrival rate $r$, climb gradients much faster than *E. coli*-like cells (green), which only have access to kinase activity $a$. In Fig. 4C, we trace this reduction in drift speed directly back to *E. coli*'s loss of behaviorally-relevant information compared to an ideal cell. Our previous theory (17) predicts that the ratio of the *E. coli* cells' drift speed, $v_a$, to the ideal cell's drift speed, $v_r$, is $v_a/v_r = \sqrt{\dot{I}^*_{s\to a}/\dot{I}^*_{s\to r}}$, and plotting these ratios against each other in Fig. 4C shows good agreement. Thus, being far from the limits of chemosensing likely has a dramatic impact on *E. coli*'s gradient-climbing performance, especially in shallow gradients.

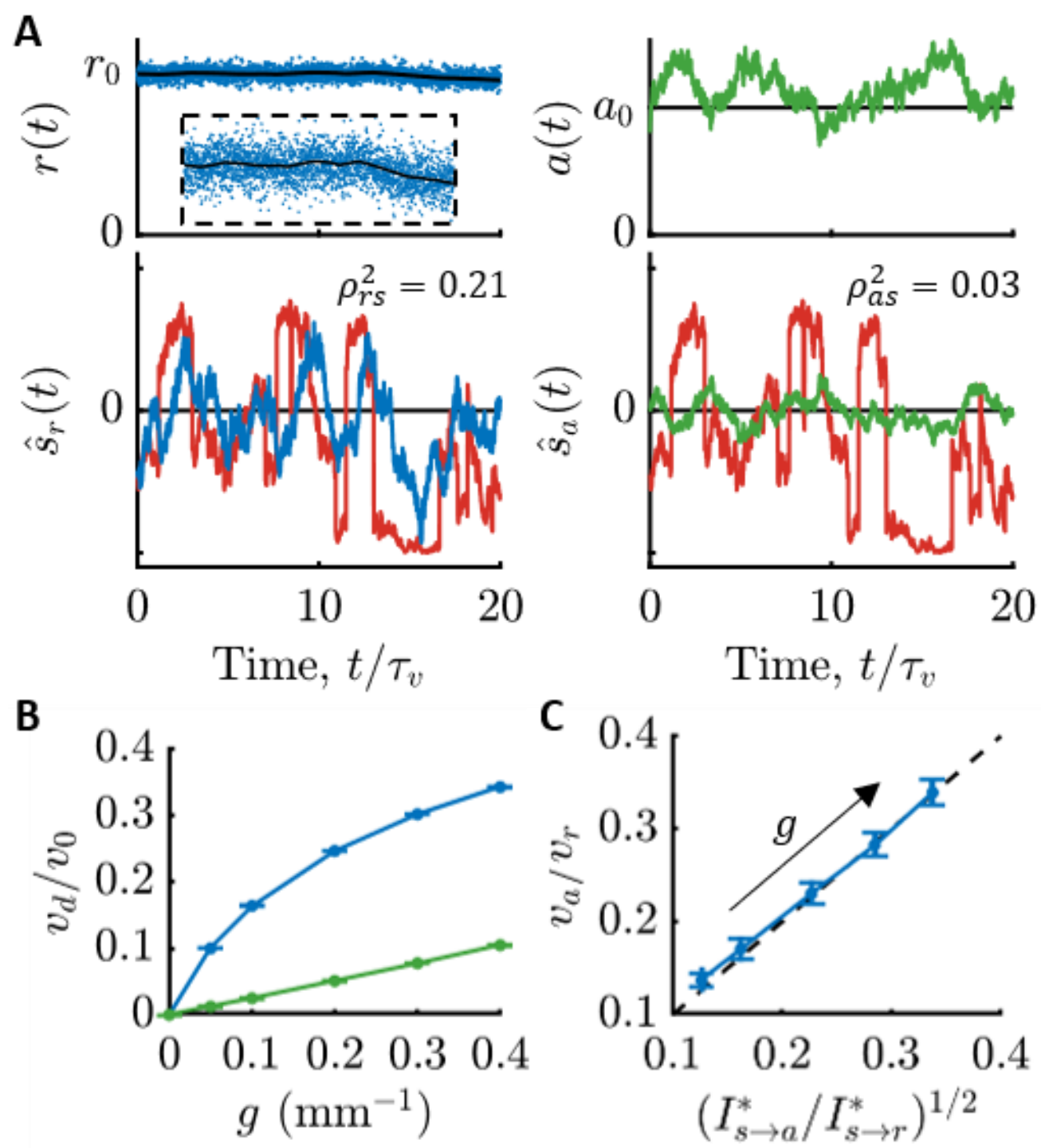

**Figure 4: Consequences of *E. coli* being far from the physical limit on sensing. A)** Simulation of run-tumble motion in a concentration gradient and optimal signal estimates using measured parameters (Fig. S1; $c_0 = 1\ \mu\text{M}$, $g = 0.4\ \text{mm}^{-1}$). Top-left: An ideal cell directly observes molecule arrival rate $r(t)$ (blue dots). Black line is the mean, $\langle r(t) \rangle = k_D\, c(t)$. Inset is the entire trajectory zoomed in to see the subtle changes in $c(t)$. Bottom-left: Optimal signal estimate from molecule arrivals, $\hat{s}_r(t)$ (blue), overlaid on the true signal, $s(t)$ (red). Top-right: Simulated *E. coli* respond to molecule arrivals with changes in kinase activity (green). Bottom-right: Optimal signal estimate from kinase activity, $\hat{s}_a(t)$ (green), overlaid on the true signal, $s(t)$ (red). $\hat{s}_a(t)$ is visibly lower-quality than $\hat{s}_r(t)$. Squared Pearson correlation coefficients, $\rho^2_{rs}$ and $\rho^2_{as}$, between the estimates and the true signal in each bottom panel quantify their accuracy. **B)** Chemotactic drift speed normalized by swimming speed, $v_d/v_0$, as a function of gradient steepness, $g$, for ideal cells (blue) and *E. coli* (green) in simulations ($c_0 = 1\ \mu\text{M}$; SI section "Simulation details"). Error bars in (B) and (C) are SEMs. **C)** Information lost between particle arrivals and kinase activity causes *E. coli* to climb gradients at speeds, $v_a$, that are smaller than those of ideal cells, $v_r$, by a factor of $v_a/v_r = \left(\dot{I}^*_{s\to a}/\dot{I}^*_{s\to r}\right)^{1/2}$. Thus, *E. coli* likely climb gradients slower than the physical limits of sensing allow. Dots are $g = \{0.05, 0.1, 0.2, 0.3, 0.4\}\ \text{mm}^{-1}$, from top-right to bottom-left.

### Discussion

Living systems process information to perform survival-relevant functions, suggesting selection might optimize information processing. Here, we asked whether chemotaxing *E. coli* approach the physical limits on information processing set by counting diffusing ligand molecules. To make this question experimentally-testable, we devised a general approach that asks: how accurately can behaviorally-relevant signals be inferred from a cell's interval variables (kinase activity) compared to the first physically-measureable quantity (ligand molecule arrivals). Accuracy in each case was quantified by an information rate, which we derived analytically. Then, we quantified these information rates in experiments using single-cell FRET measurements of cells' CheA kinase activity in multiple background concentrations. Our results show that *E. coli* encode far less information than the physical limit, and thus *E. coli*'s chemosensing is shaped by internal constraints rather than the physical limit. The functional implication is that *E. coli* likely climb chemical gradients much slower than the physical limit on chemosensing allows. Thus, with the same signal-to-noise of particle arrivals, $\gamma_r$ (set by the swimming speed, run duration, background concentration, and gradient steepness), in principle it may be possible to evolve or engineer a microswimmer that would climb gradients much faster than *E. coli.*

Our results are contrary to the belief, held in the field for nearly 50 years, that *E. coli*'s chemosensing approaches the physical limit, dramatically revising our understanding of bacterial chemotaxis. Since Berg and Purcell did not have direct access to *E. coli*'s uncertainty about ligand concentration, their argument for *E. coli*'s optimality assumed that cells must estimate the change in concentration over a *single* run, $\Delta c$, with uncertainty less than $\Delta c$ (Eqn. 57 in Ref. (4)). Using experimental measurements and their physical limit, they computed the minimum required averaging time, $T$, for this condition to be met if the cell had access to particle arrivals. They found that measured bacterial run durations were slightly longer than the minimum $T$, and argued that chemotaxis would be impossible with shorter runs. Thus, they concluded that the bacterial chemotaxis machinery is nearly optimal. The problem with this argument is its first assumption: that in order to climb gradients, *E. coli*'s sensing machinery must exceed a stringent signal-to-noise threshold, so as to accurately infer the gradient direction in each run. Instead, *E. coli*'s displacement along the gradient accumulates their inferences over many runs. Therefore, even when individual tumble decisions are inaccurate, cells still climb the gradient on average, with no hard threshold on accuracy. In fact, we can show that Berg and Purcell's assumption is too stringent: in our notation, their threshold condition can be written as $\gamma_r > 16/3$ (SI section "Berg & Purcell's SNR threshold for chemotaxis"), but both the ideal cells and the *E. coli* cells simulated in Fig. 4B are able to climb the gradient when $g = 0.05\ \mathrm{mm}^{-1}$, $c_0 = 1\ \mu\mathrm{M}$, and $\gamma_r = 0.15 \ll 16/3$.

Our results also disagree with those of Ref. (62), which argued that the marine bacterium *Vibrio ordalii* senses chemical signals with accuracy within a factor of ~6 of the physical limit, based on fits of agent-based simulations to measurements of bacteria climbing dynamic chemical gradients. We believe the reason for this difference is that their model assumed cells infer $s(t)$ in independent time windows of duration $T = 0.1\ s$. However, the signal is correlated over a time $\tau_v > T$, so an ideal agent can average out molecule arrival noise for times up to $\tau_v$. This increases the theoretical limit, and thus *V. ordalii*'s distance from it, by a factor of $(\tau_v/T)^3 = \left(\frac{0.45\ s}{0.1\ s}\right)^3 \approx 90$, due to the ${\tau_v}^3$ in $\gamma_r$ (Eqn. 3) (related to the $T^3$ in Ref (35)). This suggests that chemosensing in other bacterial species, besides *E. coli*, may also be internally-limited. Beyond bacteria, our results call for a recalibration of expectations about the extent to which biology approaches physical limits on information processing or other functions.

Why are *E. coli* so far from the physical limit? One possibility is the physical implementation of their sensory system may impose trade-offs. For example, the need to operate over a wide range of background concentrations (99–101) suppresses response gain in high backgrounds, but the noise stays constant, reducing information. Cells may need to amplify signals above downstream noise sources, such as stochastic motor switching, requiring the densely-packed arrays seen universally across bacterial species (103), but strongly-coupled CheA kinases likely also introduce noise. Indeed, the dense localization of receptors suggests that molecule counting is not limiting, since if it were, the optimal strategy would be to uniformly distribute the receptors (4). *E. coli* also need to sense amino acids, sugars, and peptides (82,104) with different receptors, but the presence of multiple receptor types in the array reduces the response to any one ligand (94). Another possibility is that *E. coli* may be, and likely are, under selection pressures to perform other tasks, such as localize at concentration peaks (76,77,105,106). Laboratory strains have long been selected for chemotaxis via collective migration assays (107–109). The steep gradients generated during migration, reaching $g \approx 1\ \mathrm{mm}^{-1}$ or steeper (110–112), might obviate the need for a high-fidelity sensor. Lastly, increasing information about signals might be possible, but too costly in resources or energy to be worth the gain in fitness (20–26,32,113–115). The mechanism of amplification is not well understood, but recent work has argued that it consumes energy (116–118). These possibilities might be distinguished by measuring information rates of single cells in an isogenic population or information rates of mutants. If any single cell approaches the physical limit, it would mean that *E. coli* are not limited by hard implementation constraints, but rather by costs or competing objectives. Answering this question will likely inform our thinking about the relevant physical constraints on information processing in other systems.

Physical limits, and whether biology approaches them, have long inspired physicists' curiosity (1–17). At first glance, our results seem to call into question the value of normative theories and physical limits for understanding biological information processing. However, our findings were only possible because we derived a physical limit that provided a reference point against which to compare. At the same time, our results motivate going beyond physical limits and taking seriously the system-specific, physical and biological *constraints* on biological information processing. Going forward, we expect *E. coli* chemotaxis will be a valuable template for studying physical limits and constraints on information processing in higher organisms.

## Methods

Strains and plasmids

All strains and plasmids used are the same as in our recent work (17). The strain used for the FRET experiments is a derivative of *E. coli* K-12 strain RP437 (HCB33), a gift of T. Shimizu, and described in detail elsewhere (49,50). The FRET acceptor-donor pair (CheY-mRFP and CheZ-mYFP) is expressed in tandem from plasmid pSJAB106 (49) under an isopropyl β-D-thiogalactopyranoside (IPTG)-inducible promoter. The glass-adhesive mutant of FliC (FliC*) was expressed from a sodium salicylate (NaSal)-inducible pZR1 plasmid (49). The plasmids are transformed in VS115, a *cheY cheZ fliC* mutant of RP437 (49) (gift of V. Sourjik). RP437, the direct parent of the FRET strain and also a gift from T. Shimizu, was used to measure swimming statistics parameters. All strains are available from the authors upon request.

Cell preparation

Single-cell FRET microscopy and cell culture was carried out essentially as described previously (17,49,50,52). Cells were picked from a frozen stock at -80°C and inoculated in 2 mL of Tryptone Broth (TB; 1% bacto tryptone, 0.5 % NaCl) and grown overnight to saturation at 30°C and shaken at 250 RPM. Cells from a saturated overnight culture were diluted 100X in 10 mL TB and grown to OD600 0.45-0.47 in the presence of 100 µg/ml ampicillin, 34 µg/ml chloramphenicol, 50 µM IPTG and 3 µM NaSal, at 33.5°C and 250 RPM shaking. Cells were collected by centrifugation (5 min at 5000 rpm, or 4080 RCF) and washed twice with motility buffer (10 mM KPO4, 0.1 mM EDTA, 1 µM methionine, 10 mM lactic acid, pH 7), and then were resuspended in 2 mL motility buffer, plus the final concentration of Asp. Cells were left at 22°C for 90 minutes before loading into the microfluidic device. All experiments, FRET and swimming, were performed at 22-23°C.

For swimming experiments, cells were prepared similarly. Saturated overnight cultures were diluted 100X in 5 mL of TB. After growing to OD600 0.45-0.47, 1 mL of cell suspension was washed twice in motility buffer with 0.05% w/v of polyvinylpyrrolidone (MW 40 kDa) (PVP-40). Washes were done by centrifuging the suspension in an Eppendorf tube at 1700 RCF (4000 RPM in this centrifuge) for 3 minutes. After the last wash, cells were resuspended with varying background concentrations of Asp.

Microfluidic device fabrication and loading for FRET measurements

Microfluidic devices for the FRET experiments (50,52,92) were constructed from polydimethylsiloxane (PDMS) on 24 x 60 mm cover glasses (#1.5) following standard soft lithography protocols (119), exactly as done before (17).

Sample preparation in the microfluidic device was conducted as follows. Five inlets of the device were connected to reservoirs (Liquid chromatography columns, C3669; Sigma Aldrich) filled with motility buffer containing various concentrations of Asp through polyethylene tubing (Polythene Tubing, 0.58 mm id, 0.96 mm od; BD Intermedic) (see SI of (17)). The tubing was connected to the PMDS device through stainless steel pins that were directly plugged into the inlets or outlet of the device (New England Tubing). Cells washed and suspended in motility buffer were loaded into the device from the outlet and allowed to attached to the cover glass surface via their sticky flagella by reducing the flow speed inside the chamber. The pressure applied to the inlet solution reservoirs was controlled by computer-controlled solenoid valves (MH1; Festo), which rapidly switched between atmospheric pressure and higher pressure (1.0 kPa) using a source of pressurized air. Only one experiment was conducted per device. *E. coli* consume Asp, so all experiments below were performed with a low dilution of cells to minimize this effect. The continuous flow of fresh media also helped ensured that consumption of Asp minimally affected the signal cells experienced.

## Single-cell FRET imaging system

FRET imaging in the microfluidic device was performed using the setup as before (17), on an inverted microscope (Eclipse Ti-E; Nikon) equipped with an oil-immersion objective lens (CFI Apo TIRF 60X Oil; Nikon). YFP was illuminated by an LED illumination system (SOLA SE, Lumencor) through an excitation bandpass filter (FF01-500/24-25; Semrock) and a dichroic mirror (FF520-Di02; Semrock). The fluorescence emission was led into an emission image splitter (OptoSplit II; Cairn) and further split into donor and acceptor channels by a second dichroic mirror (FF580-FDi01-25x36; Semrock). The emission was then collected through emission bandpass filters (F01-542/27-25F and FF02-641/75; Semrock) by a sCMOS camera (ORCA-Flash4.0 V2; Hamamatsu). RFP was illuminated in the same way as YFP except that an excitation bandpass filter (FF01-575/05-25; Semrock) and a dichroic mirror (FF593-Di03; Semorock) were used. An additional excitation filter (59026x; Chroma) was used in front of the excitation filters. To synchronize image acquisition and the delivery of stimulus solutions, a custom-made MATLAB program controlled both the imaging system (through the API provided by Micro-Manager (120)) and the states of the solenoid valves.

## Computing FRET signal and kinase activity

FRET signals were extracted from raw images using the E-FRET method (121), which corrects for different rates of photobleaching between donor and acceptor molecules. In this method, YFP (the donor) is illuminated and YFP emission images ($I_{DD}$) and RFP (the acceptor) emission images ($I_{DA}$) are captured. Periodically, RFP is illuminated and RFP emission images are captured ($I_{AA}$). From these, photobleach-corrected FRET signal is computed as before (17), which is related to kinase activity $a(t)$ by an affine transform when CheY and CheZ are overexpressed (17,93). All parameters associated with the imaging system were measured previously (17).

In each experiment, we first delivered a short saturating stimulus (1 mM MeAsp plus 100 µM serine (94)) to determine the FRET signal at minimum kinase activity, followed by motility buffer with Asp at background concentration $c_0$. Before the saturating stimulus was delivered, the donor was excited every 0.5 seconds to measure $I_{DD}$ and $I_{DA}$ (see SI of (17)) for 5 seconds. Then the stimulus was delivered for 10 seconds, and the donor was excited every 0.5 seconds during this time. Before and after the donor excitations, the acceptor was excited three times in 0.5-second intervals to measure $I_{AA}$ (see SI of (17)). After the stimulus was removed, the acceptor was excited three more times at 0.5-second intervals. Imaging was then stopped and cells were allowed to adapt to the background for 120 seconds.

Stimulus protocols for measuring kinase linear response functions and fluctuations are described below.

At the end of each experiment, we delivered a long saturating stimulus (1 mM MeAsp plus 100 µM serine) for 180 seconds to allow the cells to adapt. Then we removed the stimulus back to the background concentration, eliciting a strong response from the cells, from which we determined the FRET signal at maximum kinase activity. The donor was excited for 5 seconds before the saturating stimulus and 10 seconds after it, every 0.5 seconds. Before and after these donor excitations, the acceptor was excited three times in 0.5-second intervals. The cells were exposed to the saturating stimulus for 180 seconds. The donor was excited every 0.5 seconds for 5 seconds before cells were exposed to motility buffer with Asp at background concentration $c_0$, followed by 10 seconds of additional donor excitations. Before and after the donor excitations, the acceptor was again excited three times in 0.5-second intervals.

FRET signals were extracted as before (17). The FRET signal at minimum kinase activity, $FRET_{min}$, was computed from the average FRET signal during the first saturating stimulus. The FRET signal at maximum kinase activity, $FRET_{max}$, was computed from the average FRET signal during the first quarter (2.5 seconds) of the removal stimulus at the end of the experiment. Kinase activity was then computed from corrected FRET signal: $a(t) = \frac{FRET(t) - FRET_{min}}{FRET_{max} - FRET_{min}}$.

Kinase linear response functions

Experiments were performed in Asp background concentrations $c_0$ of 0.1, 1, and 10 µM. Measurements were made in single cells, and at least three replicates were performed per background. FRET level at minimum kinase activity was measured at the beginning of each experiment, as described above. After this, a series of stimuli were delivered to the cells in the microfluidic device. Cells were only illuminated and imaged when stimulated in order to limit photobleaching. Before each stimulus, cells were imaged for 7.5 seconds in the background concentration $c_0$. Then, the concentration of Asp was shifted up to $c_+ > c_0$ for 30 seconds and imaging continued. Donor excitation interval was 0.75 seconds and acceptor excitations were done before and after the set of donor excitations. After this, imaging was stopped and the Asp concentration returned to $c_0$ for 65 seconds to allow cells to adapt. Then, the same process was repeated, but this time shifting Asp concentration down to $c_- < c_0$. Alternating up and down stimuli were repeated 10 times each. $c_+$ and $c_-$ varied with each experiment and each background $c_0$. Finally, FRET level at maximum kinase activity was measured at the end of each experiment, as described above. The whole imaging protocol lasted <2200 seconds. In total, cells spent <60 minutes in the device, from loading to the end of imaging.

These data were analyzed as before (17) to extract linear response parameters for each cell. In brief, the responses of a cell to all steps up or steps down in concentration were averaged and the standard error of the response at each time point computed. Model parameters were extracted by maximizing the posterior probability of parameters given data, assuming a Gaussian likelihood function and log-uniform priors for the parameters. The uncertainties of single-cell parameter estimates were generated by MCMC sampling the posterior distribution. Finally, the population-median parameters were computed from all cells in experiments in a given background $c_0$. Uncertainty $\sigma^2_{\theta_i}$ of the population-median value of parameter $\theta_i$, with $\theta = (G, \tau_1, \tau_2)$, was computed using:

$$\sigma^2_{\theta_i} = \frac{1}{N}\left(1.4826 \operatorname{mad}\left(\{\theta_i^{MAP}\}\right)\right)^2 + \frac{1}{N^2}\sum_k \left(\sigma^2_{\theta_i}\right)_k . \quad (6)$$

This expression accounts both for cell-to-cell variations (first term) and uncertainties in the single-cell estimates (second term). $N$ is the number of cells. $1.4826\,\mathrm{mad}(\ )$ is an outlier-robust uncertainty estimate that coincides with the standard deviation when the samples are Gaussian-distributed, and $\mathrm{mad}(\ )$ is the median absolute deviation, used previously (17). $\{\theta_i^{MAP}\}$ are the single-cell maximum *a-posteriori* (MAP) estimates of parameter $\theta_i$. $\left(\sigma_{\theta_i}^2\right)_k$ is the uncertainty of $\theta_i^{MAP}$ in cell $k$, which was computed using

$$\left(\sigma_{\theta_i}\right)_k = 1.4826\,\mathrm{mad}\left(\{\hat{\theta}_i\}_k\right) \tag{7}$$

where $\{\hat{\theta}_i\}_k$ are the samples from the $k$th cell's posterior via Markov Chain Monte Carlo (MCMC).

Fitting the MWC kinase gain

Parameters $G_\infty$ and $K_i$ of the MWC model gain were estimated by fitting the model to estimated values of $G$ in each background $c_0$. The fit was done by minimizing the sum of squared errors between the logarithms of the measured $G$ and predicted values of $G$. Since the estimated values of $G$ varied by about an order of magnitude, taking the logarithms ensured that the smallest value of $G$ had similar weight as largest value in the objective function.

Statistics of noise in kinase activity

Fluctuations in kinase activity were measured in the same Asp background concentrations $c_0$ as above, as well as $c_0 = 0$ μM. At least three replicate experiments were performed per background. FRET level at minimum kinase activity was measured at the beginning of each experiment, as described above. After these measurements, imaging was then stopped and cells were allowed to adapt to the background for 120 seconds. After this, cells were imaged for about 1200 seconds. Throughout, donor excitations were done every 1.0 second, except when it was interrupted by acceptor excitations, which were conducted every 100 donor excitations (see SI of (17)). Finally the FRET level at maximum kinase activity was measured at the end of each experiment, as described above. The whole imaging protocol lasted <1400 seconds. In total, cells spent about < 60 minutes in the device, from loading to the end of imaging.

These data were analyzed as before (17). Bayesian filtering methods (102) were used to compute the likelihood of the parameters given the data, and the prior distribution was taken to be uniform in log. Single-cell estimates and uncertainties of the noise parameters were extracted from the posterior distribution as described above. In each background $c_0$, the population median parameter values were computed, and their uncertainties were computed as described above, with $\theta = (D_n, \tau_n)$.

Swimming velocity statistics

Cells were prepared and imaged as before (17). After the second wash step of the Cell preparation section above, cells were centrifuged again and resuspended in motility buffer containing a background concentration of Asp $c_0$. The values of $c_0$ used here were the same as in the FRET experiments, including $c_0 = 0$ μM. Then, the cell suspension was diluted to an OD600 of 0.00025. This low dilution of cells both enables tracking and minimizes the effect of cells consuming Asp. The cell suspension was then loaded into μ-Slide Chemotaxis devices (ibidi; Martinsried, Germany). Swimming cells were tracked in one of the large reservoirs. 1000-s movies of swimming cells were recorded on a Nikon Ti-E Inverted Microscope using a CFI Plan Fluor 4X objective (NA 0.13). Images were captured using a sCMOS camera (ORCA-Flash4.0 V2; Hamamatsu). Four biological replicates were performed for each background $c_0$.

Cell detection and tracking were carried out using the same custom MATLAB as we used previously (17), with the same analysis parameters (see SI of that paper for details). Tumble detection was also carried out identically as before (17). There was no minimum trajectory duration, but cells were kept only if at least two tumbles were detected in their trajectory. For each cell, we computed the fraction of time spent in the "run" state $P_{run}$. Then we constructed the distribution of $P_{run}$, correcting for biases caused by the different diffusivities of cells with different $P_{run}$ (17). As before (17), we then computed the correlation function of velocity along one spatial dimension for each cell, $V_i(t) = \langle v_x(t')v_x(t'+t)\rangle_{t'}$ among cells with $P_{run}$ within $\pm 0.01$ of the population-median value,. Finally, we computed a weighted average of the correlation functions over all cells in the population-median bin of $P_{run}$, where trajectories were weighted by their duration , giving $V(t)$. In each background $c_0$, for the median bin of $P_{run}$, the average trajectory duration was ~7.6 seconds, and the total trajectory time was $\geq 2.7 \times 10^4$ seconds.

These correlation functions $V(t)$ in each background $c_0$ and each experiment were fit to decaying exponentials $\sigma_v^2 \exp(-|t|/\tau_v)$, and the parameters and their uncertainties were extracted in two steps. First, we determined the MAP estimates of the parameters. An initial estimate of the parameters were esimated using the MATLAB *fit* function to fit exponentials to the $V(t)$ in the time rang $t \in [2\,\Delta t, 10\text{ s}]$, with $\Delta t = 50\text{ ms}$. The estimated $\tau_v$ was used to get the uncertainty of $V(t)$ in each experiment, as done before (17). Assuming a Gaussian likelihood function and parameters distributed uniformly in logarithm, the posterior distribution of parameter was constructed. In each experiment, MAP estimates of the parameters were extracted as done for the kinase parameters, and parameter uncertainties were computed from MCMC samples of the posterior distribution as above. Finally, we computed the average parameters $\sigma_v^2$ and $\tau_v$ over experimental replicates, as well as their standard errors over replicates.

Additional error analysis

Once the variance of the population-median value of parameter $i$ was computed, $\sigma_{\theta_i}^2$, we propagated the uncertainty to functions of those parameters. For some function of the parameters, $f(\theta)$, we computed the variance of $f(\theta)$, $\sigma_f^2$, as:

$$\begin{aligned}\sigma_f^2 &= \sum_i \left(\frac{\partial f}{\partial \theta_i}\right)^2 \sigma_{\theta_i}^2 \\ &= f^2 \sum_i \left(\frac{\partial \log f}{\partial \theta_i}\right)^2 \sigma_{\theta_i}^2. \qquad (8)\end{aligned}$$

The equations above neglect correlations in the uncertainties between pairs of parameters. This was used to compute the uncertainties of $\dot{I}^*_{s\to r}$, $\dot{I}^*_{s\to a}$, and $\eta$. The same formula was used to compute uncertainties of functions of time by applying the formula above pointwise at each time delay $t$ and neglecting correlations in uncertainties between time points.

**Acknowledgments**

**Funding:** This work was supported by the Alfred P. Sloan Foundation under grant G-2023-19668 (HM, TE, BB); by NIH awards R01GM106189 (TE), R01GM138533 (TE), R35GM158058 (TE), and R35GM138341 (BM); by Simons Investigator Award 624156 (BM); by the JST PRESTO grant JPMJPR21E4 (KK); and by the NSTC grant 112-2112-M-001-080-MY3 (KK). HM was supported by the Simons Foundation. KK was also supported by the Institute of Molecular Biology, Academia Sinica.

**Author contributions:** BM and HM conceived the project. KK, HM, TE, and BM designed the experiments. KK, JO, RK, and HM performed the experiments. HM and KK analyzed the data. HM and BM derived the theoretical results. HM wrote the first draft of the manuscript. HM, BM, KK, and TE edited the manuscript.

**Competing interests:** The authors declare no competing interests.

**Data availability:** Source data for the main text figures will be provided online with the manuscript. Source data for the Supplementary Figures are contained in a Supplementary Data file.

**Code availability:** Code to reproduce the main text figures will be available with the source data. All algorithms used are described in detail in the Supplementary Information.

**Supplementary Information for: Chemotaxing *E. coli* do not count single molecules**

Henry H. Mattingly†,*,1, Keita Kamino†,2,3,4, Jude Ong‡,5, Rafaela Kottou‡,5, Thierry Emonet*,5,6,7, Benjamin B. Machta*,6,7

[1] Center for Computational Biology, Flatiron Institute
[2] Institute of Molecular Biology, [3]Institute of Physics, Academia Sinica
[4]Institute of Physics, National Yang Ming Chiao Tung University
[5] Molecular, Cellular, and Developmental Biology, [6] Physics, and [7] QBio Institute, Yale University

† These authors contributed equally.
‡ These authors contributed equally.
* Correspondence to: Benjamin.machta@yale.edu, Thierry.emonet@yale.edu, Hmattingly@flatironinstitute.org.

# Contents

# Supplementary Text

## Drift speed and information rate

We recently demonstrated that a cell's drift speed $v_d$ is set by the transfer entropy rate, $\dot{I}^*_{s\to m}$, from current signal $s(t) = \frac{1}{c_0}\frac{dc}{dt}$ to (the trajectory of) swimming behavior $m(t)$ (1). The transfer entropy rate from current signal to swimming behavior is defined as:

$$\dot{I}^*_{s\to m} = \lim_{dt\to 0} \frac{1}{dt} I(m(t+dt); s(t)|\{m(t)\}) \tag{1}$$

$$= \lim_{dt\to 0} \frac{1}{dt} \left\langle \log\left(\frac{P(m(t+dt)|s(t),\{m(t)\})}{P(m(t+dt)|\{m(t)\})}\right)\right\rangle. \tag{2}$$

Here, curly brackets denote the entire past of a variable, up to and including time $t$. Angled brackets indicate an average over the joint distribution of $s(t)$, past $m(t)$, and $m(t+dt)$. This quantifies how strongly the swimming transition probabilities depend on the current signal.

The transfer entropy rate from current signal determines the cell's drift speed (1):

$$\frac{v_d}{v_0} = \frac{(1-\alpha)\,\lambda_{R0}}{(1-\alpha)\,\lambda_{R0} + 2\,D_r}\left(\frac{2}{3}\frac{\dot{I}^*_{s\to m}}{\lambda_{R0}} P_{run}\right)^{1/2} \tag{3}$$

where $v_0$ is the cell's swimming speed, $\lambda_{R0}$ is the cell's average tumble rate, $\alpha$ is the persistence of the cell's orientation upon tumbling, $D_r$ is the rotational diffusion coefficient, and $P_{run}$ is the fraction of time the cell spends in the run state. This result comes from our previous demonstration that, in shallow gradients, both the drift speed and $\dot{I}^*_{s\to m}$ are proportional to the correlation between the cell's tumble rate $\lambda(t)$ and the signal $s(t)$. Since this correlation determines the cell's chemotaxis performance, $v_d/v_0$, $s(t)$ is the behaviorally-relevant signal.

## Equivalence of transfer entropy and predictive information rates

Here we demonstrate that the transfer entropy rates above are equivalent to a predictive information rate, under some assumptions that are satisfied by bacterial chemotaxis. This relationship is useful because it allows us to derive expressions for the behaviorally-relevant information rates above.

Below, we will write transfer entropy rate from a signal $s(t)$ to a stochastic process $x(t)$, such as $r(t)$, $a(t)$, or $m(t)$. Starting from the definition above:

$$\dot{I}^*_{s\to x} = \lim_{dt\to 0} \frac{1}{dt} I(x(t+dt); s(t)|\{x(t)\})\,, \tag{4}$$

conditional mutual information can be written as a difference between two unconditioned mutual information terms:

$$= \lim_{dt\to 0} \frac{1}{dt}\Big(I\big(\{x(t+dt)\}; s(t)\big) - I\big(\{x(t)\}; s(t)\big)\Big). \tag{5}$$

This can be written as

$$= \left[\partial_T I\big(\{x(T)\}; s(t)\big)\right]_{T=t}. \tag{6}$$

Changing variables from $T$ to $\tau$, where $T = t + \tau$, we get:

$$= \left[\partial_\tau I\big(\{x(t+\tau)\}; s(t)\big)\right]_{\tau=0}. \tag{7}$$

Next, we use time stationarity to shift time $t$ by $-\tau$:

$$= \left[\partial_\tau I\big(\{x(t)\}; s(t-\tau)\big)\right]_{\tau=0}. \tag{8}$$

Finally, we can change variables to $\tau \to -\tau$, giving:

$$= -\left[\partial_\tau I\big(\{x(t)\}; s(t+\tau)\big)\right]_{\tau=0}. \tag{9}$$

This last step would not be allowed if the mutual information inside the time derivative was the entire past of $s$, i.e. $\{s(t+\tau)\}$.

Inside the time derivative above is the "predictive information" (2–4) between the entire past of the stochastic process $x(t)$ up to time $t$ and the signal $s(t)$ at some time $\tau$ into the future (if $\tau > 0$). The time derivative of this mutual information or predictive information is a monotonically decreasing function of $\tau$: the value of the signal $s$ at a time further in the future (larger $\tau$) becomes less correlated with past observations and thus harder to predict.

## Data processing inequality

The causal structure of bacterial chemotaxis has a feedback loop: $m \to s \to r \to a \to m$. Tumbles affect the dynamics of $s$ by changing the cell's heading, which affect the dynamics of $r$, which affect $a$, which then feeds back to $m$ by modulating the tumble rate. In shallow gradients, this feedback is negligible, and transfer entropies obey a data processing inequality (5,6). Thus, information about the signal available in particle arrivals sets a fundamental upper limit on how much information a cell can get, which in turn sets an upper limit its gradient climbing speed:

$$\dot{I}^*_{s\to r} \geq \dot{I}^*_{s\to a} \geq \dot{I}^*_{s\to m} \propto (v_d/v_0)^2. \tag{10}$$

We further speculate that $\dot{I}^*_{s\to x}$ defined above (involving only the current $s(t)$) obeys a data processing inequality in bacterial chemotaxis, even when the gradient is not shallow. If the link from signal to motor behavior is severed, either by having kinases that don't respond to signals or motors that don't respond to kinase activity, then we argue that $\dot{I}^*_{s\to m} = 0$. In this case, as long as post-tumble headings have no preferred direction relative to the gradient direction, then knowing when the cell tumbled (contained in $\{m\}$) provides no information about the value of the current or future signal, $s(t+\tau)$. Therefore, in this case, $I\big(\{m(t)\}; s(t+\tau)\big) = 0$, and $\dot{I}^*_{s\to m} = 0$.

Then, the only way that the past of $m$ provides information about $s(t)$ is through chemotaxis responses to signals, which induce changes in run duration—longer than average runs imply $s(t) > 0$. But tumble behavior in response to signals passes through kinase activity $a$, so $a$ must be more informative than $m$ about $s(t)$.

## Modeling concentration and molecule arrival rate

To derive the physical limit on behaviorally-relevant information for chemotaxis, we first need models for the dynamics of concentration $c(t)$, the signal $s(t) = \frac{d}{dt}\log\big(c(t)\big)$, and particle arrival rate $r(t)$. We consider a single cell navigating a shallow, static chemical gradient, $c(x) = c_0\, e^{g\,x} \sim c_0(1 + g\,x)$, that varies along one spatial dimension, $x$, in 3D space. In a static gradient, the signal is determined by the cell's motion in the gradient: $s(t) = \frac{d}{dt}\log(c) \approx \frac{1}{c_0}\frac{dc}{dt} = g\, v_x(t)$. As done before (1), we model the cell's up-gradient velocity, and thus the signal, as a Gaussian process with correlation function:

$$\langle s(t)\, s(t')\rangle = g^2\, V(t) = g^2\, \sigma_v^2 \exp\left(-\frac{|t - t'|}{\tau_v}\right). \tag{11}$$

Here, $\sigma_v^2 \approx \frac{v_0^2}{3} P_{run}$ is the variance of the cell's up-gradient velocity, $v_0$ is its swimming speed, and $P_{run}$ is the fraction of time it spends in the run state; and $\tau_v$ is the correlation time of the cell's velocity and the signal, $\tau_v^{-1} = (1-\alpha)\,\lambda_{R0} + 2\,D_r$ , where $\lambda_{R0}$ is the cell's baseline tumble rate, $\alpha$ is the directional persistence, and $D_r$ is the rotational diffusion coefficient (1).

We model concentration and particle arrival rate as follows:

$$\frac{dc}{dt} = c_0\, s(t) \tag{12}$$

$$r(t) = k_D\, c(t) + \sqrt{r_0}\,\xi(t). \tag{13}$$

$k_D = 4\,D\,l$ is the diffusion-limited rate constant of particle arrivals to a membrane patch of radius $l$ and for ligand diffusion coefficient $D$ (7–9). The particle arrival noise obeys $\langle \xi(t)\,\xi(t')\rangle = \delta(t - t')$, and $r_0 = k_D\, c_0$ is the particle arrival rate in background concentration $c_0$. Although particle arrival events follow a Poisson process, this Gaussian approximation for the arrival rate is valid when the cell encounters many particles per run, $r_0\,\tau_v \gg 1$.

We will need several power spectral densities. To compute them, we take the Fourier transforms of Eqns. 12 and 13, and then solve for the Fourier transforms of $c(t)$ and $r(t)$, denoted $c(\omega)$ and $r(\omega)$:

$$c(\omega) = \frac{c_0}{\frac{\epsilon}{\tau_v} - i\,\omega}\, s(\omega) \tag{14}$$

$$r(\omega) = k_D\, c(\omega) + \sqrt{r_0}\,\xi(\omega). \tag{15}$$

Here we have introduced a small, dimensionless parameter $\epsilon \ll 1$ that we will take to zero later. Physically, this is as if the cell experiences a weak restoring force back to regions where concentration $c(x) = c_0$. Without it, the correlation function of $c(t)$, which is proportional to the cell's mean squared displacement, would diverge at long times. Everything else remains bounded and well-defined as $\epsilon$ goes to zero.

From these and the correlation function of $s(t)$, we derive the following power spectral densities:

$$S_s(\omega) = F[C_s(T)] = \frac{2\,g^2\,\frac{\sigma_v^2}{\tau_v}}{\frac{1}{\tau_v^2} + \omega^2} \tag{16}$$

$$S_r(\omega) = F[C_r(T)] = \frac{r_0^2}{\frac{\epsilon^2}{\tau_v^2} + \omega^2} S_s(\omega) + r_0 \tag{17}$$

$$S_{rs}(\omega) = S_{sr}^*(\omega) = F[C_{rs}(T)] = \frac{r_0}{\frac{\epsilon}{\tau_v} + i\,\omega} S_s(\omega) \tag{18}$$

where $C_s(T) = \langle s(t)\, s(t+T)\rangle$ , $C_r(T) = \langle (r(t) - r_0)\,(r(t+T) - r_0)\rangle$ , and $C_{rs}(T) = \langle (r(t) - r_0)\, s(t+T)\rangle$, the Fourier transform is defined as $F[f(t)] = \int_{-\infty}^{\infty} f(t)\, e^{i\,\omega\,t}\, dt$, and the inverse transform defined as $F^{-1}[f(\omega)] = \frac{1}{2\pi}\int_{-\infty}^{\infty} f(\omega)\, e^{-i\,\omega\,t}\, d\omega$.

## Total information in particle arrival rate is a trivial upper limit on information processing

To emphasize the importance of behaviorally-relevant information, we first derive the total information about past signals, $\{s(t)\}$, encoded in past particle arrival rate, $\{r(t)\}$:

$$\dot{I}_{s\to r} = \lim_{dt\to 0} \frac{1}{dt} I(r(t+dt); \{s(t)\} | \{r(t)\}) = \dot{I}(\{s\}; \{r\}). \tag{19}$$

The last quantity is the mutual information rate (5,10,11) between $s$ and $r$, and the equality is valid when there is no feedback from $r$ to $s$ or in the regime of shallow gradients. We (1) and others (11) considered information quantities of this kind to quantify the total information about past signals encoded by *E. coli*'s chemotaxis signaling pathway (see also the section below, **Information about current versus past signals encoded in kinase activity**). The total information rate $\dot{I}_{s\to r}$ sets an upper limit on the information that *E. coli* can encode about signals. Computing it using the Gaussian approximations for $s$ and $r$, we find:

$$\begin{aligned}
\dot{I}_{s\to r} &= \lim_{\epsilon\to 0} \frac{-1}{4\pi} \int_{-\infty}^{\infty} \log\left(1 - \frac{|S_{rs}(\omega)|^2}{S_r(\omega)\, S_s(\omega)}\right) d\omega \\
&= \lim_{\epsilon\to 0} \frac{-1}{4\pi} \int_{-\infty}^{\infty} \log\left(1 - \frac{\frac{r_0^2}{(\epsilon/\tau_v)^2 + \omega^2} S_s(\omega)}{\frac{r_0^2}{(\epsilon/\tau_v)^2 + \omega^2} S_s(\omega) + r_0}\right) d\omega \\
&= \lim_{\epsilon\to 0} \frac{1}{4\pi} \int_{-\infty}^{\infty} \log\left(1 + \frac{r_0}{(\epsilon/\tau_v)^2 + \omega^2} S_s(\omega)\right) d\omega \to \infty.
\end{aligned} \tag{20}$$

Thus, we find that this total information rate provides a meaningless bound on *E. coli*'s information processing because it is dominated by information about past signals that are not relevant to chemotaxis.

## Derivation of the physical limit on behaviorally-relevant information for chemotaxis

In this section we derive the information rate from current signal $s(t)$ to past particle counts $r$, which sets a fundamental upper limit on the information rate achievable by a cell. This information rate is given by the following transfer entropy rate:

$$\begin{aligned} \dot{I}^*_{s\to r} &= \lim_{dt\to 0} \frac{1}{dt} I(r(t+dt); s(t)|\{r(t)\}) \\ &= -\big[\partial_\tau I\big(\{r(t)\}; s(t+\tau)\big)\big]_{\tau=0}. \end{aligned} \tag{21}$$

The key quantity we need to derive is the mutual information inside of the derivative:

$$I\big(\{r(t)\}; s(t+\tau)\big) = \left\langle \log\left(\frac{P(s(t+\tau)|\{r(t)\})}{P\big(s(t+\tau)\big)}\right)\right\rangle = \left\langle \log\left(\frac{P\big(\{r(t)\} \mid s(t+\tau)\big)}{P(\{r(t)\})}\right)\right\rangle. \tag{22}$$

In general, it is difficult to derive the conditional distributions above. However, we can make a few simplifying assumptions. First, as mentioned above, the distribution of particle arrival rate $P\big(\{r(t)\} \mid s(t+\tau)\big)$ has Poisson statistics, but if a sufficient number of particles arrive at the cell's receptor array per unit time, the Poisson statistics are approximately Gaussian.

Even with this approximation, $P(\{r(t)\}) = \int P\big(\{r(t)\} \mid s(t+\tau)\big)\, P\big(s(t+\tau)\big)\, ds$ is not Gaussian because $P\big(s(t+\tau)\big)$ is not Gaussian. However, in shallow gradients (small $s$), the (roughly) Gaussian particle arrival noise described by $P\big(\{r(t)\} \mid s(t+\tau)\big)$ blurs the non-Gaussian structure in $P\big(s(t+\tau)\big)$, making $P(\{r(t)\})$ nearly Gaussian, as well. As a result, we can approximate the mutual information in Eqn. 22 by approximating all distributions as Gaussian, as shown rigorously by others (12–14).

Since all distributions are approximately Gaussian, the posterior distribution of $s(t+\tau)$ given past $r(t)$ is Gaussian as well, with mean $\hat{s}_r(t+\tau)$ and variance $\sigma^2_{s|r}(\tau)$ : $P(s(t+\tau)|\{r(t)\}) = \mathcal{N}\left(\hat{s}_r(t+\tau), \sigma^2_{s|r}(\tau)\right)$. With this, the mutual information can be computed from:

$$I\big(\{r(t)\}; s(t+\tau)\big) \approx \frac{1}{2}\log\left(\frac{\sigma_s^2}{\sigma^2_{s|r}(\tau)}\right) = -\frac{1}{2}\log\big(1-\rho^2_{rs}(\tau)\big), \tag{23}$$

and the information rate is:

$$\dot{I}^*_{s\to r} = \frac{1}{2}\left[\frac{-\partial_\tau \rho^2_{rs}(\tau)}{1-\rho^2_{rs}(\tau)}\right]_{\tau=0}. \tag{24}$$

Above, $\sigma_s^2 = g^2\, \sigma_v^2$ is the marginal variance of $s(t+\tau)$ or $s(t)$, i.e. the variance of the distribution $P\big(s(t+\tau)\big) = P\big(s(t)\big)$ (by time-translation invariance). Then, $\rho^2_{rs}(\tau) = 1 - \frac{\sigma^2_{s|r}(\tau)}{\sigma_s^2}$ is a generalized correlation coefficient between $s(t+\tau)$ and past $r$, or the fraction reduction of variance in $s(t+\tau)$ upon observing past $r$.

To calculate $\dot{I}^*_{s\to r}$, we now need to calculate the generalized correlation coefficient $\rho^2_{rs}(\tau)$, or the posterior variance of $s(t+\tau)$, $\sigma^2_{s|r}(\tau)$, using the models for the dynamics of $s$ and $r$ above. To do this, we first note that the posterior mean of $s(t+\tau)$, $\hat{s}_r(t+\tau)$, can be computed using the causal Wiener filter, $M_r(T)$, which minimizes the following mean squared error $\langle e^2(\tau)\rangle$:

$$\langle e^2(\tau)\rangle = \left\langle \left( s(t+\tau) - \int_{-\infty}^{t} M_r(t-t')\, r(t')\, dt' \right)^2 \right\rangle \tag{25}$$

Once the optimal kernel $M_r(T)$ is obtained, the posterior mean is $\hat{s}_r(t+\tau) = \int_{-\infty}^{t} M_r(t-t')\, r(t')\, dt'$ and the posterior variance is $\sigma^2_{s|r}(\tau) = \langle e^2(\tau)\rangle$. Therefore, to derive the mutual information $I\big(\{r(t)\}; s(t+\tau)\big)$, and thus the information rate $\dot{I}^*_{s\to r}$, we need to derive this Wiener filter. The main challenge in deriving $M_r(T)$ is that it must satisfy the constraint that it is causal: that is, we require that $M_r(T) = 0$ for $T < 0$. In Appendix A, we derive the necessary equations and explain where they come from, but here we will just apply them to get $M_r(t)$. See also references (4,15,16).

The optimal kernel can be expressed in Fourier space terms of the power spectra of the signal $s(t)$ and the particle arrival rate $r(t)$ as (Appendix A):

$$M_r(\omega) = \frac{1}{\phi_r(\omega)} \left[ \frac{S_{rs}(\omega)}{\phi_r^*(\omega)}\, e^{-i\,\omega\,\tau} \right]^+. \tag{26}$$

$M_r(\omega)$ is the Fourier transform of $M_r(T)$. $\phi_r(\omega)$ is the causal part of the spectral decomposition of $S_r(\omega)$ (defined below and in Appendix A), where $S_r(\omega)$ is the power spectrum of $r$. $\phi_r^*(\omega)$ is its (anti-causal) complex conjugate. $S_{rs}(\omega)$ is the cross-spectra of $r$ and $s$, equivalent to the Fourier transform of $C_{rs}(\tau)$, where $C_{rs}(\tau) = \langle (r(t) - r_0) s(t+\tau)\rangle$ is the cross-correlation of $s$ and $r$ in the time domain. Finally, $[f(\omega)]^+$ indicates the causal part of the inverse Fourier transform of $f(\omega)$, which can be found by taking the inverse Fourier transform of $f(\omega)$, multiplying the result by a Heaviside step function in the time domain, and then taking the Fourier transform.

As explained in Appendix A, to find the optimal causal kernel, we need to decompose $S_r(\omega)$ into the product of a causal and an anti-causal part. This requires finding the zeros and poles of $S_r(\omega)$. The zeros satisfy $S_r(\omega = i\, z_r) = 0$, and therefore are the complex solutions to the equation:

$$2\, r_0\, g^2\, \sigma_v^2\, \tau_v^3 + (\epsilon^2 + \tau_v^2\, \omega^2)\,(1 + \tau_v^2\, \omega^2) = 0 \tag{27}$$

or, defining $\gamma_r = 2\, r_0\, g^2\, \sigma_v^2\, \tau_v^3$:

$$\gamma_r + (\epsilon^2 + \tau_v^2\, \omega^2)\,(1 + \tau_v^2\, \omega^2) = 0. \tag{28}$$

$\gamma_r$ is a dimensionless signal-to-noise ratio parameter, where the signal is $r_0^2\, g^2\, \sigma_v^2\, \tau_v^3$ (the prefactor of the first term in $S_r(\omega)$ when $\omega$ is rescaled by $1/\tau_v$) and the noise is $r_0$ (the second term in $S_r(\omega)$).

The zeros of $S_r(\omega)$ are:

$$i\, z_{r,1} = i \frac{1}{\sqrt{2}\,\tau_v} \sqrt{1 + \epsilon^2 + \sqrt{(1-\epsilon^2)^2 - 4\,\gamma_r}}, \qquad i\, z_{r,2} = i \frac{1}{\sqrt{2}\,\tau_v} \sqrt{1 + \epsilon^2 - \sqrt{(1-\epsilon^2)^2 - 4\,\gamma_r}}, \tag{29}$$

as well as their complex conjugates, $z_1^*$ and $z_2^*$. As $\epsilon \to 0$, these will simplify to:

$$i\, z_{r,1} = i\, \frac{1}{\sqrt{2}\,\tau_v} \sqrt{1 + \sqrt{1 - 4\,\gamma_r}}, \qquad i\, z_{r,2} = i\, \frac{1}{\sqrt{2}\,\tau_v} \sqrt{1 - \sqrt{1 - 4\,\gamma_r}}, \tag{30}$$

Note that there are several equivalent forms for these zeros, and they change from being fully imaginary to complex when $\gamma_r > 1/4$.

The poles of $S_r(\omega)$ satisfy $\frac{1}{S_r(\omega = i\,p_r)} = 0$ and are given by $i\,p_{r,1} = i\frac{\epsilon}{\tau_v}$ and $i\,p_{r,2} = i\frac{1}{\tau_v}$, as well as their complex conjugates $p_{r,1}^*$ and $p_{r,2}^*$.

Power spectral densities of real, stable, causal systems can generally be decomposed into causal and anti-causal parts ("Wiener-Hopf factorization") (17–19):

$$S_r(\omega) = \phi_r(\omega)\,\phi_r^*(\omega) \tag{31}$$

where

$$\phi_r(\omega) = \sqrt{r_0}\,\frac{\left(z_{r,1} - i\,\omega\right)\left(z_{r,2} - i\,\omega\right)}{\left(p_{r,1} - i\,\omega\right)\left(p_{r,2} - i\,\omega\right)} \tag{32}$$

has zeros and poles with negative imaginary parts, and $\phi_r^*(\omega)$ is its complex conjugate.

Next, we need the causal part of (see Appendix A):

$$\frac{S_{rs}(\omega)}{\phi_r^*(\omega)} e^{-i\,\omega\,\tau} = \frac{\sqrt{r_0}}{\frac{\epsilon}{\tau_v} + i\,\omega} S_s(\omega) \frac{\left(p_{r,1} + i\,\omega\right)\left(p_{r,2} + i\,\omega\right)}{\left(z_{r,1} + i\,\omega\right)\left(z_{r,2} + i\,\omega\right)} e^{-i\,\omega\,\tau} \tag{33}$$

$$= \frac{\sqrt{r_0}}{\frac{\epsilon}{\tau_v} + i\,\omega} \frac{2\,g^2\,\frac{\sigma_v^2}{\tau_v}}{\frac{1}{\tau_v^2} + \omega^2} \frac{\left(\frac{\epsilon}{\tau_v} + i\,\omega\right)\left(\frac{1}{\tau_v} + i\,\omega\right)}{\left(z_{r,1} + i\,\omega\right)\left(z_{r,2} + i\,\omega\right)} e^{-i\,\omega\,\tau} \tag{34}$$

$$= \sqrt{r_0}\,\frac{2\,g^2\,\frac{\sigma_v^2}{\tau_v}}{\left(\frac{1}{\tau_v} - i\,\omega\right)} \frac{1}{\left(z_{r,1} + i\,\omega\right)\left(z_{r,2} + i\,\omega\right)} e^{-i\,\omega\,\tau} \tag{35}$$

$$= \frac{\gamma_r}{\sqrt{r_0}\,\tau_v^4} \frac{1}{\left(\frac{1}{\tau_v} - i\,\omega\right)} \frac{1}{\left(z_{r,1} + i\,\omega\right)\left(z_{r,2} + i\,\omega\right)} e^{-i\,\omega\,\tau} \tag{36}$$

One approach would be to compute the inverse Fourier transform of $\frac{S_{rs}(\omega)}{\phi_r^*(\omega)}$, apply the time shift forward by $\tau$ implied by $e^{-i\,\omega\,\tau}$, multiply the result by a Heaviside step function $\Theta(T)$, and compute the Fourier transform of the result. An alternative approach is to compute the partial fraction decomposition of the expression above and keep only the terms with poles and zeros that have negative imaginary part:

$$\frac{S_{rs}(\omega)}{\phi_r^*(\omega)} e^{-i\,\omega\,\tau} = \frac{A}{\left(\frac{1}{\tau_v} - i\,\omega\right)} + \frac{B}{\left(z_{r,1} + i\,\omega\right)} + \frac{C}{\left(z_{r,2} + i\,\omega\right)} \tag{37}$$

for unknown $A$, $B$, and $C$. Only the pole of the first term ($\omega = -i\frac{1}{\tau_v}$) has negative imaginary part, so we only need to compute $A$ to get the causal part of this expression. With some algebra, this is:

$$A = \left[ \frac{\gamma_r}{\sqrt{r_0}\, \tau_v^4} \frac{1}{\left(z_{r,1} + i\,\omega\right)\left(z_{r,2} + i\,\omega\right)} e^{-i\,\omega\,\tau} \right]_{\omega = -i\frac{1}{\tau_v}} \tag{38}$$

$$= \frac{\gamma_r}{\sqrt{r_0}\, \tau_v^2} \frac{1}{\left(1 + \tau_v\, z_{r,1}\right)\left(1 + \tau_v\, z_{r,2}\right)} e^{-\frac{\tau}{\tau_v}}, \tag{39}$$

and the causal part of $\frac{S_{rs}(\omega)}{\phi_r^*(\omega)} e^{-i\,\omega\,\tau}$ is then:

$$\left[ \frac{S_{rs}(\omega)}{\phi_r^*(\omega)} e^{-i\,\omega\,\tau} \right]^+ = \frac{\gamma_r}{\sqrt{r_0}\, \tau_v^2} \frac{1}{\left(1 + \tau_v\, z_{r,1}\right)\left(1 + \tau_v\, z_{r,2}\right)} \frac{1}{\left(\frac{1}{\tau_v} - i\,\omega\right)} e^{-\frac{\tau}{\tau_v}}. \tag{40}$$

Finally, the optimal kernel that computes the mean of $p(s(t+\tau)|\{r(t)\})$, $\mu_{s|r}(\tau)$, is (Appendix A):

$$M_r(\omega) = \frac{1}{\phi_r(\omega)} \left[ \frac{S_{rs}(\omega)}{\phi_r^*(\omega)} e^{-i\,\omega\,\tau} \right]^+ \tag{41}$$

$$= e^{-\frac{\tau}{\tau_v}} \frac{\gamma_r}{r_0\, \tau_v^2} \frac{1}{\left(1 + \tau_v\, z_{r,1}\right)\left(1 + \tau_v\, z_{r,2}\right)} \frac{-i\,\omega}{\left(z_{r,1} - i\,\omega\right)\left(z_{r,2} - i\omega\right)}, \tag{42}$$

after taking $\epsilon$ to zero. We convert this kernel to the time domain and discuss its properties in the next section.

The variance of $p(s(t+\tau)|\{r(t)\})$, $\sigma_{s|r}^2(\tau)$, is (Appendix A, Eqn. 181):

$$\sigma_{s|r}^2(\tau) = \sigma_s^2 - \frac{1}{2\pi} \int_{-\infty}^{\infty} S_{rs}^*(\omega)\, e^{i\,\omega\,\tau}\, M_r(\omega)\, d\omega \tag{43}$$

$$= \sigma_s^2 \left( 1 - e^{-2\frac{\tau}{\tau_v}} \frac{\gamma_r}{\left(1 + \tau_v\, z_{r,1}\right)^2 \left(1 + \tau_v\, z_{r,2}\right)^2} \right) \tag{44}$$

where we used $\sigma_s^2 = g^2 \sigma_v^2 = \gamma/(2\, r_0\, \tau_v^3)$. Then the correlation coefficient $\rho_{rs}^2(\tau)$ is:

$$\rho_{rs}^2(\tau) = 1 - \frac{\sigma_{s|r}^2(\tau)}{\sigma_s^2} = e^{-2\frac{\tau}{\tau_v}} \frac{\gamma_r}{\left(1 + \tau_v\, z_{r,1}\right)^2 \left(1 + \tau_v\, z_{r,2}\right)^2}. \tag{45}$$

Finally, using Eqn. 24 from above, we find that the behaviorally-relevant information available in particle counts is:

$$\dot{I}_{s\to r}^* = \frac{1}{\tau_v} \frac{\rho_{rs}^2(\tau=0)}{1 - \rho_{rs}^2(\tau=0)} = \frac{1}{\tau_v} \frac{\gamma_r \frac{1}{\left(1 + \frac{1}{\sqrt{2}}\sqrt{1 + \sqrt{1 - 4\,\gamma_r}}\right)^2 \left(1 + \frac{1}{\sqrt{2}}\sqrt{1 - \sqrt{1 - 4\,\gamma_r}}\right)^2}}{1 - \gamma_r \frac{1}{\left(1 + \frac{1}{\sqrt{2}}\sqrt{1 + \sqrt{1 - 4\,\gamma_r}}\right)^2 \left(1 + \frac{1}{\sqrt{2}}\sqrt{1 - \sqrt{1 - 4\,\gamma_r}}\right)^2}}. \tag{46}$$

Expanding around small SNR, $\gamma_r$, gives:

$$\dot{I}^*_{s\to r} \approx \frac{1}{\tau_v}\frac{\gamma_r}{4} = \frac{1}{2} r_0\, g^2\, \sigma_v^2\, \tau_v^2. \tag{47}$$

Note that for small signals, Eqn. 46 can be written $\dot{I}^*_{s\to r} \approx \frac{1}{\tau_v}\rho_{rs}^2(\tau = 0) \approx \frac{2}{\tau_v} I\big(\{r(t)\}; s(t)\big)$.

The optimal kernel in the time domain and the information rate remain real when $\gamma_r > 1/4$, even though $z_{r,1}$ and $z_{r,2}$ become complex. In this regime, they can be written:

$$z_{r,1} = \frac{1}{2\,\tau_v}\left(\sqrt{2\sqrt{\gamma_r} + 1} - i\sqrt{2\sqrt{\gamma_r} - 1}\right), \qquad z_{r,2} = \frac{1}{2\,\tau_v}\left(\sqrt{2\sqrt{\gamma_r} + 1} + i\sqrt{2\sqrt{\gamma_r} - 1}\right). \tag{48}$$

The optimal kernel in frequency space can be written:

$$M_r(\omega) = e^{-\frac{\tau}{\tau_v}}\,\frac{\gamma_r}{r_0\,\tau_v^2}\frac{1}{(1 + \tau_v\, z_{r,1})\,(1 + \tau_v\, z_{r,2})}\frac{-i\,\omega}{(z_{r,1} - i\,\omega)\,(z_{r,2} - i\omega)} \tag{49}$$

$$= e^{-\frac{\tau}{\tau_v}}\,\frac{\gamma_r}{r_0\,\tau_v^2}\frac{1}{\left(1 + \sqrt{1 + 2\sqrt{\gamma_r}} + \sqrt{\gamma_r}\right)}\frac{-i\,\omega}{(z_{r,1} - i\,\omega)\,(z_{r,2} - i\omega)}, \tag{50}$$

the correlation coefficient at $\tau = 0$ can be written:

$$\rho_{rs}^2(\tau = 0) = \frac{\gamma_r}{(1 + z_{r,1}\,\tau_v)^2\,(1 + z_{r,2}\,\tau_v)^2} \tag{51}$$

$$= \frac{\gamma_r}{|1 + z_{r,1}\,\tau_v|^4} \tag{52}$$

$$= \frac{\gamma_r}{\left(1 + \sqrt{1 + 2\sqrt{\gamma_r}} + \sqrt{\gamma_r}\right)^2}, \tag{53}$$

and the information rate is:

$$\dot{I}^*_{s\to r} = \frac{1}{\tau_v}\frac{\dfrac{\gamma_r}{\left(1 + \sqrt{1 + 2\sqrt{\gamma_r}} + \sqrt{\gamma_r}\right)^2}}{1 - \dfrac{\gamma_r}{\left(1 + \sqrt{1 + 2\sqrt{\gamma_r}} + \sqrt{\gamma_r}\right)^2}}. \tag{54}$$

For small $\gamma_r$, this reduces again to Eqn. 47.

## Correlation coefficients

As described above, in shallow gradients, the posterior distribution $P(s(t)|\{r\})$ is approximately Gaussian with constant variance, $\sigma^2_{s(t)|r}$, and time-varying mean, $\hat{s}_r(t)$, that depends on $\{r\}$. Once $\hat{s}_r(t)$ is computed, the posterior $P(s(t)|\{r\}) = P\big(s(t)\big|\hat{s}_r(t)\big)$ does not depend on $\{r\}$ in any other way, so $\hat{s}_r(t)$ is a sufficient statistic. This in turn means that the posterior variance of $s(t)$ given $\hat{s}_r(t)$ is the same as the posterior variance of $s(t)$ given $\{r\}$, and thus the reduction in uncertainty about $s(t)$ by observing $\{r\}$ or $\hat{s}_r(t)$ is the same. Since the generalized correlation coefficient $\rho_{rs}^2$ quantifies this reduction in uncertainty about $s(t)$ by observing $\{r\}$, $\rho_{rs}^2$ must also be the squared correlation between $s(t)$ and $\hat{s}_r(t)$. The same applies if we replace $r$ with $a$.

## Optimal kernel for estimating signal from particle arrivals

To get the time-domain kernel mapping past particle arrival rate $r(t)$ to signal $s(t+\tau)$, $M_r(T)$,we take the inverse Fourier transform of $M_r(\omega)$. $M_r(T)$ has the form of a sum of two exponentials, with real exponents when $\gamma_r \leq 1/4$ and complex ones when $\gamma_r > 1/4$. For $\gamma_r < 1/4$, the kernel in the time domain is:

$$M_r(T) = e^{-\frac{\tau}{\tau_v}} \frac{\gamma_r}{r_0\, \tau_v^2} \frac{1}{\left(1+z_{r,1}\,\tau_v\right)\left(1+z_{r,2}\,\tau_v\right)} \frac{\left(z_{r,1}\, e^{-z_{r,1}\,T} - z_{r,2}\, e^{-z_{r,2}\,T}\right)}{z_{r,1}-z_{r,2}}\,\Theta(T), \tag{55}$$

where $\Theta(T)$ is the Heaviside step function, indicating that the kernel is indeed causal.

The optimal kernel $M_r(T)$ essentially computes the time derivative of concentration, while also averaging out shot noise from particle arrivals. It has several notable features. First, it is biphasic and exhibits perfect adaptation, a hallmark of the chemotaxis pathway. Any derivative operation should adapt perfectly because it should only respond to *changes* in the input.

It is interesting to examine how the time scales of the optimal kernel are set by the signal-to-noise ratio $\gamma_r = 2\, r_0\, g^2\, \sigma_v^2\, \tau_v^3$. The initial response time scale is set by $z_{r,1}^{-1}$ and its adaptation time scale is set by $z_{r,2}^{-1}$. When the inputs are very noisy, i.e. as $\gamma_r \to 0$, $z_{r,1}^{-1}$ gets longer but saturates at $\tau_v$:

$$z_{r,1}^{-1}(\gamma_r \to 0) \;\approx\; \tau_v\left(1-\frac{\gamma_r}{2}\right). \tag{56}$$

This makes sense because it maximally averages out shot noise, but only for as long as past signals are correlated with the current signal. As the SNR increases, this initial averaging time gets shorter.

As the inputs get noisier, i.e. as $\gamma_r \to 0$, the adaptation time approaches:

$$z_{r,2}^{-1}(\gamma_r \to 0) \;\approx\; \tau_v\left(\frac{1}{\sqrt{\gamma_r}}-\frac{\sqrt{\gamma_r}}{2}\right). \tag{57}$$

This shows that the adaptation time can become long compared to $\tau_v$ when $\gamma_r < 1/4$.

Interestingly, in this regime, the kernel $M_r(T)$ has the same functional form as the phenomenological kernel we measured previously (1) (after transforming the input quantity from $s(t)$ to $c(t)$).

When signal and noise have similar strength $\gamma_r = 1/4$, $z_{r,1} = z_{r,2} = z_r = \frac{1}{\sqrt{2}}\,\tau_v^{-1}$, and the optimal kernel becomes:

$$M_r(T) = e^{-\frac{\tau}{\tau_v}} \frac{\gamma_r}{r_0\, \tau_v^2} \frac{1}{\left(1+\frac{1}{\sqrt{2}}\right)^2} e^{-z_r\,T}(1-z_r\,T)\,\Theta(T). \tag{58}$$

When SNR is high $\gamma_r > 1/4$, $z_{r,1}$ and $z_{r,2}$ become complex. However, since they are complex conjugates of each other, the kernel remains real:

$$M_r(T) = e^{-\frac{\tau}{\tau_v}} \frac{\gamma_r}{r_0\, \tau_v^2} \frac{1}{(1 + z_{r,1}\, \tau_v)\,(1 + z_{r,2}\, \tau_v)} e^{-\mathrm{Re}[z_{r,2}]\,T} \left(\cos(\mathrm{Im}[z_{r,2}]\,T) - \frac{\mathrm{Re}[z_{r,2}]}{\mathrm{Im}[z_{r,2}]} \sin(\mathrm{Im}[z_{r,2}]\,T)\right) \Theta(T)$$

$$= e^{-\frac{\tau}{\tau_v}} \frac{\gamma_r}{r_0\, \tau_v^2} \frac{1}{\left(1 + \sqrt{1 + 2\sqrt{\gamma_r}} + \sqrt{\gamma_r}\right)} e^{-\frac{1}{2}\sqrt{2\sqrt{\gamma_r}+1}\frac{T}{\tau_v}} \times$$

$$\left(\cos\left(\frac{1}{2}\sqrt{2\sqrt{\gamma_r} - 1}\ \frac{T}{\tau_v}\right) - \sqrt{\frac{2\sqrt{\gamma_r} + 1}{2\sqrt{\gamma_r} - 1}} \sin\left(\frac{1}{2}\sqrt{2\sqrt{\gamma_r} - 1}\ \frac{T}{\tau_v}\right)\right) \Theta(T) \qquad (59)$$

The optimal kernel, $M_r(T)$, is plotted in Fig. S4 for varying values of $\gamma_r$.

As the SNR $\gamma_r$ increases, the initial response time and the adaptation time both get shorter. Although the kernel oscillates, its decay rate is faster than the period of oscillations. The time scales of decay and oscillation are closest to each other, and thus the oscillation amplitude is largest, when $\gamma_r$ is large: in the limit that $\gamma_r \to \infty$, $\mathrm{Re}[z_{r,2}] = \mathrm{Im}[z_{r,2}] = {\gamma_r}^{1/4}$. Even in this limit, the peak of the kernel following the first negative lobe occurs at time $T = 3\pi\, {\gamma_r}^{-1/4}$ and is smaller than the kernel's maximum value ($M_r(T = 0)$) by a factor of $e^{-3\pi/2} \sim 0.009$. Thus, the oscillations are small. $M_r(T)$ transitions continuously between the forms above as $\gamma_r$ varies.

The results of this and previous section could also be derived using the continuous-time Kalman-Bucy filter (20,21). That approach provides a pair of ODEs for the estimator of $s$ (i.e. conditional mean $\mu_{s|r}$) and its uncertainty (i.e. the conditional variance $\sigma_{s|r}^2$) that are driven by the observations, $r(t)$. Once $\sigma_{s|r}^2$ reaches steady state in that formulation (consistent with our assumption of stationarity here), the ODE for $\mu_{s|r}$ can be solved in terms of a kernel convolved with past $r(t)$, which is identical to the optimal kernel above.

## Comparing temporal and spatial sensing

Berg and Purcell argued that bacteria are too small to accurately infer concentration differences across their body. We can estimate the effect of measuring concentration differences across the cell body by comparing its SNR to that of temporal comparisons. For temporal comparisons, we showed above that the SNR is $\gamma_r = 2\, r_0\, g^2\, \sigma_v^2\, \tau_v^3 = 2\,(r_0\, \tau_v)\,(g^2\, \sigma_v^2\, \tau_v^2) = 2\,(r_0\, \tau_v)\,\left(L_{run}/L_{gradient}\right)^2$. Here, we defined the run length scale $L_{run} \approx \sigma_v\, \tau_v$ and the gradient length scale $L_{gradient} = g^{-1}$. When making spatial comparisons instead of temporal ones, the run length is replaced by the length over which spatial comparisons are made, i.e. the cell body length, $L_{body}$: $\gamma_r^{spatial} = 2\,(r_0\, \tau_v)\,\left(L_{body}/L_{gradient}\right)^2$. The ratio of these two SNR's is $\gamma_r^{spatial}/\gamma_r = \left(L_{body}/L_{run}\right)^2$. The length of the cell is roughly 2 μm and its run length is roughly 20 μm, making $\gamma_r^{spatial}/\gamma_r \approx 1/100$. Therefore, consistent with Berg and Percell's arguments, spatial sensing has a negligible effect on sensing accuracy for bacteria with run lengths much longer than the body length.

## Berg & Purcell's SNR threshold for chemotaxis

Berg and Purcell claimed that *E. coli* chemosensing approaches the physical limit (7), based on the following argument. They postulated that, in order to perform chemotaxis, a cell must estimate the change in concentration over a single run with uncertainty smaller than the concentration change itself.

They envisioned the cell making two measurements of concentration, $c$, during a run, each of duration $T$. Their condition for chemotaxis to be possible, with a slight change in notation, reads (Eqn. 57 in Ref. (7)):

$$\tau_v \frac{dc}{dt} > \sqrt{2}\, \delta c(T). \tag{60}$$

The left-hand side is the concentration change over a single run of duration $\tau_v$, and the right-hand side is the standard deviation of the estimated concentration change (i.e. the uncertainty of the difference of two independent measurements of concentration, $c_1$ and $c_2$, is $\sqrt{\delta c_1^2 + \delta c_2^2} \approx \sqrt{2}\, \delta c$), which depends on the measurement time $T$. Equating the two sides above sets a minimum time for each measurement, $T^{min}$, and the argument is that chemotaxis would not be possible if the cell's run duration is shorter than the time needed to make the two measurements. Thus, chemotaxis requires $\tau_v > 2\, T^{min}$. Conversely, if a cell's run duration is much longer than $2\, T^{min}$, that would imply that the cell's sensing accuracy is far below the physical limit. Thus, they argue that an ideal agent would exhibit an optimal run duration of:

$$\tau_v^* \approx 2\, T^{min}. \tag{61}$$

Plugging realistic parameters, they argued that this equality approximately holds. Based on this, they concluded, that "[t]he chemotactic sensitivity of *Escherichia coli* approaches that of the cell of optimal design."

The problem with their argument is the initial postulate: to perform chemotaxis, *E. coli* do not have to detect concentration changes with significance above a strict threshold *in a single run*. Instead, they only need to increase their run duration when the concentration increases, and decrease it when concentration decreases, *on average*. This is because chemotaxis drift speed depends on an *average* over many runs. This is consistent with our previous result (1) that the total amount of information that *E. coli* acquire in a single run ($\sim 1$ second) is much smaller than 1 bit, the amount required to unambiguously distinguish whether concentration increased or decreased over a single run.

Next, we provide a counterexample to the argument above using simulations, showing that chemotaxis is possible even when Eqn. 60 is not satisfied. First, we express their condition in terms of our signal-to-noise ratio of particle arrivals, $\gamma_r$. For an ideal agent with a circular sensor of radius $l$ that uptakes arrival particles, the particle arrival flux is $J = r_0$ and the average receptor occupancy is $\bar{p} = 0$ because the receptors are always free to sense particle arrivals. Then Berg and Purcell showed that the ideal agent's uncertainty about a measurement of concentration is (Eqn. 55 of (7)):

$$\delta c = \frac{c_0}{\sqrt{\frac{1}{2}\, r_0\, T}}. \tag{62}$$

If the cell makes two consecutive measurements in a single run, each of duration $T = \frac{\tau_v}{2}$, the condition for chemotaxis in Eq. 60 becomes:

$$\tau_v \frac{dc}{dt} > 2\sqrt{2} \frac{c_0}{\sqrt{r_0 \tau_v}}. \tag{63}$$

In a single up-gradient run, $\frac{dc}{dt} \approx v_0 \frac{dc}{dx}$, and the equation above can be written as:

$$\tau_v\, v_0\, \frac{dc}{dx} > \ 2\sqrt{2}\frac{c_0}{\sqrt{r_0 \tau_v}}. \tag{64}$$

Then, since $\frac{dc}{dx} \approx c_0 \frac{d}{dx}\log(c) = c_0\, g$, we get:

$$\tau_v\, v_0\, g > \frac{2\sqrt{2}}{\sqrt{r_0 \tau_v}} \tag{65}$$

$$r_0\, v_0^2\, g^2\, \tau_v^3 > 8 \tag{66}$$

$$2\, r_0 \frac{v_0^2}{3} g^2\, \tau_v^3 > \frac{16}{3} \tag{67}$$

$$2\, r_0\, \sigma_v^2\, g^2\, \tau_v^3 > \frac{16}{3} \tag{68}$$

$$\gamma_r > \frac{16}{3}. \tag{69}$$

Eqn. 68 used $\sigma_v^2 = \frac{1}{3} v_0^2$ when tumbles are instantaneous. Thus, Berg and Purcell's condition for chemotaxis to be possible in Eqn. 60 is equivalent to requiring that $\gamma_r > 16/3$. However, we show in simulations (Fig. 4B of the main text) that ideal cells, and even *E. coli*, can clearly climb the gradient when $g = 0.05\ \mathrm{mm}^{-1}$, $c_0 = 1\ \mu\mathrm{M}$, and $\gamma_r = 0.015 \ll 16/3$.

## Modeling kinase activity

In shallow gradients, CheA kinases respond approximately linearly to recent signals. We model kinase responses, $a(t)$, to past particle arrival rates, $r(t)$, in background particle arrival rate $r_0$, as:

$$a(t) = a_0 - \int_{-\infty}^{t} K_r(t - t')\,(r(t') - r_0)\, dt' + \eta(t). \tag{70}$$

The response function to particle arrival rate, $K_r(T)$, is:

$$K_r(T) = G_r \left( \left(\frac{1}{\tau_1} + \frac{1}{\tau_2}\right) \exp\left(-\left(\frac{1}{\tau_1} + \frac{1}{\tau_2}\right) T\right) - \frac{1}{\tau_2} \exp\left(-\frac{T}{\tau_2}\right) \right) \Theta(T), \tag{71}$$

where $\tau_1$ and $\tau_2$ above have the same meaning as in our previous work (1). In the main text, we replaced $\left(\frac{1}{\tau_1} + \frac{1}{\tau_2}\right) \to \frac{1}{\tau_1}$ for space, since $\tau_2 \gg \tau_1$.

The Fourier transform of this kernel is:

$$K_r(\omega) = \frac{G_r}{\tau_1} \frac{(-i\,\omega)}{\left(\frac{1}{\tau_2} - i\,\omega\right)\left(\frac{1}{\tau_1} + \frac{1}{\tau_2} - i\,\omega\right)}. \tag{72}$$

In our previous work (1), we modeled responses of kinase activity to past signals $s$ instead of past particle arrival rate $r$. These two descriptions are equivalent in the regime of shallow gradients. We show this below by starting from average responses of kinase activity to particle arrival rate:

$$\langle a(t)\rangle = a_0 - \int_{-\infty}^{t} K_r(t-t')\,(\langle r(t')\rangle - r_0)\,dt'\,, \tag{73}$$

where angled brackets indicate averaging over repeated presentation of the same signal trajectory $\{s\}$, and thus they average out particle noise and kinase noise. From here, we will derive a response kernel to past signals that gives identical kinase responses.

First, we note that:

$$\langle r(t)\rangle - r_0 = k_D\,(c(t)-c_0) = r_0 \int_{-\infty}^{t} s(t')\,dt'\,, \tag{74}$$

where we used $s(t) \approx \frac{1}{c_0}\frac{dc}{dt}$ in shallow gradients. Then, it is convenient to transform the expressions above to Fourier space, where $\delta a(\omega) = F[\langle a(t)\rangle - a_0]$ , $\delta r(\omega) = F[\langle r(t)\rangle - r_0]$ , and $K_r(\omega) = F[K_r(T)]$, and again $F[f(t)] = \int_{-\infty}^{\infty} f(t)\,e^{i\,\omega\,t}\,dt$ is the Fourier transform. Then we have:

$$\delta a(\omega) = -K_r(\omega)\,\delta r(\omega), \tag{75}$$

$$\delta r(\omega) = r_0 \frac{s(\omega)}{-i\,\omega}. \tag{76}$$

With this, we get:

$$\delta a(\omega) = -K_r(\omega)\,r_0 \frac{s(\omega)}{-i\,\omega} = -K(\omega)\,s(\omega) \tag{77}$$

where $K(\omega) = r_0 \frac{K_r(\omega)}{-i\,\omega}$ is the Fourier transform of the linear response function to signals. Thus, we can either write down average kinase responses to particle arrival rate $r(t)$, with linear response function $K_r(T)$, or responses to signals $s(t)$, with linear response function $K(T)$ (1):

$$K(T) = r_0 \int_0^T K_r(t')\,dt' = G \exp\left(-\frac{T}{\tau_2}\right)\left(1 - \exp\left(-\frac{T}{\tau_1}\right)\right)\Theta(T). \tag{78}$$

Here, we have defined the MWC model gain $G = r_0\,G_r$ (22,23). In the MWC model, kinase-receptor complexes can be in active or inactive states. The dissociation constants for the attractant in each state, $K_i$ and $K_a$, are different, with $K_i \ll K_a$, which causes attractant concentration to influence the fraction of kinases in the active state. When the background concentration $c_0 \ll K_a$, the gain of the kinase response to changes in log-concentration of attractant can be written:

$$G(c_0) \approx G_\infty \frac{c_0}{c_0 + K_i}. \tag{79}$$

where $G_\infty$ is the "log-sensing" gain (when $c_0 \gg K_i$).

We can use the response function to particle arrivals, $K_r(T)$, to compute the power spectrum of particle counting noise filtered through the kinase response kernel, $K_r(T)$, but expressed it in terms of the response kernel $K(T)$ to signals $s$. Since we model particle arrival noise as shot noise, its power spectrum is constant and equal to $r_0$. Filtering this noise through the response kernel $K_r(\omega)$ gives:

$$N_r(\omega) = r_0\, |K_r(\omega)|^2 = r_0\, \left|\frac{-\,i\,\omega}{r_0}\, K(\omega)\right|^2 = \frac{1}{r_0}\, \omega^2\, |K(\omega)|^2. \tag{80}$$

In experiments, we measure responses to absolute changes in concentration $c(t)$, with response kernel $K_c(t)$, which has the same form as $K_r(T)$ above, but with gain $G_c$. Then, we convert $G_c$ to $G_r$ via $G_r = G_c/k_D$, and thus convert $K_c(T)$ to $K_r(T)$. With this, the intensity of filtered particle noise in Eqn. 80 is proportional to $G_r^2\, r_0 = G_c^2\, c_0/k_D$. This conversion implies that *E. coli* respond to every particle arriving at their receptors, which is unlikely. Instead, one might use an effective $k_D^{eff} < k_D$ to do the conversion above, which would increase our estimate for the intensity of filtered particle noise, being proportional to $G_c^2\, c_0/k_D^{eff}$. However, modeling the filtered particle noise with $k_D^{eff} = k_D$ maximizes our estimate of *E. coli*'s information rate. Since we find that *E. coli* are far from the physical limit, this is a conservative modeling choice. Furthermore, we believe $k_D^{eff}$ can't be much smaller than $k_D$, otherwise kinase activity noise would exhibit significant concentration-dependence, which we do not observe experimentally (SI Fig. S1C).

Next, we consider modeling noise in kinase activity. As explained in the main text and in Fig. S3, the FRET system we use for measuring kinase activity has limited time resolution, about $0.3\ s$. This allows us to constrain slow fluctuations in kinase activity, whose correlation function is characterized by a single decaying exponential function (1,24):

$$\langle \eta(t)\eta(t')\rangle = N_n(t-t') = \sigma_n^2 \exp\left(-\frac{|t-t'|}{\tau_n}\right) = D_n\, \tau_n \exp\left(-\frac{|t-t'|}{\tau_n}\right). \tag{81}$$

The parameters here are the long-time variance $\sigma_n^2$ and the correlation time $\tau_n$, which are related to the diffusivity of the noise by $D_n = \sigma_n^2/\tau_n$. The power spectrum of this noise is

$$N_n(\omega) = \frac{2\, D_n}{\frac{1}{\tau_n^2} + \omega^2}. \tag{82}$$

There can also be noise at higher frequencies that we don't observe. Kinase responses to particle arrival noise set a minimum noise level at all frequencies. At high frequencies, simply extrapolating the power spectrum in Eqn. 82 drops below the implied filtered particle noise in Eqn. 80 if we take $\tau_1$ to be the value measured previously in biochemical studies (25,26), $\tau_1 \approx 1/60\ s$. One possibility is that cooperativity of the receptor-kinase lattice slows down $\tau_1$ to a value closer to what we measure in FRET, $\tau_1 \approx 0.35\ s$. In this case, extrapolating the slow noise to high frequencies does not cause any problems.

To avoid having unphysical noise power at high frequencies, we take the total noise in kinase activity to be a sum of the measured slow noise in Eqn. 82 plus the filtered particle arrival noise in Eqn. 80. There are likely other noise sources at high frequencies, so this modeling choice maximizes our estimate of *E. coli*'s information rate. Since we find that *E. coli* are far from the physical limit, this is a conservative modeling choice. Ultimately, even if we only model noise in kinase activity as being the slow, measurable noise, the effects on the numerical values of the information rate are small.

We will make an additional simplifying assumption. The adaptation time of kinase responses, $\tau_2$, and the correlation time of kinase noise, $\tau_n$, are each roughly $\sim 10\ s$. Therefore, below we will also assume $\tau_2 \approx$ $\tau_n$, which also has small quantitative effects on the results. These simplifications also allow us to derive interpretable analytical expressions.

## Derivation of the behaviorally-relevant information rate in kinase activity

In this section, we derive the information about current signal encoded in the kinase activity of a typical *E. coli* cell. Here, we seek an expression for the following transfer entropy rate:

$$\dot{I}^{*}_{s\to a} = \lim_{dt\to 0} \frac{1}{dt} I(a(t+dt); s(t)|\{a(t)\})$$
$$= -\left[\partial_\tau I\big(\{a(t)\}; s(t+\tau)\big)\right]_{\tau=0}. \tag{83}$$

Again, the calculation centers on calculating the mutual information between past kinase activity $a$ and signal at some time $\tau$ into the future, $I\big(\{a(t)\}; s(t+\tau)\big)$. The quantity we need to derive this is the posterior distribution of signal given past kinase activity, $P(s(t+\tau)|\{a(t)\})$. Past measurements by us and others (1,24,27) have shown that kinase activity in wild type cells (i.e. cells with all receptor types and with their adaptation system intact) is well-approximated by a Gaussian process. Because of this, and because we consider shallow gradients, we only need the variance of $P(s(t+\tau)|\{a(t)\})$ to compute the mutual information to leading order in $g$ (see the section **Derivation of the behaviorally-relevant information rate in particle arrivals**, above). Thus, we can approximate $s$ and $a$ as jointly Gaussian distributed.

With the approximation that $s$ and $a$ are also jointly Gaussian distributed, $P(s(t+\tau)|\{a(t)\})$ is Gaussian, and therefore we again need to compute a mean $\hat{s}_a(t+\tau)$ and a variance $\sigma^2_{s|a}(\tau)$. Then, the mutual information can then be computed from:

$$I\big(\{a(t)\}; s(t+\tau)\big) = \frac{1}{2}\log\left(\frac{\sigma_s^2}{\sigma^2_{s|a}(\tau)}\right) = -\frac{1}{2}\log\big(1-\rho^2_{as}(\tau)\big), \tag{84}$$

and the predictive information rate is

$$\dot{I}^{*}_{s\to a} = \frac{1}{2}\left[\frac{-\partial_\tau \rho^2_{as}(\tau)}{1-\rho^2_{as}(\tau)}\right]_{\tau=0} \tag{85}$$

Here, $\rho^2_{as}(\tau) = 1 - \frac{\sigma^2_{s|a}(\tau)}{\sigma_s^2}$ is the generalized correlation between $s(t+\tau)$ and past $a$, or the fraction reduction of variance in $s(t+\tau)$ upon observing past $a$.

To compute the rate of information transfer from current signal $s(t)$ to kinase activity $a(t)$, we need the conditional mean and variance of $s(t+\tau)$, $\hat{s}_a(t+\tau)$ and $\sigma^2_{s|a}(\tau)$. These in turn require deriving the kernel $M_a(T)$ that maps past kinase activity $a$ to the conditional mean, $\hat{s}_a(t+\tau)$. This can again be derived using Wiener filtering theory and expressed in terms of the power spectra of $s$ and $a$. These are:

$$S_s(\omega) = F[C_s(T)] = \frac{2\, g^2 \frac{\sigma_v^2}{\tau_v}}{\frac{1}{\tau_v^2} + \omega^2} \tag{86}$$

$$S_a(\omega) = F[C_a(T)] = |K_r(\omega)|^2\, S_r(\omega) + N_{n(\omega)} \tag{87}$$

$$= |K_r(\omega)|^2\, \left( r_0^2 \frac{S_s(\omega)}{\omega^2} + r_0 \right) + \frac{2\, D_n}{\frac{1}{\tau_2^2} + \omega^2} \tag{88}$$

$$= \left(\frac{G_r}{\tau_1}\right)^2 \frac{\omega^2}{\left(\frac{1}{\tau_2^2} + \omega^2\right)\left(\left(\frac{1}{\tau_1} + \frac{1}{\tau_2}\right)^2 + \omega^2\right)} \left( r_0^2 \frac{S_s(\omega)}{\omega^2} + r_0 \right) + \frac{2\, D_n}{\frac{1}{\tau_2^2} + \omega^2} \tag{89}$$

$$S_{as}(\omega) = S_{sa}^*(\omega) = F[C_{as}(T)] = -K_r^*(\omega)\, S_{rs}(\omega) \tag{90}$$

$$= -\frac{G_r}{\tau_1} \frac{r_0}{\left(\frac{1}{\tau_2} + i\,\omega\right)\left(\frac{1}{\tau_1} + \frac{1}{\tau_2} + i\,\omega\right)}\, S_s(\omega) \tag{91}$$

where $C_s(T) = \langle s(t)\, s(t+T)\rangle$ , $C_a(T) = \langle (a(t) - a_0)\, (a(t+T) - a_0)\rangle$ , and $C_{as}(T) = \langle (a(t) - a_0)\, s(t+T)\rangle$. The first term in $S_a(\omega)$ comes from responses to signals, the second term comes from filtered particle arrival noise, and the third term comes from internal kinase noise. For convenience, we will define $\tau_3^{-1} = \tau_1^{-1} + \tau_2^{-1}$.

We now need to decompose $S_a(\omega)$ into the product of a causal and an anti-causal part by finding its zeros and poles. The zeros satisfy $S_a(\omega = i\, z_a) = 0$ are complex solutions to the equation:

$$\frac{G_r^2}{\tau_1^2} \left( 2\, r_0^2\, g^2\, \frac{\sigma_v^2}{\tau_v} + r_0\, \omega^2 \left( \frac{1}{\tau_v^2} + \omega^2 \right) \right) + 2\, D_n \left( \frac{1}{\tau_v^2} + \omega^2 \right) \left( \frac{1}{\tau_3^2} + \omega^2 \right) = 0. \tag{92}$$

This can be written in terms of the particle arrival SNR, $\gamma_r = 2\, r_0\, g^2\, \sigma_v^2\, \tau_v^3$, and the ratio of the diffusivity of filtered particle noise and the diffusivity of slow kinase noise, $R = \frac{1}{2} \frac{G_r^2}{\tau_1^2} \frac{r_0}{D_n}$:

$$R\, \left(\gamma_r + \tau_v^2\, \omega^2\, (1 + \tau_v^2\, \omega^2)\right) + (1 + \tau_v^2\, \omega^2) \left( \frac{\tau_v^2}{\tau_3^2} + \tau_v^2\, \omega^2 \right) = 0 \tag{93}$$

The zeros of $S_a(\omega)$ are:

$$i\, z_{a,1} = i \frac{1}{\tau_v} \frac{1}{\sqrt{2\,(1+R)}} \sqrt{\left(\frac{\tau_v}{\tau_3}\right)^2 + (1+R) - \sqrt{(1+R)\big(1 + R\,(1 - 4\,\gamma_r)\big) - 2\,(1+R)\left(\frac{\tau_v}{\tau_3}\right)^2 + \left(\frac{\tau_v}{\tau_3}\right)^4}},$$

$$i\, z_{a,2} = i \frac{1}{\tau_v} \frac{1}{\sqrt{2\,(1+R)}} \sqrt{\left(\frac{\tau_v}{\tau_3}\right)^2 + (1+R) + \sqrt{(1+R)\big(1 + R\,(1 - 4\,\gamma_r)\big) - 2\,(1+R)\left(\frac{\tau_v}{\tau_3}\right)^2 + \left(\frac{\tau_v}{\tau_3}\right)^4}}, \tag{94}$$

as well as their complex conjugates.

The poles of $S_a(\omega)$ satisfy $\frac{1}{S_a(\omega=i\,p_a)}=0$ and are $i\,p_{a,1}=i\frac{1}{\tau_v}$, $i\,p_{a,2}=i\frac{1}{\tau_2}$, and , $i\,p_{a,3}=i\frac{1}{\tau_3}$, as well as their complex conjugates.

We decompose $S_a(\omega)$ as:

$$S_a(\omega)=\phi_a(\omega)\,\phi_a^*(\omega) \tag{95}$$

where

$$\phi_a(\omega)=\sqrt{2\,D_n\,(1+R)}\frac{\left(z_{a,1}-i\,\omega\right)\left(z_{a,2}-i\,\omega\right)}{\left(p_{a,1}-i\,\omega\right)\left(p_{a,2}-i\,\omega\right)\left(p_{a,3}-i\,\omega\right)}. \tag{96}$$

Next, we need the causal part of the following (see Appendix A):

$$\frac{S_{as}(\omega)}{\phi_a^*(\omega)}e^{-i\,\omega\,\tau}=-\frac{G_r/\tau_1}{\sqrt{2\,D_n\,(1+R)}}\frac{2\,r_0\,g^2\,\sigma_v^2\,\frac{1}{\tau_v}}{\left(\frac{1}{\tau_v}-i\,\omega\right)\,\left(z_{a,1}+i\,\omega\right)\left(z_{a,2}+i\,\omega\right)}e^{-i\,\omega\,\tau} \tag{97}$$

Again, we find the causal part of this expression by doing a partial fraction decomposition and keeping only the terms with poles and zeros that have negative imaginary part:

$$\frac{S_{as}(\omega)}{\phi_a^*(\omega)}e^{-i\,\omega\,\tau}=\frac{A}{\left(\frac{1}{\tau_v}-i\,\omega\right)}+\frac{B}{\left(z_{a,1}+i\,\omega\right)}+\frac{C}{\left(z_{a,2}+i\,\omega\right)}, \tag{98}$$

for unknown $A$, $B$, and $C$. Only the pole of the first term (at $\omega=-i\frac{1}{\tau_v}$) has negative imaginary part, so we only need to compute $A$ to get the causal part of this expression. This is:

$$A=\left[-\frac{G_r/\tau_1}{\sqrt{2\,D_n\,(1+R)}}\frac{2\,r_0\,g^2\,\sigma_v^2\,\frac{1}{\tau_v}}{\left(z_{a,1}+i\,\omega\right)\left(z_{a,2}+i\,\omega\right)}e^{-i\,\omega\,\tau}\right]_{\omega=-i\frac{1}{\tau_v}} \tag{99}$$

$$=-\frac{G_r/\tau_1}{\sqrt{2\,D_n\,(1+R)}}\frac{2\,r_0\,g^2\,\sigma_v^2\,\tau_v}{\left(1+z_{a,1}\,\tau_v\right)\left(1+z_{a,2}\,\tau_v\right)}e^{-\frac{\tau}{\tau_v}}, \tag{100}$$

and the causal part of $S_{as}(\omega)/\phi_a^*(\omega)$ is then:

$$\left[\frac{S_{as}(\omega)}{\phi_a^*(\omega)}e^{-i\,\omega\,\tau}\right]^+=-\frac{G_r/\tau_1}{\sqrt{2\,D_n\,(1+R)}}\frac{2\,r_0\,g^2\,\sigma_v^2\,\tau_v}{\left(1+z_{a,1}\,\tau_v\right)\left(1+z_{a,2}\,\tau_v\right)}\frac{e^{-\frac{\tau}{\tau_v}}}{\left(\frac{1}{\tau_v}-i\,\omega\right)}. \tag{101}$$

Finally, like $C_{rs}(\tau)$ in the section above, $C_{as}(\tau)\propto\exp\left(-\frac{\tau}{\tau_v}\right)$ when $\tau\geq 0$.

With these expressions, the optimal kernel that computes the mean of $p(s(t+\tau)|\{a\})$ is (Appendix A):

$$M_a(\omega)=\frac{1}{\phi_a(\omega)}\left[\frac{S_{as}(\omega)}{\phi_a^*(\omega)}e^{-i\,\omega\,\tau}\right]^+ \tag{102}$$

$$= -e^{-\frac{\tau}{\tau_v}} \frac{2\frac{G_r}{\tau_1}\, r_0\, g^2\, \sigma_v^2\, \tau_v}{2\, D_n\, (1+R)\, \left(1+z_{a,1}\, \tau_v\right)\left(1+z_{a,2}\, \tau_v\right)} \frac{\left(\frac{1}{\tau_2} - i\,\omega\right)\left(\frac{1}{\tau_3} - i\,\omega\right)}{\left(z_{a,1} - i\,\omega\right)\left(z_{a,2} - i\,\omega\right)} \tag{103}$$

We discuss this kernel in the following section.

The variance of $P(s(t+\tau)|\{a\})$, $\sigma_{s|a}^2(\tau)$, is (Appendix A, Eqn. 181):

$$\sigma_{s|a}^2(\tau) = \sigma_s^2 - \frac{1}{2\pi}\int_{-\infty}^{\infty} S_{as}^*(\omega)\, e^{i\,\omega\,\tau}\, M_a(\omega)\, d\omega \tag{104}$$

$$= \sigma_s^2 \left(1 - e^{-2\frac{\tau}{\tau_v}} \frac{2\frac{G_r^2}{\tau_1^2}\, r_0^2\, g^2\, \sigma_v^2\, \tau_v^3}{2\, D_n\, (1+R)\left(1+z_{a,1}\, \tau_v\right)^2 \left(1+z_{a,2}\, \tau_v\right)^2}\right), \tag{105}$$

where $\sigma_s^2 = g^2\, \sigma_v^2$. Therefore, the correlation coefficient $\rho_{as}^2(\tau)$ is:

$$\rho_{as}^2(\tau) = 1 - \frac{\sigma_{s|a}^2(\tau)}{\sigma_s^2} = e^{-2\frac{\tau}{\tau_v}} \frac{2\frac{G_r^2}{\tau_1^2}\, r_0^2\, g^2\, \sigma_v^2\, \tau_v^3}{2\, D_n\, (1+R)\left(1+z_{a,1}\, \tau_v\right)^2 \left(1+z_{a,2}\, \tau_v\right)^2}, \tag{106}$$

or in terms of $\gamma_r = 2\, r_0\, g^2\, \sigma_v^2\, \tau_v^3$ and $R = \frac{1}{2}\frac{G_r^2}{\tau_1^2}\frac{r_0}{D_n}$:

$$= e^{-2\frac{\tau}{\tau_v}} \frac{R}{(1+R)} \frac{\gamma_r}{\left(1+z_{a,1}\, \tau_v\right)^2 \left(1+z_{a,2}\, \tau_v\right)^2}. \tag{107}$$

Finally, using Eqn. 85 above, we find that the information about current signal encoded in *E. coli*'s kinase activity is:

$$\dot{I}_{s\to a}^* = \frac{1}{\tau_v} \frac{\rho_{as}^2(\tau=0)}{1-\rho_{as}^2(\tau=0)} = \frac{1}{\tau_v} \frac{\frac{R}{(1+R)} \frac{\gamma_r}{\left(1+z_{a,1}\, \tau_v\right)^2 \left(1+z_{a,2}\, \tau_v\right)^2}}{1 - \frac{R}{(1+R)} \frac{\gamma_r}{\left(1+z_{a,1}\, \tau_v\right)^2 \left(1+z_{a,2}\, \tau_v\right)^2}}. \tag{108}$$

In shallow gradients, $\dot{I}_{s\to a}^* \approx \frac{1}{\tau_v}\rho_{as}^2(\tau=0) \approx \frac{2}{\tau_v} I\left(\{a(t)\}; s(t)\right)$, and only the leading order $g^2$ term of $\dot{I}_{s\to a}^*$ contributes to the final expression. Since $\gamma_r \propto g^2$ in Eqn. 107, we can get the shallow-gradient expression for $\dot{I}_{s\to a}^*$ by evaluating $z_{a,1}$ and $z_{a,2}$ at $g=0$. This is equivalent to taking $\gamma_r \to 0$, which gives:

$$z_{a,1} \approx \frac{1}{\tau_3\sqrt{1+R}}, \qquad z_{a,2} \approx \frac{1}{\tau_v}. \tag{109}$$

Thus, in shallow gradients, we get:

$$\dot{I}_{s\to a}^* \approx \frac{1}{\tau_v}\frac{1}{4}\frac{R}{1+R}\frac{\gamma_r}{\left(1+\frac{\tau_v}{\tau_3}\frac{1}{\sqrt{1+R}}\right)^2} = \frac{1}{\tau_v}\frac{1}{4}\frac{\frac{1}{2}\frac{G_r^2}{\tau_1^2}\frac{r_0}{D_n}}{1+\frac{1}{2}\frac{G_r^2}{\tau_1^2}\frac{r_0}{D_n}} \frac{2\, r_0\, g^2\, \sigma_v^2\, \tau_v^3}{\left(1+\frac{\tau_v}{\tau_3}\left(1+\frac{1}{2}\frac{G_r^2}{\tau_1^2}\frac{r_0}{D_n}\right)^{-1/2}\right)^2}, \tag{110}$$

where again $\tau_3^{-1} = \tau_1^{-1} + \tau_2^{-1}$. Furthermore, since $\tau_1 \ll \tau_v$, taking $\tau_1 \to 0$ only slightly increases the information rate and gives a simpler expression in terms of a kinase signal to noise ratio, $\gamma_a = \frac{G_r^2}{D_n} r_0^2\, g^2\, \sigma_v^2\, \tau_v$, and the particle arrival signal to noise ratio, $\gamma_r = 2\, r_0\, g^2 \sigma_v^2\, \tau_v^3$:

$$\dot{I}_{s\to a}^* \approx \frac{1}{\tau_v}\frac{1}{4}\,\gamma_a \frac{\frac{\gamma_r}{\gamma_a}}{\left(1+\sqrt{\frac{\gamma_r}{\gamma_a}}\right)^2} = \frac{1}{\tau_v}\frac{1}{4}\frac{G_r^2}{D_n} r_0^2\, g^2\, \sigma_v^2\, \tau_v \frac{\frac{2\, D_n\, \tau_v^2}{G_r^2\, r_0}}{\left(1+\sqrt{\frac{2\, D_n\, \tau_v^2}{G_r^2\, r_0}}\right)^2}. \tag{111}$$

We also note that for finite $g$ but $\tau_1 \to 0$, $z_{a,1}$ and $z_{a,2}$ are:

$$z_{a,1} \approx \frac{1}{\tau_v}\frac{1}{\sqrt{2}}\sqrt{1+\frac{\gamma_r}{\gamma_a} - \sqrt{1 - 4\,\gamma_r - 2\frac{\gamma_r}{\gamma_a} + \left(\frac{\gamma_r}{\gamma_a}\right)^2}},$$

$$z_{a,2} \approx \frac{1}{\tau_v}\frac{1}{\sqrt{2}}\sqrt{1+\frac{\gamma_r}{\gamma_a} + \sqrt{1 - 4\,\gamma_r - 2\frac{\gamma_r}{\gamma_a} + \left(\frac{\gamma_r}{\gamma_a}\right)^2}}. \tag{112}$$

We plugged these expressions into Eqn. 108, with $\frac{R}{1+R} \to 1$ as $\tau_1 \to 0$, to generate the plots in Fig. 3 of the main text.

Eqns. 46, 54, and 108 for the information rates $\dot{I}_{s\to r}^*$ and $\dot{I}_{s\to a}^*$ are nearly exact, but make several assumptions. They require $r_0\, \tau_v \gg 1$ so that we can approximate particle arrivals as Gaussian. They also use Gaussian approximations for the mutual information quantities $I(s(t); \{r\})$ and $I(s(t); \{a\})$, which are valid when these quantities are small (shallow gradients, small $g$). We used linear theory to model kinase responses, which is valid if deviations in kinase activity from baseline are small—i.e. when $g$ is small. And we ignored feedbacks in which responses to signals change the signal statistics that the cell experiences, again valid when $g$ is small. Each of these assumptions can break at a different characteristic value of $g$: for particle arrival rate, small $g$ means $\gamma_r \ll 1$; for kinase activity, small $g$ means $\gamma_a \ll 1$. That all said, Eqns. 46, 54, and 108 currently provide our best analytical insight into information transfer during chemotaxis.

## Optimal kernel for estimating signal from kinase activity

To understand the kernel $M_a(\omega)$ that constructs an estimate of the current signal, $s(t)$, from past kinase activity, $\{a\}$, we first multiply it by the kinase response function of particle arrivals, $K_r(\omega)$. This gives a composite kernel that effectively maps the past of particle arrivals $r$, corrupted by kinase noise, to an estimate of the signal $s(t)$:

$$-M_a(\omega)\, K_r(\omega) = -\frac{2\left(\frac{G_r}{\tau_1}\right)^2 r_0\, g^2\, \sigma_v^2\, \tau_v}{2\, D_n\,(1+R)\,\big(1+z_{a,1}\,\tau_v\big)\big(1+z_{a,2}\,\tau_v\big)} \frac{(-i\,\omega)}{\big(z_{a,1} - i\,\omega\big)\big(z_{a,2} - i\,\omega\big)}. \tag{113}$$

In the time domain, this is:

$$IFT[-M_a(\omega)\, K_r(\omega)] = \frac{2\left(\frac{G_r}{\tau_1}\right)^2 r_0\, g^2\, \sigma_v^2\, \tau_v}{2\, D_n\, (1+R)} \frac{\left(z_{a,2}\, \exp(-z_{a,2}\, t) - z_{a,1}\, \exp(-z_{a,1}\, t)\right)}{(1+z_{a,1}\, \tau_v)(1+z_{a,2}\, \tau_v)(z_{a,1}-z_{a,2})}\, \Theta(t), \quad (114)$$

This composite kernel that effectively acts on particle arrivals has the same structure as the optimal kernel $M_r(T)$ (Eqn. 55) for directly constructing $s(t)$ from particle arrivals. It's biphasic and adapts perfectly, although with different time scales than $M_r(T)$. This means that $M_a(T)$ attempts to invert the kinase response function $K_r(T)$, to the extent possible given the kinase noise $N_n(T)$, and then apply something as close as possible to the optimal kernel for particle counts, $M_r(T)$.

Taking this further, we argue that the optimal kernel acting on particle counts, $M_r(T)$, is the response kernel that the cell should *try* to implement as its behavioral output using its intracellular signaling network (up to changes of units). However, the cell has to communicate information about the signal $s(t)$ through multiple chemical species in order to send them from the kinases at one location to the motors at various other locations. These steps impose constraints on the cell's signaling pathway, and they add noise. Despite this, the cell should be attempting to make its composite kernel from input (particle counts) to output (tumble rate) look like $M_r(T)$.

## Simulation details

In Figs. 4A and 4B of the main text, we performed two types of simulations in which a run-tumble particle moved in three dimensions with rotational diffusion through a concentration gradient. In both cases, we compare "ideal" cells, which sense particle arrivals directly, to "*E. coli*," which respond to particle arrivals with changes in kinase activity. In Fig. 4A, we compute the optimal estimates of the signal in each case, $\hat{s}_r(t)$ and $\hat{s}_a(t)$, but the cells' tumble behaviors do not depend on the signal. Optimal estimates of $s(t)$ were computed using the kernels above. In Fig. 4B, the cells' tumble rates are modulated by their optimal estimates of the signal. We detail these two simulations below.

In both Fig. 4A and 4B, the velocity correlation time $\tau_v$ has contributions from the cell's average tumble rate, $\lambda_0$, the persistence of tumbles, $\alpha$, and rotational diffusion, $D_r$: $\tau_v^{-1} = (1-\alpha)\, \lambda_{R0} + 2\, D_r$. We took $\alpha = 0$ and $D_r = 0.044\ \mathrm{rad}^2/\mathrm{s}$ based on our previous measurements (1), and then computed the tumble rate $\lambda_0$ from the value of $\tau_v$ measured here. Assuming nearly instantaneous tumbles, we computed the swimming speed to use in simulations from $\sigma_v^2 = v_0^2/3$ in 3D.

Tumbles were generated by sampling a uniform random variable $u$, and a tumble occurred if $u < \lambda_0\, \Delta t$. Tumbles lasted one time step. Rotational diffusion during runs was simulated as before (28,29). The simulation was performed by stepping forward in time by discrete steps of size $\Delta t = \min(\tau_v,\ k_1^{-1})\,/100$.

Cells were initialized at position $x(t=0) = 0$ in a concentration gradient $c(x) = c_0 \exp(g\, x)$. At each time step, stochastic molecule arrival rate was computed from $r(t) = k_D\, c\big(x(t)\big) + \sqrt{r_0/dt}\, \eta(t)$, where $\eta(t)$ was a standard Gaussian-distributed random variable simulating shot noise. Kinase activity was computed after the simulation using Eqn. 4 of the main text, with $\tau_1 \to 0$ and otherwise measured parameters (Fig. S1), and the kinase noise $\eta_n(t)$ was simulated separately. Note that any nonlinear model of kinase activity must behave like the measured linear response kernel and noise when the signals are small.

**Simulations without tumble responses to signals (Fig. 4A)**

In Fig. 4A (no tumble response to signals), we simulated run-and-tumble motion with rotational diffusion in a static gradient, as described above, to get $s(t)$, $c(t)$, and $r(t)$. After the simulation, we computed $a(t)$ and the optimal estimates of the signal $s(t)$, $\hat{s}_r(t) = \int_{-\infty}^{t} M_r(t - t')\,(r(t') - r_0)\,dt'$ and $\hat{s}_a(t) = \int_{-\infty}^{t} M_a(t - t')\,(a(t') - a_0)\,dt'$, using expressions above. We used these simulations to check our expressions for the correlation coefficients between the optimal signal estimates and the true signals, $\rho_{rs}^2$ and $\rho_{as}^2$, and thus validate our expressions for the information rates. Specifically, we numerically computed these correlations in the simulations as $\rho_{rs}^2 = \langle s(t)\,\hat{s}_r(t)\rangle^2 / \left(\sigma_s^2\,\sigma_{\hat{s}_r}^2\right)$ and $\rho_{as}^2 = \langle s(t)\,\hat{s}_a(t)\rangle^2 / \left(\sigma_s^2\,\sigma_{\hat{s}_a}^2\right)$ and compared to the theoretical expressions. For this, we simulated 5000 cells for $T = 100\,\tau_v$ and excluded the initial transient ($10\,\tau_v$) from the analysis. Fig. S5 below shows excellent agreement between the theory and simulations.

**Simulations with tumble responses to signals (Fig. 4BC)**

Our main goals were to construct a model that allowed fair comparison between "ideal" and "realistic" cells, and to demonstrate that the loss of information due to kinase noise translates into a reduction of drift speed.

To simulate tumbling in response to estimated signals in Fig. 4B, we needed to simulate kinase activity $a(t)$ and the optimal estimates of the signal. To achieve this, we next derive dynamical systems for $a(t)$, $\hat{s}_r(t)$, and $\hat{s}_a(t)$. In general, if a quantity is the convolution of some variable, $h(t)$, with a kernel $g(T)$:

$$f(t) = \int_{-\infty}^{t} g(t - t')\,h(t')\,dt', \tag{115}$$

then its time derivative is:

$$\frac{df}{dt} = g(0)\,h(t) + \int_{-\infty}^{t} g'(t - t')\,h(t')\,dt'. \tag{116}$$

First consider kinase activity,

$$a(t) = a_0 - \int_{-\infty}^{t} K_a(t - t')\,(r(t') - r_0)\,dt' + \eta_n(t). \tag{117}$$

As $\tau_1 \to 0$, this becomes

$$a(t) = a_0 - G_r\left((r(t) - r_0) - \frac{1}{\tau_2}\int_{-\infty}^{t} \exp\left(-\frac{(t - t')}{\tau_2}\right)(r(t') - r_0)\,dt'\right) + \eta_n(t). \tag{118}$$

We can write down a dynamical system for the deterministic part of $a(t)$ by defining the convolution term above as a new variable:

$$a_r(t) = \frac{G_r}{\tau_2}\int_{-\infty}^{t} \exp\left(-\frac{(t - t')}{\tau_2}\right)(r(t') - r_0)\,dt'. \tag{119}$$

Then, using Eqn. 116, its time derivative is:

$$\frac{d}{dt}a_r(t) = \frac{G_r}{\tau_2}\,(r(t) - r_0) - \frac{1}{\tau_2}a_r(t). \tag{120}$$

The internal noise $\eta_n(t)$ is an Ornstein-Uhlenbeck process with diffusivity $D_n$ and time scale $\tau_n \approx \tau_2$, which we simulated using standard methods. Then $a(t)$ at each time step was computed from the dynamical variables $r(t)$, $a_r(t)$, and $\eta(t)$ using:

$$a(t) = a_0 - G_r(r(t) - r_0) + a_r(t) + \eta_n(t). \tag{121}$$

To simulate $\hat{s}_r(t)$, we first use its definition, $\hat{s}_r(t) = \int_{-\infty}^{t} M_r(t-t')\,(r(t') - r_0)\,dt'$, where the optimal kernel $M_r(T)$ is (Eqn. 59, $\gamma > 1/4$):

$$M_r(t) = A\exp(-k_1\,t)\,\left(\cos(k_2\,t) - \frac{k_1}{k_2}\sin(k_2\,t)\right),$$

and

$$A = \frac{\gamma_r}{r_0\,\tau_v^2}\frac{1}{\left(1 + \sqrt{1 + 2\sqrt{\gamma_r}} + \sqrt{\gamma_r}\right)}, \tag{122}$$

$$k_1 = \frac{1}{\tau_v}\frac{1}{2}\sqrt{2\sqrt{\gamma_r} + 1}, \tag{123}$$

$$k_2 = \frac{1}{\tau_v}\frac{1}{2}\sqrt{2\sqrt{\gamma_r} - 1}. \tag{124}$$

We can write down a dynamical system to simulate $\hat{s}_r(t)$ by defining the following two variables:

$$\hat{s}_r^1(t) = A\int_{-\infty}^{t} \exp\big(-k_1\,(t-t')\big)\cos\big(k_2\,(t-t')\big)\,\,(r(t') - r_0)\,dt'\,, \tag{125}$$

$$\hat{s}_r^2(t) = A\int_{-\infty}^{t} \exp\big(-k_1\,(t-t')\big)\sin\big(k_2\,(t-t')\big)\,\,(r(t') - r_0)\,dt'\,, \tag{126}$$

and using $\hat{s}_r(t) = \hat{s}_r^1(t) - \frac{k_1}{k_2}\hat{s}_r^2(t)$.

We can derive a closed system of equations for the dynamics of these variables using Eqn. 116:

$$\frac{d}{dt}\hat{s}_r^1(t) = A\,(r(t) - r_0) - k_1\,\hat{s}_r^1(t) - k_2\,\hat{s}_r^2(t), \tag{127}$$

$$\frac{d}{dt}\hat{s}_r^2(t) = k_2\,\hat{s}_r^1(t) - k_1\,\hat{s}_r^2(t). \tag{128}$$

A similar approach can be used to derive dynamics $\hat{s}_a(t) = \int_{-\infty}^{t} M_a(t-t')\,(a(t') - a_0)\,dt'$. For $\tau_1 \ll \tau_v$ and $\tau_n \approx \tau_2$, which are consistent with our measurements, $M_a(t)$ has the form (inverse Fourier transform of Eqn. 103, after taking parameter limits):

$$M_a(t) = -(B_3\exp(-k_3\,t) - B_4\exp(-k_4\,t)), \tag{129}$$

where $k_3 = z_{a,1}$, $k_4 = z_{a,2}$ (Eqn. 94), and

$$B_3 = \frac{2\,g^2\,\sigma_v^2\,\tau_v}{G_r}\frac{\left(z_{a,1}-\frac{1}{\tau_2}\right)}{\left(z_{a,1}-z_{a,2}\right)\left(1+z_{a,1}\,\tau_v\right)\left(1+z_{a,2}\,\tau_v\right)}, \tag{130}$$

$$B_4 = \frac{2\,g^2\,\sigma_v^2\,\tau_v}{G_r}\frac{\left(z_{a,2}-\frac{1}{\tau_2}\right)}{\left(z_{a,1}-z_{a,2}\right)\left(1+z_{a,1}\,\tau_v\right)\left(1+z_{a,2}\,\tau_v\right)}. \tag{131}$$

Next, we define the following variables:

$$\hat{s}_a^1(t) = -B_3 \int_{-\infty}^{t} \exp\left(-k_3\,(t-t')\right)(a(t')-a_0)\,dt'\,, \tag{132}$$

$$\hat{s}_a^2(t) = -B_4 \int_{-\infty}^{t} \exp\left(-k_4\,(t-t')\right)\,(a(t')-a_0)\,dt'\,, \tag{133}$$

and use $\hat{s}_a(t) = \hat{s}_a^1(t) - \hat{s}_a^2(t)$.

By similar steps as above, the dynamics of these variables are:

$$\frac{d}{dt}\hat{s}_a^1(t) = -B_3\,(a(t')-a_0) - k_3\,\hat{s}_a^1(t), \tag{134}$$

$$\frac{d}{dt}s_a^2(t) = -B_4\,(a(t')-a_0) - k_4\,\hat{s}_a^2(t). \tag{135}$$

Next, the optimal estimates of signals, $\hat{s}_r(t)$ and $\hat{s}_a(t)$, modulate the cells' tumble rates. For both ideal and realistic cells, we again modeled tumbling as a Poisson process, but with time-varying rate depending on the estimated signal from their available observations, $\lambda(t) = \lambda\left(\hat{s}_r(t)\right)$ or $\lambda(t) = \lambda\left(\hat{s}_a(t)\right)$. Tumbling is a stochastic process, and the "signal to noise" ratio of this process depends on the variance of the input that drives the tumble rate. To allow fair comparisons between the ideal and realistic cells, we modeled the tumble rates as functions of rescaled optimal estimators $\hat{s}_r(t)$ and $\hat{s}_a(t)$ such that the two estimators had the same variance, which was also independent of $g$:

$$\lambda(t) = \lambda\left(\frac{\hat{s}(t)}{v_0\,g\,\rho}\right), \tag{136}$$

where $\hat{s}(t) = \hat{s}_r(t)$ and $\rho = \rho_{rs}$ for ideal cells, and $\hat{s}(t) = \hat{s}_a(t)$ and $\rho = \rho_{as}$ for realistic cells. Rescaling the optimal estimates of the signal does not change the information rates. The factor of $\rho$ corrects for the fact the posterior mean, $\hat{s}(t)$, is shrunken towards zero by a factor of $\rho$. Note that our goal here is not to develop a mechanistic biochemical model, which has been pursued elsewhere (28,30–35). Rather, we aim to examine the effects of information on gradient climbing in a minimal setting that allows fair comparison between ideal cells, with access to information $\dot{I}^*_{s\to r}$, and *E. coli*-like cells, with access to information $\dot{I}^*_{s\to a}$.

Defining $y(t) = \frac{\hat{s}(t)}{v_0\,g\,\rho}$ for short, the tumble rate was parameterized as follows:

$$\lambda(t) = \lambda_0 \frac{A(G_m, \Delta)}{1+\exp(B(G_m,\Delta)\,G_m\,y(t)+\Delta)}. \tag{137}$$

The constants $A(G_m, \Delta)$ and $B(G_m, \Delta)$ were determined by enforcing the baseline tumble rate in the absence of signal $\lambda(y = 0) = \lambda_0$, and by enforcing that the linearized tumble rate for small signals was always

$$\lambda(t) \approx \lambda_0 \left(1 - G_m\, y(t)\right). \quad (138)$$

Thus, $G_m$ is the gain of the tumble response to the input $y(t)$. The offset parameter $\Delta$ determined where on the "tumble curve" cells lie at baseline, and also determined the maximum possible tumble rate when $y \ll 0$. For example, $\Delta \to \infty$ gives $\lambda(t) = \lambda_0 \exp\left(-G_m\, y(t)\right)$, a functional form that has been used before by others (e.g. (30)). In general,

$$A = 1 + \exp(\Delta) \quad (139)$$

$$B = \frac{1 + \exp(\Delta)}{\exp(\Delta)} \quad (140)$$

We chose $\Delta = 1$ and $G_m = 4$ so that the chemotaxis coefficient, $\chi$, of the "realistic" cells was comparable to values we measured previously for *E. coli* climbing gradients of methyl-aspartate (1).

We simulated $N_{cells} = 10{,}000$ cells for $T = 100\,\tau_v$ and excluded the initial transient ($10\,\tau_v$) from the analysis. For the smallest value of $g = 0.05\ \mathrm{mm}^{-1}$, we simulated $N_{cells} = 50{,}000$ cells. Drift speeds were computed as the average up-gradient velocity over time and cells, $v_d = \langle v_x(t) \rangle$, and standard errors were computed as $\sigma_{v_d}^2 = Var\left(v_x(t)\right)/(N_{cells}\, T/\tau_v)$, where the denominator is approximately the number of independent samples. Information rates were computed from simulations by computing the correlation coefficient between the true and estimated signals and plugging into Eqn. $46$ or Eqn. $108$, e.g. $\dot{I} = \frac{1}{\tau_v} \frac{\rho^2}{1-\rho^2}$. The numerically-computed information rates agreed with our theoretical expressions in shallow gradients, where feedback from signals onto behavior was weak (29). In steeper simulated gradients, the information rate in simulations was larger than in our theory, up to a factor of 2 for ideal cells in $g = 0.4\ \mathrm{mm}^{-1}$, mostly because the actual $\tau_v$ in simulation dynamically became longer than the value we input to the simulation.

## Estimating population variability in $\boldsymbol{\eta}$

Although the median phenotype in our strain's population is far from the molecule-counting limit on chemical sensing, individual cells with the same genes exhibit large non-genetic differences. Therefore, it's possible that a subset of the population could approach the bound. To tentatively test this possibility, we use maximum likelihood estimation.

Due to variations in swimming, kinase response, and kinase noise parameters among cells, there is a distribution of $\eta = \dot{I}^*_{s\to a}/\dot{I}^*_{s\to r}$ in the population when exposed to background concentration $c_0$ and gradient steepness $g$. If we could measure all parameters in the same individual cells, we could try to construct this distribution directly. Instead, our experimental setup gives single-cell swimming, response, and noise parameters in different cells. With these single-cell parameters, we can make tentative inferences about the variation in $\eta$ in the population if we assume that swimming, response, and noise parameters are uncorrelated in single cells. This is not necessarily true—for example, cells with higher kinase gain, $G_r$, might also have larger kinase fluctuations, $\sigma_n^2$—but it gives a first estimate. Our single-cell parameters also have uncertainties, and we want to account for this in our estimation.

Using the data we have, we can draw $N$ sample parameters and their uncertainties from single cells in our data set. We chose $N \approx 100$ in a given $c_0$ condition to be the smaller of the number of cells measured in our kinase response experiment in that condition and the number of cells measured in our kinase noise experiment in that condition. In particular, we draw all kinase response parameters are taken from one measured cell and all noise parameters from another measured cell. For swimming parameters, we sample a value of tumble bias from the TB distribution and take the average run duration and swimming speed associated with that tumble bias.

For each sample parameter set $i$, we compute a noise-corrupted estimate of $\eta_i$, where noise comes from the uncertainties in the single-cell parameters. When we transform the noisy observations $\eta_i \to y_i = \log\left(\frac{\eta_i}{1-\eta_i}\right)$, empirically the collection of sampled values of $y_i \in (-\infty, \infty)$ is roughly Gaussian distributed. Therefore, we assume a model in which $y_i = x_i + \xi_i$, where $x_i$ is the true single-cell value, which is Gaussian-distributed due to phenotypic variation in the population with mean $\mu$ and variance $S$, and $\xi_i$ is zero-mean Gaussian noise due to our parameter uncertainties with variance $\sigma_i^2$. Since we know the parameter uncertainties, we know $\sigma_i^2$ for each sample $y_i$, from which we seek to infer $\mu$ and $S$.

Mathematically, we have:

$$P(\{y\},\{x\}|\mu,S) = \prod_{i=1}^{N} P(y_i|x_i)\, P(x_i|\mu,S) \tag{141}$$

$$= \prod_{i=1}^{N} \frac{1}{\sqrt{2\pi\sigma_i^2}} \exp\left(-\frac{1}{2}\frac{(y_i - x_i)^2}{\sigma_i^2}\right) \frac{1}{\sqrt{2\pi S}} \exp\left(-\frac{1}{2}\frac{(x_i-\mu)^2}{S}\right). \tag{142}$$

$P(x_i|\mu,S)$ is the probability density of a single cell from the population having phenotype $x_i$. $P(y_i|x_i)$ is the probability distribution of the noise added due to our parameter uncertainties, resulting in observation $y_i$.

Since we only observe $\{y\}$, and not the true phenotypes $\{x\}$, we marginalize out $\{x\}$:

$$P(\{y\}|\mu,S) = \int P(\{y\},\{x\}|\mu,S)\, d\{x\} = \prod_{i=1}^{N} P(y_i|\mu,S) \tag{143}$$

$$= \prod_{i=1}^{N} \frac{1}{\sqrt{2\pi\left(\sigma_i^2 + S\right)}} \exp\left(-\frac{1}{2}\frac{(y_i-\mu)^2}{\sigma_i^2 + S}\right). \tag{144}$$

This is the likelihood of seeing noise-corrupted phenotypes $\{y\}$ if the true parameters were $\mu$ and $S$. With this, we can estimate the population mean $\mu$ and variance $S$ by maximum likelihood estimation. The log-likelihood of the data $\{y\}$ is:

$$L(\mu,S)=\log\big(P(\{y\}|\mu,S)\big)=-\frac{1}{2}\sum_{i=1}^{N}\left(\log\left(2\pi\left(\sigma_i^2+S\right)\right)+\frac{(y_i-\mu)^2}{\sigma_i^2+S}\right). \tag{145}$$

To find the parameter values that maximize the log likelihood, we first take derivative with respect to $\mu$ set it to zero:

$$\frac{d}{d\mu}L(\mu,S)=\sum_{i=1}^{N}\frac{(y_i-\mu)}{\sigma_i^2+S}=0. \tag{146}$$

Solving for $\mu$:

$$\mu=\frac{\sum_{i=1}^{N}\frac{y_i}{\sigma_i^2+S}}{\sum_{i=1}^{N}\frac{1}{\sigma_i^2+S}}. \tag{147}$$

The derivative of the log likelihood with respect to $S$ gives a second equation:

$$\frac{d}{dS}L(\mu,S)=-\frac{1}{2}\sum_{i=1}^{N}\left(\frac{1}{\sigma_i^2+S}-\frac{(y_i-\mu)^2}{\left(\sigma_i^2+S\right)^2}\right)=0. \tag{148}$$

This equation cannot be solved for $S$ analytically, but we can numerically solve it together with Eqn. 147 to get estimates of $\mu$ and $S$. Table S1 shows estimates of $\mu$ and $S$ in each $(c_0,g)$ condition. Point estimates are the mean estimated parameters from 100 bootstrapped data sets of single-cell phenotypes $\{y\}$, and error bars are the standard deviations.

| $g\rightarrow$ $c_0\downarrow$ | $0^+$ mm$^{-1}$ | 0.1 mm$^{-1}$ | 0.2 mm$^{-1}$ | 0.3 mm$^{-1}$ | 0.4 mm$^{-1}$ |
|---|---|---|---|---|---|
| 0.1 μM | $\mu=-5.8\pm0.1$ $S=1.6\pm0.2$ | $\mu=-5.4\pm0.1$ $S=1.7\pm0.2$ | $\mu=-5.3\pm0.1$ $S=1.7\pm0.2$ | $\mu=-4.9\pm0.2$ $S=1.7\pm0.3$ | $\mu=-4.7\pm0.1$ $S=1.6\pm0.3$ |
| 1 μM | $\mu=-4.4\pm0.1$ $S=1.0\pm0.1$ | $\mu=-3.4\pm0.1$ $S=1.0\pm0.2$ | $\mu=-2.8\pm0.1$ $S=1.0\pm0.2$ | $\mu=-2.4\pm0.1$ $S=1.1\pm0.2$ | $\mu=-2.1\pm0.1$ $S=1.0\pm0.2$ |
| 10 μM | $\mu=-6.1\pm0.1$ $S=0.8\pm0.1$ | $\mu=-4.1\pm0.1$ $S=0.8\pm0.2$ | $\mu=-3.3\pm0.1$ $S=0.8\pm0.2$ | $\mu=-2.8\pm0.1$ $S=0.8\pm0.1$ | $\mu=-2.5\pm0.1$ $S=0.9\pm0.2$ |

**Table S1: Parameters characterizing population variability in $\eta$ from maximum-likelihood estimation.**

To estimate population variation in $\eta$, we transform the Gaussian distribution of $x=\log\left(\frac{\eta}{1-\eta}\right)$ to the distribution of $\eta=\frac{\exp(x)}{1+\exp(x)}$ (abusing notation in the sense that this $\eta$ is noise-free):

$$P(\eta)=P\big(x(\eta)\big)\left|\frac{dx(\eta)}{d\eta}\right|=\frac{1}{\eta\,(1-\eta)}\frac{1}{\sqrt{2\pi S}}\exp\left(-\frac{1}{2}\frac{\left(\log\left(\frac{\eta}{1-\eta}\right)-\mu\right)^2}{S}\right). \tag{149}$$

The larger error bars in Fig. 3C of the main text were computed from the 5% and 95% percentiles of this distribution in each $(c_0, g)$ condition.

## Information about current versus past signals encoded in kinase activity

We previously quantified the information about all past signals encoded in kinase activity, $\dot{I}_{s\to a}$, and found that *E. coli* use this information efficiently: they climb gradients at speeds near the information-performance limit (1). There are two possible inefficiencies that prevent *E. coli* from reaching the limit: first, cells might encode information about past signals, which don't contribute to gradient-climbing; and second, information about current signal can be lost in communication to the motor behavior. Now that we have an expression for the information about current signal $s(t)$ in kinase activity, we can distinguish between these two effects.

We previously defined the information about all past signals encoded in kinase activity using the following transfer entropy rate:

$$\dot{I}_{s\to a} \equiv \lim_{dt\to 0} \frac{1}{dt} I(a(t+dt); \{s\}|\{a\}) \,. \tag{150}$$

The subset of this information that is relevant to chemotaxis is:

$$\dot{I}^{*}_{s\to a} \equiv \lim_{dt\to 0} \frac{1}{dt} I(a(t+dt); s(t)|\{a\}) \,, \tag{151}$$

which is the information we have considered here. How do these information rates compare to each other for the kinase response function and noise correlation function that we measured here and previously?

First, note that if kinase activity $a$ were Markovian in $s(t)$, then we would have

$$\begin{aligned} \dot{I}_{s\to a} &= \lim_{dt\to 0} \frac{1}{dt} I(a(t+dt); \{s\}|\{a\}) \\ &= \lim_{dt\to 0} \frac{1}{dt} I(a(t+dt); s(t)|\{a\}) \\ &= \dot{I}^{*}_{s\to a}, \end{aligned} \tag{152}$$

and all information about signals encoded in kinase activity is relevant to gradient climbing. Surprisingly, this means that a long response adaptation time does not necessarily degrade information about the current signal.

We can evaluate both of these information rates for the response and noise models used here. In the regime of shallow gradients and $\tau_2 \approx \tau_n$, the information about past and present signals is (1,11):

$$\dot{I}_{s\to a} \approx \frac{1}{4\pi} \int_{-\infty}^{\infty} \frac{S(\omega)\frac{r_0^2}{\omega^2}|K_r(\omega)|^2}{N_n(\omega) + r_0\,|K_r(\omega)|^2}\, d\omega \tag{153}$$

$$= \frac{\frac{G_r^2}{\tau_1^2}\, r_0^2\, g^2\, \sigma_v^2\, \tau_3^2}{4\, D_n \left(1 + \frac{\tau_3}{\tau_v}\sqrt{1 + \frac{G_r^2}{\tau_1^2}\frac{r_0}{2\, D_n}}\right)} = \frac{1}{\tau_v}\frac{1}{4}\frac{R\, \gamma_r \left(\frac{\tau_3}{\tau_v}\right)^2}{\left(1 + \frac{\tau_3}{\tau_v}\sqrt{1+R}\right)} \tag{154}$$

which we have expressed in terms of the ratio of the diffusivity of filtered particle noise and the diffusivity of slow kinase noise, $R = \frac{G_r^2}{\tau_1^2}\frac{r_0}{2\, D_n}$; the particle arrival signal-to-noise ratio, $\gamma_r = 2\, r_0\, g^2\, \sigma_v^2\, \tau_v^3$; and $\tau_3^{-1} = \tau_1^{-1} + \tau_2^{-1}$.

We compare this to the information about current signal only, derived in the previous section, Eqn. 110, reproduced below:

$$\dot{I}_{s\to a}^{*} = \frac{1}{\tau_v}\frac{1}{4}\frac{R}{1+R}\frac{\gamma_r}{\left(1 + \frac{\tau_v}{\tau_3}\frac{1}{\sqrt{1+R}}\right)^2} \tag{155}$$

The ratio of these two information rates has a simple form:

$$\frac{\dot{I}_{s\to a}^{*}}{\dot{I}_{s\to a}} \approx \frac{\tau_v}{\tau_3\sqrt{1+R} + \tau_v} = \frac{\tau_v}{\tau_3\sqrt{1 + \frac{G_r^2}{\tau_1^2}\frac{r_0}{2\, D_n}} + \tau_v}. \tag{156}$$

Thus, for $\dot{I}_{s\to a}$ to mostly carry information about current signal and be close to $\dot{I}_{s\to a}^{*}$, 1) the time scale of initial kinase response must be short compared to the signal correlation time, $\tau_1 \ll \tau_v$; and 2) the diffusivity of filtered particle noise must be small compared to that of internal kinase noise, $G_r^2\, r_0 \ll 2\, D_n$. Using $\tau_1 = 1/60\ s$ from biochemistry studies Refs. (25,26), we estimate that $\frac{\dot{I}_{s\to a}^{*}}{\dot{I}_{s\to a}} \approx 0.88 \pm 0.01$ in $c_0 = 1\ \mu\text{M}$, and increases as $c_0$ gets larger or smaller. This suggests that *E. coli*'s main source of "inefficiency" is that relevant information in kinase activity is lost in communication with the motors.

This result might appear to be in contradiction with the results of Ref. (16), which found that the fraction of predictive information about signals compared to past information about signals was very small (about 1%) in a model of *E. coli*'s kinase activity, $a$, and downstream readout molecules, $x$ (CheYp). Our $\dot{I}_{s\to a}^{*}$, being a predictive information rate, is very similar to their predictive information, while $\dot{I}_{s\to a}$ is very similar to their past information. However, that study considered predictive and past information encoded in the *current* value of the readout molecule, $x(t)$, instead of the entire history of readout molecules $\{x\}$. This difference in how our information quantities are defined explains the large difference in result. Kinase activity $a$ and even CheY phosphorylation level, $x$, are not the final outputs of the chemotaxis system. Downstream pathway dynamics can act on the entire past of $a$ or $x$ to extract information and make behavioral decisions (tumble rate). Therefore, the current values of $a(t)$ and $x(t)$ do not need to be faithful estimates of the current (or future) signal $s(t)$; they just need to carry decodable information about $s(t)$ in their past trajectories. Our information measures above account for this.

## Summary of information inequalities and results

In summary, we have two sets of inequalities. The first set of inequalities,

$$\dot{I}_{s\to a} \geq \dot{I}^*_{s\to a} \geq \dot{I}^*_{s\to m} \propto \left(\frac{v_d}{v_0}\right)^2, \quad (157)$$

was the focus of our previous work (1), and it quantifies how efficiently *E. coli* use the information *that they have* at the level of kinase activity, $\dot{I}_{s\to a}$, to climb gradients. The main result of that work was that $\dot{I}_{s\to a} \approx 2\,\dot{I}^*_{s\to m}$. The analysis above adds to this: $\dot{I}_{s\to a} \approx \dot{I}^*_{s\to a} \approx 2\,\dot{I}^*_{s\to m}$.

The second set of inequalities,

$$\dot{I}^*_{s\to r} \geq \dot{I}^*_{s\to a} \geq \dot{I}^*_{s\to m} \propto \left(\frac{v_d}{v_0}\right)^2, \quad (158)$$

particularly the left-most one, is the focus of this work. It quantifies how much information *E. coli* get compared to the physical limit. The main result of this manuscript is that $\dot{I}^*_{s\to r} \gg \dot{I}^*_{s\to a}$.

## Appendix A: Causal Wiener filter derivation

Causal Wiener filtering theory seeks a linear estimator of an unknown quantity $s(t+\tau)$ at time $\tau$ in the future, from past observations of a quantity $x$ that is correlated with $s$ (36). The past of $x$ is denoted $\{x(t)\}$. Both $s$ and $x$ are assumed to be stationary stochastic processes with zero means: $\langle x(t)\rangle = \langle s(t)\rangle = 0$. The Wiener filter, $M_x(T)$, is the kernel that minimizes the mean squared error of the estimator:

$$M_x(T) = \underset{K(T)}{\operatorname{argmin}}\ \langle e^2(\tau)\rangle = \underset{K(T)}{\operatorname{argmin}} \left\langle \left( s(t+\tau) - \int_{-\infty}^{t} K(t-t')\,x(t')\,dt' \right)^2 \right\rangle. \quad (159)$$

In general, the estimator of $s(t+\tau)$ that minimizes the mean squared error is the conditional mean $\langle s(t+\tau)|\{x(t)\}\rangle$. In the case of Gaussian-distributed $s$ and $x$, the conditional mean $\langle s(t+\tau)|\{x(t)\}\rangle$ is exactly a linear function of $\{x(t)\}$, so the linear estimator above is the global optimum. The minimum error $\langle e^*(\tau)^2\rangle = \sigma^2_{s|x}(\tau)$ is the conditional variance of $s(t+\tau)$ given past $x$. The main technical challenge of finding the optimal kernel is the constraint that it must be causal: $M_x(T) = 0$ for $T < 0$.

To derive the optimal kernel, first we expand the square in the objective function:

$$\langle e^2(\tau)\rangle = \left\langle s(t+\tau)^2 - 2\,s(t+\tau)\int_{-\infty}^{\infty} K(t-t')\,x(t')\,dt' + \int_{-\infty}^{\infty} K(t-t')\,x(t')\,dt' \int_{-\infty}^{\infty} K(t-t'')\,x(t'')\,dt'' \right\rangle, \quad (160)$$

and move the expectation inside of the integrals:

$$= \sigma_s^2 - 2\int_{-\infty}^{\infty} K(t-t')\,\langle s(t+\tau)\,x(t')\rangle\,dt' + \int_{-\infty}^{\infty}\int_{-\infty}^{\infty} K(t-t')K(t-t'')\,\langle x(t')\,x(t'')\rangle\,dt'\,dt''\,. \quad (161)$$

Here we used time-translation invariance of $s$: $\langle s(t+\tau)^2\rangle = \langle s(t)^2\rangle = \sigma_s^2$. Next, change variables to $t' \to \tau' = t - t'$ and $t'' \to \tau'' = t - t''$, replacing absolute time with time delays. $\tau'$ and $\tau'' > 0$ correspond to time delay into the past.

$$= \sigma_s^2 - 2\int_{-\infty}^{\infty} K(\tau')\,\langle s(t+\tau)\,x(t-\tau')\rangle\,d\tau' + \int_{-\infty}^{\infty}\int_{-\infty}^{\infty} K(\tau')K(\tau'')\,\langle x(t-\tau')\,x(t-\tau'')\rangle\,d\tau'\,d\tau''. \qquad (162)$$

Defining the cross-correlation function $C_{xs}(t-t') = \langle x(t')\,s(t)\rangle$ and autocorrelation function $C_x(t-t') = \langle x(t')\,x(t)\rangle$:

$$= \sigma_s^2 - 2\int_{-\infty}^{\infty} K(\tau')\,C_{xs}(\tau'+\tau)\,d\tau' + \int_{-\infty}^{\infty}\int_{-\infty}^{\infty} K(\tau')K(\tau'')\,C_x(\tau''-\tau')\,d\tau'\,d\tau''. \qquad (163)$$

When $\tau > 0$ in $C_{xy}(\tau)$, $y$ is evaluated at a time point in the future relative to $x$. Note that $C_{sx}(\tau) = C_{xs}(-\tau)$.

Next, we take the functional derivative of the mean squared error with respect to $K(T)$:

$$\frac{\delta\langle e^2(\tau)\rangle}{\delta K} = -2\,C_{xs}(\tau'+\tau) + \int_{-\infty}^{\infty} K(\tau'')\,C_x(\tau''-\tau')\,d\tau'' + \int_{-\infty}^{\infty} K(\tau'')\,C_x(\tau'-\tau'')\,d\tau'' \qquad (164)$$

Since $C_x(\tau) = C_x(-\tau)$, this is:

$$= -2\,C_{xs}(\tau'+\tau) + 2\int_{-\infty}^{\infty} K(\tau'')\,C_x(\tau''-\tau')\,d\tau''. \qquad (165)$$

Now we need to consider the causal constraint on $K(T)$. For optimality with this constraint, the equation above must equal zero for $\tau' \geq 0$ (at times when $x$ precedes $s$ in $C_{xs}$). Otherwise, for $\tau' < 0$, the derivative is not necessarily zero. Therefore, at the optimum we can write (37):

$$\int_{-\infty}^{\infty} M_x(\tau'')\,C_x(\tau'-\tau'')\,d\tau'' - C_{xs}(\tau'+\tau) = A(\tau') \qquad (166)$$

where

$$A(\tau') = \begin{cases} 0, & \tau' \geq 0 \\ a(\tau'), & \tau' < 0 \end{cases} \qquad (167)$$

and $a(\tau')$ is some unspecified function. $A(\tau')$ is therefore anti-causal – it is only non-zero at times in the future ($\tau' < 0$). At first glance, this optimality condition might seem less constrained than if $A(\tau')$ were zero for all $\tau'$ (the optimality condition for the optimal non-causal filter). However, the fact that $A(\tau')$ is nonzero for $\tau' < 0$ actually limits the space of filters $M_x(\tau)$ that keep $A(\tau') = 0$ for $\tau' \geq 0$, as we will see below.

Next, we take the Fourier transform of both sides, defined as $f(\omega) = F[f(t)] = \int_{-\infty}^{\infty} f(t)\,e^{i\,\omega\,t}\,dt$. Convolutions in the time domain become element-wise products in the Fourier domain:

$$M_x(\omega)\,C_x(\omega) = C_{xs}(\omega)\,e^{-i\,\omega\,\tau} + A(\omega). \qquad (168)$$

On the left-hand side, we have the product of a causal function and a function that is nonzero for positive and negative time delays, the result of which is also nonzero for positive and negative time delays. On the right-hand side, we have a function that is nonzero for positive and negative time delays and an anti-causal function. How do we get the optimal causal kernel $M_x(\omega)$ out of this?

Naively, one might divide both sides by $C_x(\omega)$ and then multiply element-wise by a Heaviside step function in the time domain to get a causal kernel $M_x(\omega)$. However, although the resulting kernel is causal, it does not satisfy the optimality condition. Plugging that kernel back into Eqn. 168, it multiplies the non-causal $C_x(\omega)$, and the result is non-causal. Thus, $A(\omega)$ is non-causal, so that kernel does not satisfy the optimality condition, $A(\tau') = 0$ for $\tau' \geq 0$.

Instead, we need to split $C_x(\omega)$ into causal and anti-causal parts, called a spectral factorization or Wiener-Hopf factorization (17–19):

$$C_x(\omega) = \phi(\omega)\,\phi^*(\omega), \tag{169}$$

where $\phi(\omega)$ is a causal function in the time domain and its complex conjugate $\phi^*(\omega)$ is anti-causal. $\phi(\omega)$ is constructed by putting all poles and zeros of $C_x(\omega)$ with negative real part into $\phi(\omega)$ and those with positive real part into $\phi^*(\omega)$.

Plugging this into the optimality condition:

$$M_x(\omega)\,\phi(\omega)\,\phi^*(\omega) = C_{xs}(\omega)\,e^{-i\,\omega\,\tau} + A(\omega) \tag{170}$$

$$M_x(\omega)\,\phi(\omega) = \frac{C_{xs}(\omega)}{\phi^*(\omega)}\,e^{-i\,\omega\,\tau} + \frac{A(\omega)}{\phi^*(\omega)}. \tag{171}$$

The left-hand side is now a causal function in the time domain, being the product of causal functions, and the right-hand side contains a non-causal function and an anti-causal function.

Multiplying both sides of Eqn. 171 by a Heaviside function in the time domain and then transforming back to Fourier space eliminates the anti-causal term $\frac{A(\omega)}{\phi^*(\omega)}$ and leaves the left-hand side unaffected:

$$M_x(\omega)\,\phi(\omega) = \left[\frac{C_{xs}(\omega)}{\phi^*(\omega)}e^{-i\,\omega\,\tau}\right]^+. \tag{172}$$

Now the right-hand side is causal, and dividing by $\phi(\omega)$ gives the optimal causal filter:

$$M_x(\omega) = \frac{1}{\phi(\omega)}\left[\frac{C_{xs}(\omega)}{\phi^*(\omega)}e^{-i\,\omega\,\tau}\right]^+. \tag{173}$$

To check that this filter satisfies the optimality condition (Eqn. 168), we can plug it in:

$$\frac{1}{\phi(\omega)}\left[\frac{C_{xs}(\omega)}{\phi^*(\omega)}e^{-i\,\omega\,\tau}\right]^+\ \phi(\omega)\,\phi^*(\omega) = C_{xs}(\omega)\,e^{-i\,\omega\,\tau} + A(\omega) \tag{174}$$

$$\left[\frac{C_{xs}(\omega)}{\phi^*(\omega)}e^{-i\,\omega\,\tau}\right]^+\ \phi^*(\omega) = C_{xs}(\omega)\,e^{-i\,\omega\,\tau} + A(\omega) \tag{175}$$

$$\left[\frac{C_{xs}(\omega)}{\phi^*(\omega)}e^{-i\,\omega\,\tau}\right]^+ = \frac{C_{xs}(\omega)}{\phi^*(\omega)}\,e^{-i\,\omega\,\tau} + \frac{A(\omega)}{\phi^*(\omega)}. \tag{176}$$

Now the left-hand side is the causal part of the first term on the right-hand side. Therefore, their difference is anti-causal and $A(\omega)$ is thus anti-causal, as desired:

$$A(\omega) = -\phi^*(\omega)\left[\frac{C_{xs}(\omega)}{\phi^*(\omega)} e^{-i\,\omega\,\tau}\right]^- . \tag{177}$$

At the optimum, the mean square error $\langle e^2(\tau)\rangle = \sigma_{s|x}^2(\tau)$ is:

$$\sigma_{s|x}^2(\tau) = \sigma_s^2 - 2\int_0^\infty M_x(\tau')\, C_{xs}(\tau'+\tau)\, d\tau' + \int_0^\infty \int_0^\infty M_x(\tau')\, M_x(\tau'')\, C_x(\tau''-\tau')\, d\tau'\, d\tau'' , \tag{178}$$

where we have set the lower limit to zero because the kernel $M_x(T)$ is zero for $T < 0$. Using the optimality condition $\int_0^\infty M_x(\tau'')\, C_x(\tau''-\tau')\, d\tau'' - C_{xs}(\tau'+\tau) = 0$ for $\tau' \geq 0$, we get:

$$= \sigma_s^2 - \int_0^\infty M_x(\tau')\, C_{xs}(\tau'+\tau)\, d\tau' . \tag{179}$$

Since $C_{xs}(\tau) = C_{sx}(-\tau)$, this is:

$$= \sigma_s^2 - \int_{-\infty}^\infty M_x(\tau')\, C_{sx}(-\tau-\tau')\, d\tau' , \tag{180}$$

which is the convolution of $M_x(T)$ and $C_{sx}(T)$, with the result evaluated at $-\tau$. This can be expressed using their Fourier transforms (note the minus sign in front of tau in equation (112) leads to a plus sign in the exponent below) as:

$$\sigma_{s|x}^2(\tau) = \sigma_s^2 - \frac{1}{2\pi}\int_{-\infty}^\infty M_x(\omega)\, C_{sx}(\omega)\, e^{i\,\omega\,\tau}\, d\omega . \tag{181}$$

Plugging in the optimal kernel:

$$= \sigma_s^2 - \frac{1}{2\pi}\int_{-\infty}^\infty \frac{C_{sx}(\omega)}{\phi(\omega)}\left[\frac{C_{xs}(\omega)}{\phi^*(\omega)} e^{-i\,\omega\,\tau}\right]^+ e^{i\,\omega\,\tau}\, d\omega . \tag{182}$$

Finally, the correlation coefficient is:

$$\rho_{xs}^2(\tau) = 1 - \frac{\sigma_{s|x}^2(\tau)}{\sigma_s^2} = \frac{1}{\sigma_s^2}\frac{1}{2\pi}\int_{-\infty}^\infty \frac{C_{sx}(\omega)}{\phi(\omega)}\left[\frac{C_{xs}(\omega)}{\phi^*(\omega)} e^{-i\,\omega\,\tau}\right]^+ e^{i\,\omega\,\tau}\, d\omega . \tag{183}$$

## Supplementary Figures

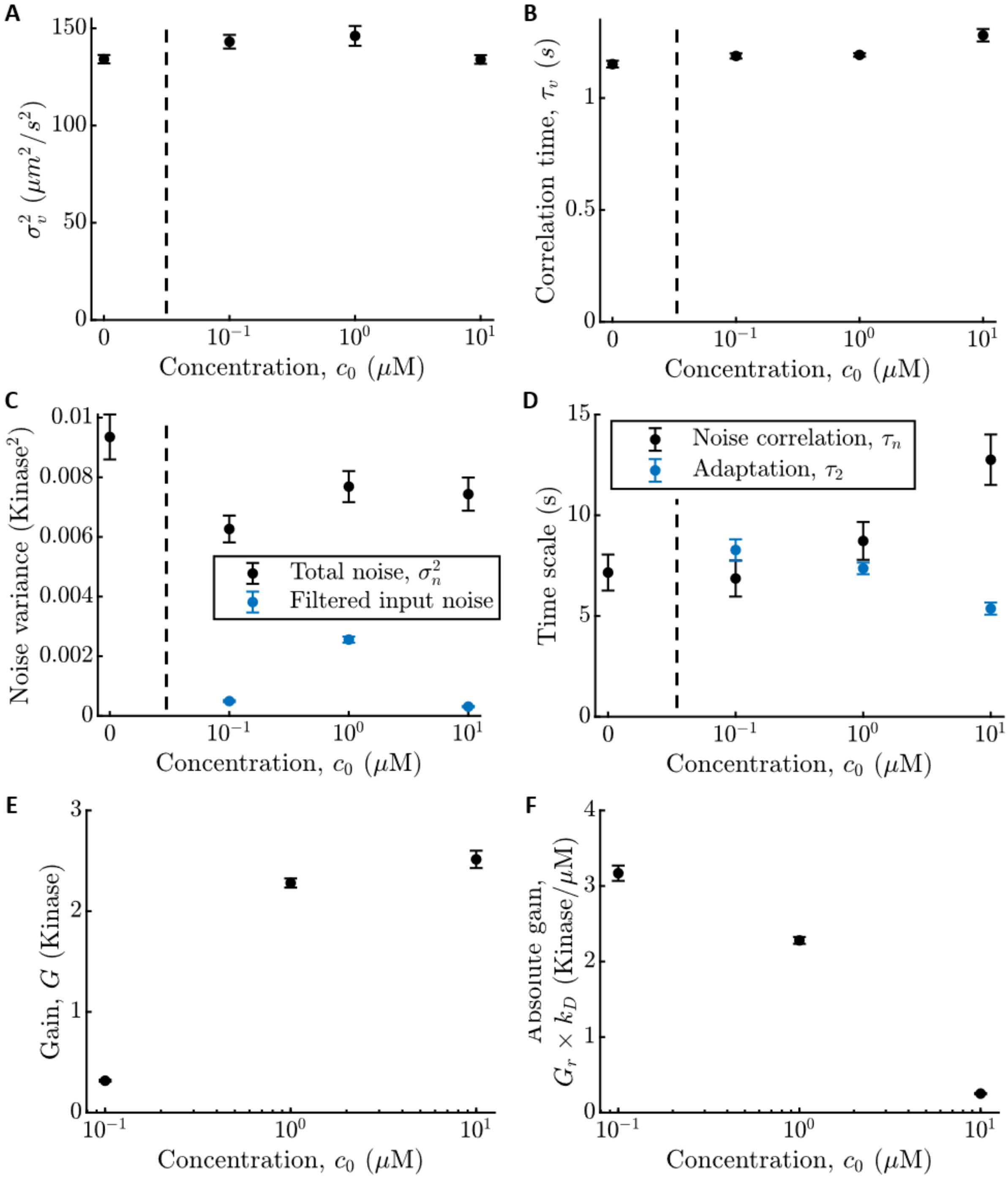

**Figure S1: Measured signal, response, and noise parameter values in different background concentrations. A)** Estimated variance of up-gradient velocity, $\sigma_v^2$, as a function of background $c_0$, which together with the gradient steepness $g$ sets the signal strength. Horizontal axes are on log-scale, and vertical dashed lines throughout separate parameters measured at $c_0 = 0$ from those measured at finite

$c_0$. Error bars throughout are standard error of the mean (see Methods section of the main text). **B)** Correlation time of up-gradient velocity, $\tau_v$, which sets the signal correlation time. Parameters in (A) and (B) are those of the median phenotype in Fig. S2, with tumble bias $TB \approx 0.09$. **C)** Variance of the total noise in kinase activity, $\sigma_n^2$ (black), and the estimated variance of particle arrival noise filtered through the kinase response kernel (blue) with $\tau_1 = 1/60\ s$ (25,26). **D)** Kinase noise correlation time, $\tau_n$, and kinase response adaptation time, $\tau_2$ (blue). **E)** Gain of kinase response to signal or log-concentration, $G$. **D)** Gain of kinase response to absolute concentration, $G_c = k_D\ G_r = G/c_0$, where $G_r$ is the gain of kinase responses to particle arrival rate.

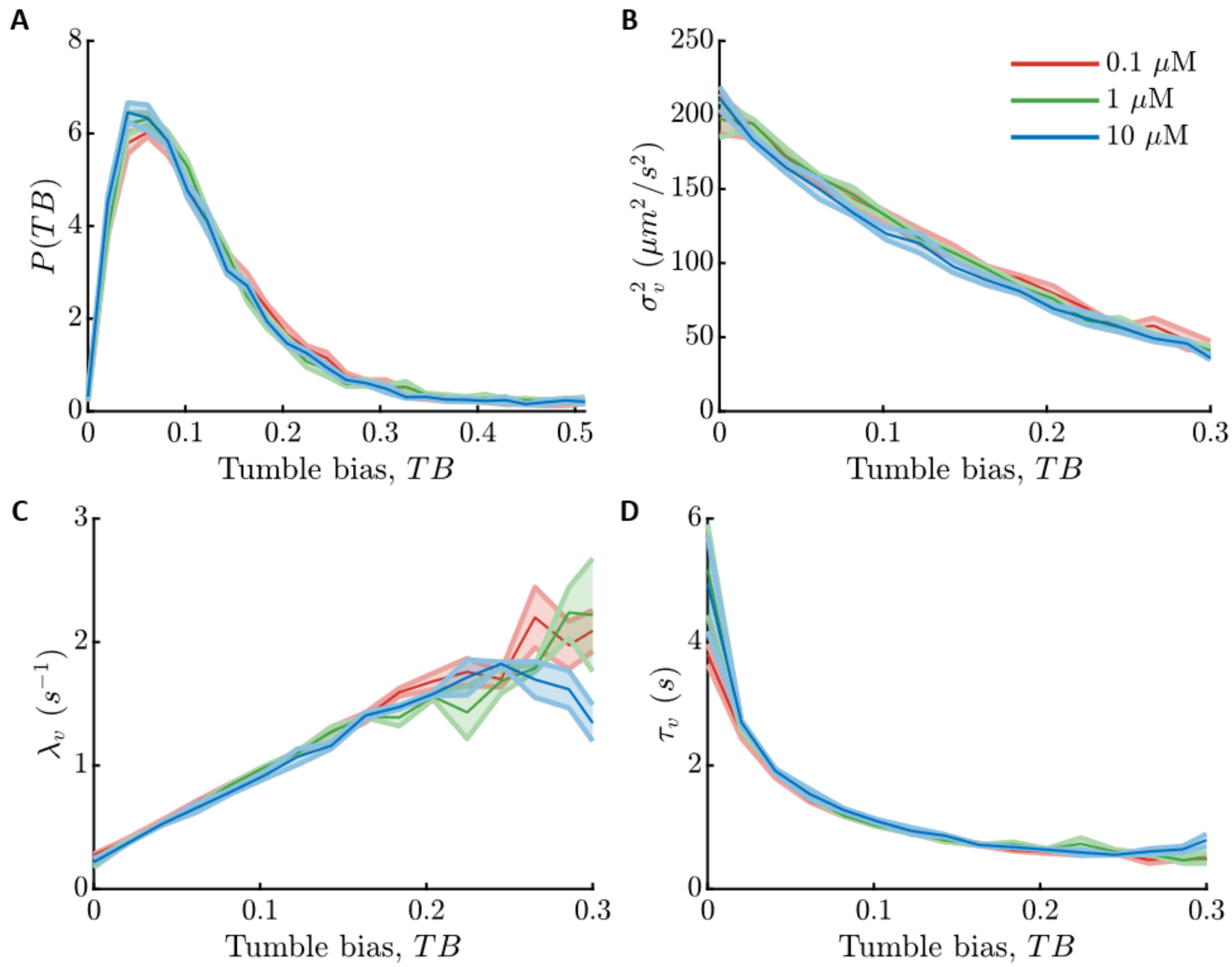

**Figure S2: Swimming parameters as a function of tumble bias in different background concentrations.** **A)** Distribution of tumble bias, $TB = 1 - P_{run}$, or fraction of time cells spend in the tumble state, among cells in an isogenic population. Throughout: red is $c_0 = 0.1$ μM, green is $c_0 = 1$ μM, and blue is $c_0 = 10$ μM. Shading is standard error of the mean (Methods). **B)** Variance of up-gradient velocity, $\sigma_v^2$, versus tumble bias, $TB$. **C)** Velocity decorrelation rate, $\lambda_v = \tau_v^{-1} \approx (1-\alpha)\,\lambda_{R0} + 2\,D_r$, versus $TB$. $\alpha$ quantifies how correlated heading is before and after a tumble; $\lambda_{R0}$ is the average tumble rate; and $D_r$ is the rotational diffusion coefficient (1). **D)** Velocity correlation time, $\tau_v$.

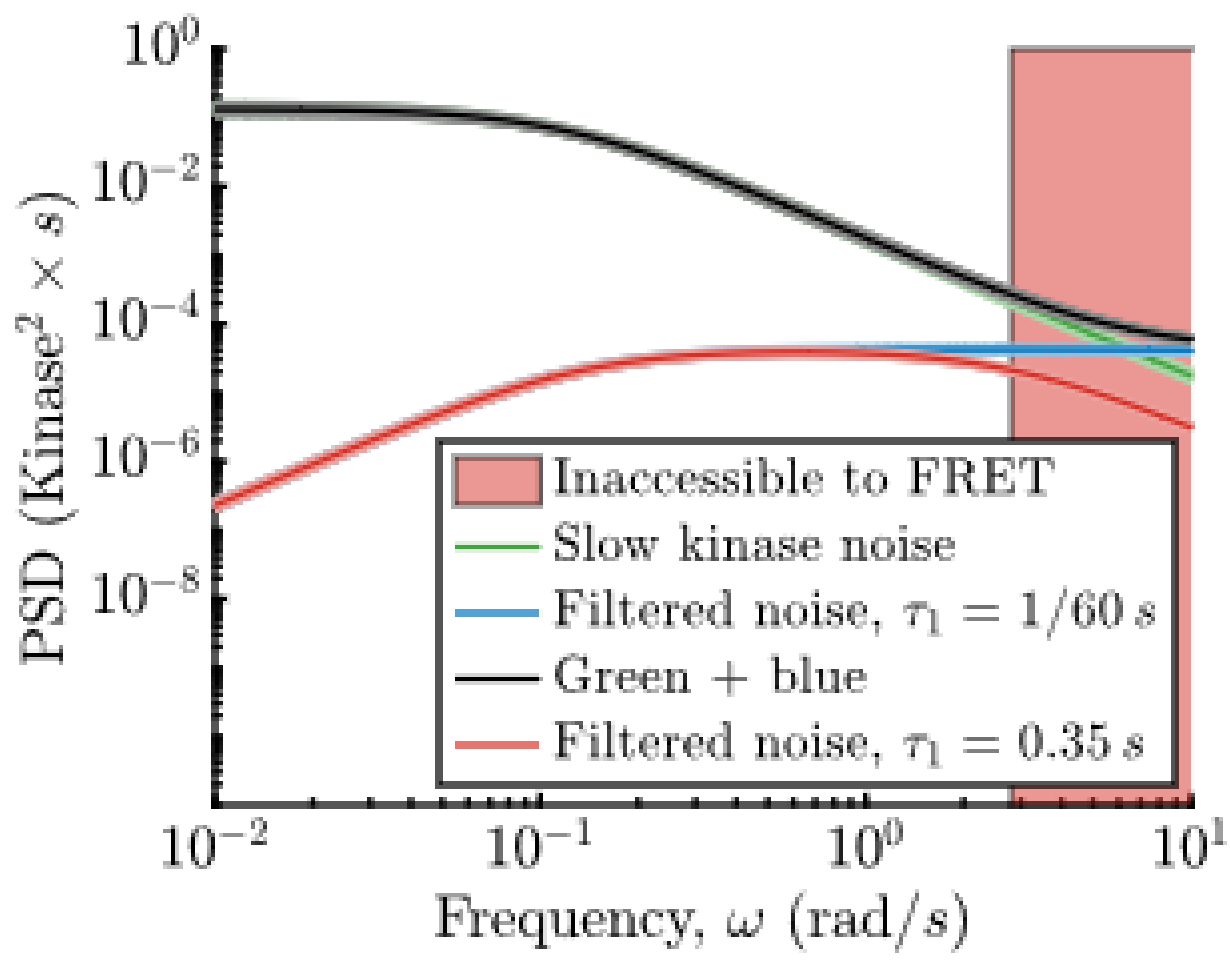

**Figure S3: Noise power spectra.** In frequency space, kinase responses to particle arrivals implies that the noise in kinase activity must be larger than filtered particle arrival noise (blue, using $\tau_1 = 1/60\ s$ from biochemistry studies Refs. (25,26)). At low frequencies where we can measure noise and responses with our FRET system (green), this bound is far from saturated. Naively extrapolating to higher frequencies (red shaded region, marked by the value of $1/\tau_1$ measured in FRET experiments) violates this bound (the green line goes below the blue line). This implies either additional noise at high frequencies that is not captured by a single exponential (black line is slow noise, green, plus filtered particle noise, blue) or a slower kinase response time $\tau_1$ (red line is filtered particle noise with $\tau_1 \approx 0.35\ s$ measured in FRET experiments), which could be a necessary by product of the coupling between kinases that creates large gain (thus raising the red line) but also slows down the response. The behaviorally-relevant information rates computed in the main text are insensitive to these choices.

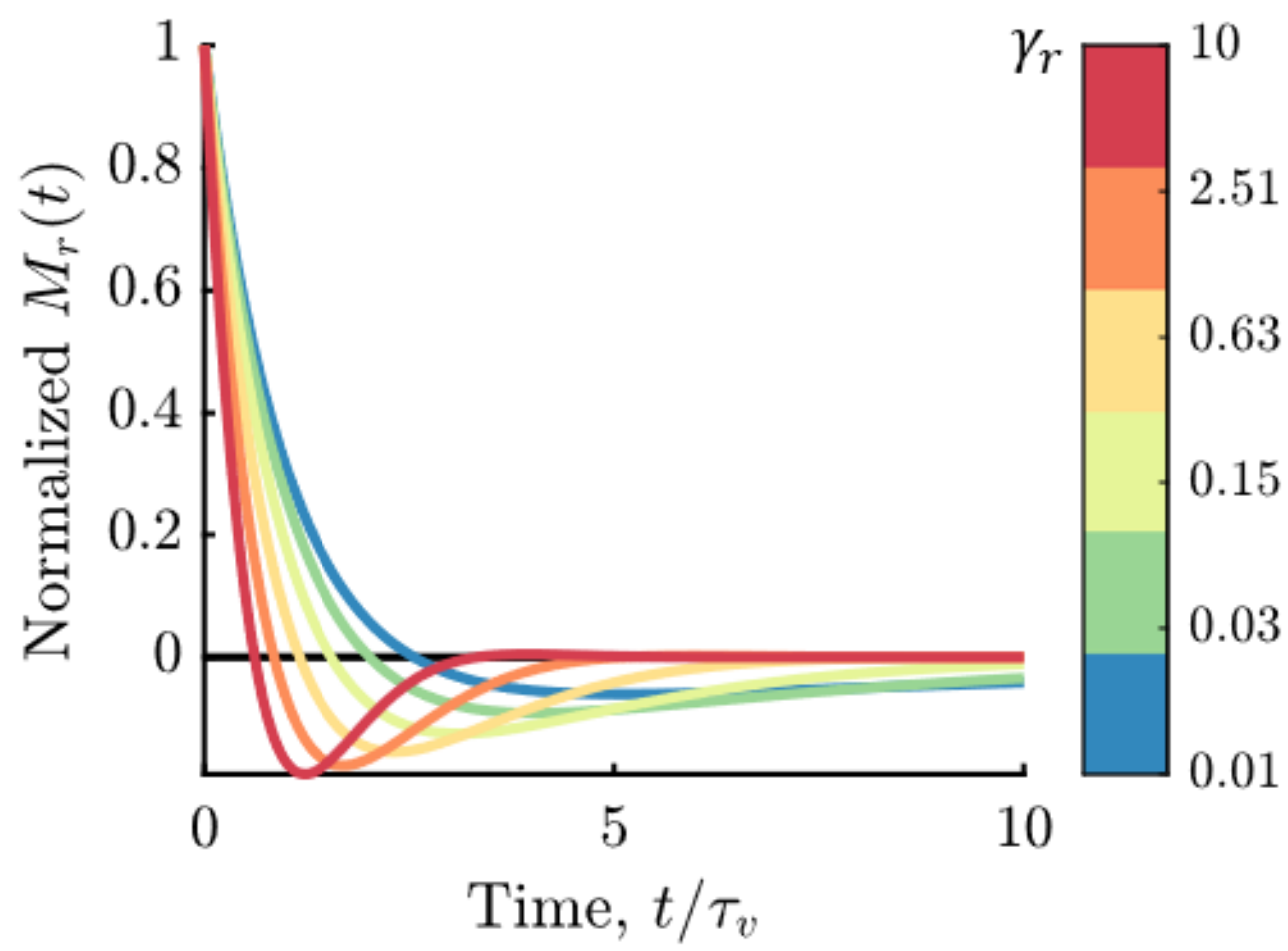

**Figure S4: Optimal kernel for inferring current signal, $s(t)$, from past particle arrivals, $r$.** Colors indicate different values of the signal-to-noise ratio $\gamma_r$, marked on the right. Each kernel is normalized so that $M_r(0) = 1$.

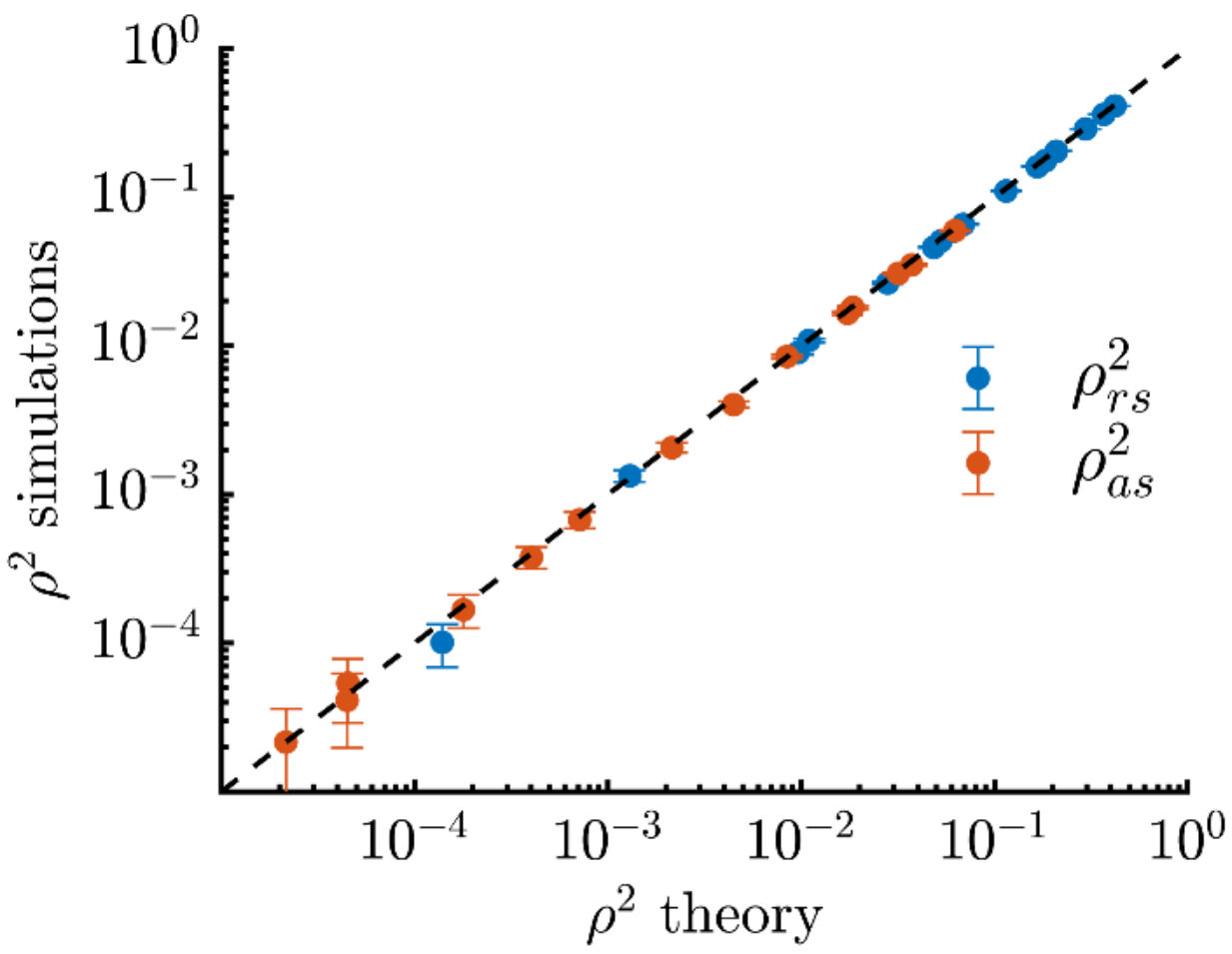

**Figure S5: Squared Pearson correlation coefficients, $\rho^2_{rs}$ and $\rho^2_{as}$, between true signals $s(t)$ and optimal estimates, $\hat{s}_r(t)$ and $\hat{s}_a(t)$, computed from simulations and theory.** Dots are varying $c_0 \in \{0.1, 1, 10\}$ μM and $g \in \{0.01, 0.1, 0.2, 0.3, 0.4\}$ $\mathrm{mm}^{-1}$. Simulations are described in SI section **Simulation details**. Values of the squared correlation coefficients range from $10^{-4.6}$ to $0.42$. Error bars are SEMs and were computed by bootstrapping trajectories. The theoretical expressions show excellent agreement with the correlations computed from simulations.

## Supplementary References